\documentclass[aps,prx,floats,floatfix,superscriptaddress,nofootinbib,twocolumn,showkeys,longbibliography]{revtex4-2}
\usepackage{bbm}
\usepackage{amsmath,amssymb,amsthm}
\usepackage{xcolor}
\usepackage{graphicx}
\usepackage{caption}
\usepackage{url}
\usepackage{answers}
\usepackage[mathscr]{euscript}
\usepackage{eufrak}
\usepackage{calligra}
\usepackage{algpseudocode,algorithm}
\usepackage{subcaption}

\usepackage{amsthm}

\newtheorem{theorem}{Theorem}

\newtheorem{proposition}[theorem]{Proposition}

\usepackage[colorlinks=true, linkcolor=blue, citecolor=blue, urlcolor=blue]{hyperref}

\usepackage{bm}%
\usepackage{booktabs}

\newcommand{\ZZZ}[2]{}

\begin{document}

\title{\textbf{Where are the shortest-path nodes in spatial networks?} 
}%

\author{Xinhan~Liu}
\email[Contact author: ]{X.Liu-22@tudelft.nl}
\affiliation{
Faculty of Electrical Engineering, Mathematics and Computer Science,
Delft University of Technology,
2600 GA Delft, The Netherlands}

\author{Robert~Kooij}
\email[Contact author: ]{R.E.Kooij@tudelft.nl}
\affiliation{
Faculty of Electrical Engineering, Mathematics and Computer Science, 
Delft University of Technology, 
2600 GA Delft, The Netherlands}
\affiliation{
Unit ICT, Strategy and Policy, TNO, The Netherlands}

\author{Piet~Van~Mieghem}
\affiliation{
Faculty of Electrical Engineering, Mathematics and Computer Science, 
Delft University of Technology, 
2600 GA Delft, The Netherlands}

\date{\today}%

\begin{abstract}
To predict the spatial distribution of shortest-path nodes, we propose
a Gaussian estimate of the shortest-path node probability
\(\pi_{uv}(w)\), the probability that a node \(w\) lies on a shortest
path between a source \(u\) and a destination \(v\). According to the
Gaussian estimate, three variables affect the shortest-path node
probability \(\pi_{uv}(w)\) and combine into one variable, the
standardized detour \(X_{uvw}\). In our
simulations of hard and soft random geometric graphs in Euclidean,
hyperbolic and spherical space, the standardized detour \(X_{uvw}\)
alone approximately determines how fast the shortest-path node
probability \(\pi_{uv}(w)\) decreases as the node \(w\) moves away from the
geodesic between the source \(u\) and the destination \(v\). We find and
explain that,
when the hopcount variance \(\sigma^2(H_{uv})\) is very small, the
shortest-path nodes have a discrete spatial distribution, a quantized
distribution of shortest-path nodes. We further extend
the study to spatially inhomogeneous networks, for which we propose a method to
find the shortest-path nodes.
\end{abstract}

\maketitle
\section{Introduction}
\label{sec.introduction}

After the wooden horse had tricked the gates of Troy open, Odysseus
spent ten years returning to Ithaca, a mere five hundred kilometers
away in a straight line. When he finally returned, his beautiful queen
was no longer young, his son had grown up without him and
108 suitors were consuming his wealth~\cite{homer1900odyssey}. His
journey offers a memorable reminder that knowing the destination does
not make the way easy to find.

In many communication networks, the number of hops rather
than the geometric length sets the cost of a path. In a wireless
multihop network, for example, a signal crosses a link of a few hundred
meters in about one microsecond, whereas every relay node needs a much
longer time to receive, process and forward the
packet~\cite{kurose2021computer,stamatiou2014delay}. The delay along a
path thus grows with the number of hops. Every hop also costs
transmission energy and adds a chance to lose the packet. Routing
protocols such as RIP~\cite{rfc2453} and OLSR~\cite{rfc3626} choose
paths with the fewest hops.

With the full map in hand, algorithms such as Dijkstra's~\cite{dijkstra1959note} find the
shortest path. Complete information about a network is often
unavailable, however. Links may be missing or unobserved. The network may also be too
large for a search between every source and destination. One would
then still like to know which nodes are likely to lie on a shortest
path, for example to place relay nodes or to restrict the search for a
path to a small region of the network~\cite{ko2000location}. Finding paths without the full map is
a well-known problem. Papadimitriou and
Yannakakis~\cite{papadimitriou1991shortest} studied shortest paths on a
map that is revealed only during the trip. Here we ask three questions. How likely is a node at a given position to
lie on a shortest path between a given source and destination? What
controls that likelihood? How can one choose a small region that contains
most of the shortest-path nodes?

Much of the previous work concerns the length of shortest paths. In
hard and soft random geometric
graphs~\cite{gilbert1961random,penrose2003random,dettmann2016random,penrose2016connectivity},
the hopcount \(H_{uv}\), the number of links on a shortest path between
two nodes \(u\) and \(v\), grows linearly with the distance \(d_{uv}\)
between the two nodes. Shape theorems and explicit bounds make the linear
growth precise for the hopcount~\cite{yao2011large,diaz2016relation} and
for the Euclidean length of shortest paths~\cite{hirsch2015first}. The
hopcount \(H_{uv}\) grows at least linearly when long links are
rare~\cite{luchtrath2024all} and as a power of the logarithm of the
distance \(d_{uv}\) in long-range percolation~\cite{biskup2004scaling}.
Betweenness centrality counts the shortest paths through a node over all
source--destination
pairs~\cite{freeman1977betweenness,brandes2001faster,barthelemy2004betweenness,kirkley2018betweenness},
whereas the deviation of shortest paths from the straight
line~\cite{kartun2019shape} and the hopcount
distribution~\cite{vanmieghem2014performance} describe the shortest path
as a whole. No such result gives the shortest-path node probability
\(\pi_{uv}(w)\), the probability that a node \(w\) at a given position
lies on at least one shortest path between a source \(u\) and a
destination \(v\).

Several studies relate shortest paths to the geometry of the network.
In spatial networks~\cite{barthelemy2011spatial,barthelemy2018morphogenesis},
nearby nodes are more likely to be linked, so that the node positions
carry information about where shortest paths go. Node coordinates help
to find short paths from local
information~\cite{kleinberg2000navigation,boguna2009navigability,boguna2010sustaining,krioukov2010hyperbolic,boguna2021network,serrano2022shortest}.
Kitsak et al.~\cite{kitsak2023finding} found shortest-path nodes in
substantially incomplete networks with a hyperbolic embedding. We have
shown~\cite{qiu2026geometric} that shortest paths in soft random
geometric graphs align along the geodesic between their end nodes. In
scale-free hyperbolic network models, the length of the projection of
shortest paths agrees with the geodesic distance only when the power-law
exponent of the degree distribution is close to
two~\cite{cannistraci2022geometrical}. Closest to the present work,
Verbavatz and Barthelemy~\cite{verbavatz2022betweenness} expressed the
betweenness centrality of dense spatial networks through the probability
that a node lies on a shortest path weighted by Euclidean length. The
shortest-path node probability \(\pi_{uv}(w)\) depends on the position
of the node \(w\), on the distance \(d_{uv}\) and on the hopcount
variance \(\sigma^2(H_{uv})\). A description that combines the three
effects and compares network models and spaces with one quantity is
missing.

We consider networks in which every link has weight one, so that the
length of a path equals the hopcount, the number of links in the path.
From the statistics of the hopcount in random spatial networks, we
derive a Gaussian estimate of the shortest-path node probability
\(\pi_{uv}(w)\). The Gaussian estimate assumes that the hopcount between
two nodes has a Gaussian distribution, with a mean that grows linearly
and a variance that grows sublinearly with the distance between the two
nodes. The Gaussian estimate needs the positions of the three nodes and
the hopcount statistics. The Gaussian
estimate predicts three variables that affect the shortest-path node
probability \(\pi_{uv}(w)\): the relative geometric deviation
\(\rho_{uvw}\) of the node \(w\), the mean hopcount \(\mu(H_{uv})\) and
the variance \(V_{uvw}\) of the hopcount excess. The three variables enter
the exponent of the Gaussian estimate only through one variable, the
standardized detour \(X_{uvw}\). In our simulations of hard and soft
random geometric graphs in the Euclidean plane, the hyperbolic plane and
the sphere, each variable changes the shortest-path node probability
\(\pi_{uv}(w)\) as the Gaussian estimate predicts. Plotted against the
standardized detour \(X_{uvw}\), the normalized profiles of the
shortest-path node probability \(\pi_{uv}(w)\) from all simulated
networks and spaces approximately follow one curve. The Gaussian
estimate ignores the correlations between the hopcounts. Our simulations show
that the correlations increase the shortest-path node probability
\(\pi_{uv}(w)\) more near the source and the destination than near the
midpoint of the geodesic.

When the hopcount variance \(\sigma^2(H_{uv})\) is very small, the
shortest-path node probability \(\pi_{uv}(w)\) tends to one inside
separate regions along the geodesic and to zero outside, a phenomenon
that we call the quantized distribution of shortest-path nodes. In
spatially inhomogeneous networks, the velocity \(\nu(x)\), the distance that a
shortest path near the position \(x\) advances per hop, varies with the
position \(x\). We then count the extra hops in the travel time
\(\tau(u,v)\) instead of the distance \(d_{uv}\), as in propagation
through media of varying
speed~\cite{born1999principles,sethian1996fast,sethian1999level}. The
velocity field \(\nu(x)\) is estimated from local structure alone,
without shortest-path data. In our simulations of an inhomogeneous model
network and in our tests on 22 empirical networks, the travel-time
region \(C^\tau\), built from the travel time \(\tau(u,v)\), contains
more shortest-path nodes than the near-geodesic region \(C^E\) of the
same area, built from the distance \(d_{uv}\).

Compared with previous work, the present results change the question
from how long shortest paths are to where their nodes lie. Previous
studies give the hopcount of shortest paths or how far the paths wander
from the straight line~\cite{yao2011large,kartun2019shape}, whereas the
Gaussian estimate relates the shortest-path node probability
\(\pi_{uv}(w)\) at every position to the node positions and two
hopcount statistics. Verbavatz and
Barthelemy~\cite{verbavatz2022betweenness} fitted the probability as a
function of one angle, whereas the Gaussian estimate shows that three
variables control the probability and that one variable, the
standardized detour \(X_{uvw}\), brings the normalized profiles of hard
and soft random geometric graphs in three spaces approximately onto one
curve. Shortest paths are known to align along the
geodesic~\cite{qiu2026geometric}, whereas a small hopcount variance
turns the alignment into the quantized distribution of shortest-path
nodes, a phenomenon that rests on the integer hopcount. Kitsak
et al.~\cite{kitsak2023finding} located shortest-path nodes by their
proximity to geodesics, whereas in our simulations of an inhomogeneous
model network and in our tests on 22 empirical networks, the
travel-time region \(C^\tau\) contains more shortest-path nodes than the
near-geodesic region \(C^E\) of the same area.

Section~\ref{sec:spatial} defines shortest paths, the shortest-path node
probability \(\pi_{uv}(w)\) and the hopcount statistics of random
spatial networks. Section~\ref{sec:gspne} derives the Gaussian estimate
of the shortest-path node probability \(\pi_{uv}(w)\), explains the
probability field of the shortest-path nodes with the Gaussian estimate
and verifies the explanation in our simulations of hard and soft random
geometric graphs in the Euclidean plane, the hyperbolic plane and the
sphere. The section also analyzes the effect of the correlations between
the hopcounts \(H_{uw}\), \(H_{wv}\) and \(H_{uv}\) on the shortest-path
node probability \(\pi_{uv}(w)\). Section~\ref{sec:quantization} shows
and explains the quantized distribution of shortest-path nodes that
appears when the hopcount variance \(\sigma^2(H_{uv})\) is very small.
Section~\ref{sec:heterogeneous} studies shortest paths in spatially
inhomogeneous networks. The section proposes a method that estimates the
velocity field \(\nu(x)\) from local information, replaces the distance
\(d_{uv}\) by the travel time \(\tau(u,v)\) and compares the travel-time
region \(C^\tau\) with the near-geodesic region \(C^E\) of the same area
in an inhomogeneous model network and in 22 empirical networks.
Section~\ref{sec_conclusions} summarizes the results. The Appendixes
give the details of the simulations and the estimates.

\section{Shortest paths in random spatial networks}
\label{sec:spatial}

\subsection{Shortest paths}
\label{sec:shortest_paths}

A graph \(G=(\mathcal N,\mathcal L)\) consists of the set \(\mathcal N\)
of \(N\) nodes and the set \(\mathcal L\) of \(L\)
links~\cite[Chap.~2]{vanmieghem2023graph}. Every link has weight one.
The length of a path is the sum of the weights of the links in the
path. With unit weights, the length of a path equals the hopcount of
the path, the number of links in the path. A shortest path between two
nodes \(u\) and \(v\) is a path of minimum length between the nodes
\(u\) and \(v\). The hopcount \(h_{uv}\) is the number of links in a
shortest path between the nodes \(u\) and \(v\). We assume that the
graph \(G\) is connected. Every node \(u\) has a position \(x_u\) in a
space with a metric. The distance \(d_{uv}\) between the nodes \(u\) and
\(v\) is the geodesic distance between the positions \(x_u\) and
\(x_v\).

\subsection{Random spatial networks}
\label{sec:models}

Following Kartun-Giles et al.~\cite{kartun2019shape}, a random spatial
network (RSN) is a graph \(G=(\mathcal N,\mathcal L)\) whose nodes are
random points in a space with a metric and whose links follow a rule
that depends on the node positions. We study two
RSNs, the hard and the soft random geometric
graph (RGG)~\cite{gilbert1961random,penrose2003random}, in three metric spaces: the
Euclidean plane, the hyperbolic plane~\cite{krioukov2010hyperbolic} and
the sphere. Random geometric graphs serve as models of wireless multihop
networks~\cite{haenggi2009stochastic,dettmann2016random}. In our simulations, a fixed number \(N\) of nodes is placed
uniformly in a domain \(\Omega\). For large \(N\), these nodes look
locally like a Poisson point process~\cite{penrose2003random}. In the
hard RGG, two nodes \(u\) and \(v\) at the distance
\(d_{uv}\) are linked if \(d_{uv}\le r\), with a given connection radius
\(r\). In the soft
RGG~\cite{dettmann2016random,penrose2016connectivity}, the two nodes are
linked with a probability \(p(d_{uv})\) that is a function of the
distance \(d_{uv}\).
In this study, we use the link probability
\(p(d)=1/[1+(\alpha d)^{\beta}]\) of Ref.~\cite{qiu2026geometric}, with
the scale \(\alpha\) and the steepness \(\beta\). The function is
inspired by Fermi--Dirac statistics: the steepness \(\beta\) acts as an
inverse temperature and sets the range of the links.

\subsection{Shortest-path node probability}
\label{sec:probability}

A node \(w\) lies on at least one shortest path from a source \(u\) to
a destination \(v\) if and only if
\begin{equation}
h_{uw}+h_{wv}=h_{uv}.
\label{eq:membership_event}
\end{equation}
We call such a node \(w\) a shortest-path node of the pair \((u,v)\).
We know the positions \(x_u\), \(x_v\) and \(x_w\) of the three nodes
and the type of the RSN, but not the positions of the other nodes or the
links. We define the shortest-path node probability \(\pi_{uv}(w)\) as
the probability that the node \(w\) is a shortest-path node of the pair
\((u,v)\),
\begin{equation}
\pi_{uv}(w)=\Pr\left[\,H_{uw}+H_{wv}=H_{uv}\,\right],
\label{eq:membership_probability}
\end{equation}
where \(H_{uv}\) denotes the random hopcount between the nodes \(u\) and
\(v\). The hopcount excess
\begin{equation}
Z_{uvw}=H_{uw}+H_{wv}-H_{uv}
\label{eq:hopcount_excess}
\end{equation}
is the number of extra hops of the shortest path from \(u\) to \(v\)
through the node \(w\). The hopcount excess \(Z_{uvw}\) is never
negative, because the hopcount obeys the triangle inequality
\(H_{uv}\le H_{uw}+H_{wv}\). By Eq.~\eqref{eq:membership_event}, the
node \(w\) is a shortest-path node when the hopcount excess \(Z_{uvw}\)
is zero, so that \(\pi_{uv}(w)=\Pr[Z_{uvw}=0]\). Varying the position \(x_w\) of the node \(w\) turns the
shortest-path node probability \(\pi_{uv}(w)\) into a probability field
over the space.

\subsection{Hopcount statistics}
\label{sec:hopcount_stats}

The hopcount \(H_{uv}\) between two nodes \(u\) and \(v\) at the
distance \(d_{uv}\) concentrates around the mean hopcount
\(\mu(H_{uv})\), which grows linearly with the distance \(d_{uv}\),
\begin{equation}
\mu(H_{uv})\approx\frac{d_{uv}}{\nu},
\label{eq:linear_law}
\end{equation}
as shown in
Refs.~\cite{yao2011large,diaz2016relation,luchtrath2024all}. The
velocity \(\nu\) is the mean distance that a shortest path advances
toward the end node per hop. For two points at the distance \(d\), the
length \(\ell\) of a shortest path between the two points is the sum of
the weights of the links in the path, not the distance \(d\). The weight
of a link is a random passage time in first-passage percolation, one in
percolation clusters and the Euclidean length of the link in random
geometric graphs. The variance \(\sigma^2(\ell)\) of the length \(\ell\) grows sublinearly in
first-passage percolation~\cite{benjamini2003first,damron2015sublinear},
in percolation clusters~\cite{dembin2024variance} and in random
geometric graphs~\cite{kartun2019shape}: the variance
\(\sigma^2(\ell)\) divided by the distance \(d\) tends to zero for
\(d\to\infty\). In random geometric graphs, the fluctuations of the
length \(\ell\) are Gaussian~\cite{kartun2019shape}. The variance
\(\sigma^2(\ell)\) is usually described by a power
law~\cite{auffinger2017fifty,kartun2019shape} of the distance \(d\).
In our networks, every link has weight one, so that the length \(\ell\)
of a shortest path between two nodes \(u\) and \(v\) is numerically
equal to the hopcount \(H_{uv}\), the number of links in the path. For
the variance
\(\sigma^2(H_{uv})\) of the hopcount \(H_{uv}\) at the fixed distance
\(d_{uv}\), the power law reads
\begin{equation}
\sigma^2(H_{uv})\propto d_{uv}^{\gamma},
\label{eq:power_law}
\end{equation}
with an exponent \(\gamma<1\).

In our simulations (Appendix~\ref{app:simulations}),
the mean hopcount \(\mu(H_{uv})\) grows linearly with the distance
\(d_{uv}\) (Appendix~\ref{app:hopcount}). Because the mean hopcount
\(\mu(H_{uv})\) grows linearly with the distance \(d_{uv}\), we write
the power law~\eqref{eq:power_law} as the variance law
\begin{equation}
\sigma^2(H_{uv})=\sigma_0^2\,\bigl[\mu(H_{uv})/\mu_0\bigr]^{\gamma},
\label{eq:variance_law}
\end{equation}
where \(\sigma_0^2\) is the hopcount variance at a reference mean
hopcount \(\mu_0\). In our simulations, the fitted exponent \(\gamma\)
is smaller than one and the hopcount \(H_{uv}\) at a fixed distance
\(d_{uv}\) has a unimodal distribution, i.e., a distribution with a
single peak (Appendix~\ref{app:hopcount}).

\section{Gaussian estimate of the shortest-path node probability}
\label{sec:gspne}

\begin{flushright}
\parbox{\linewidth}{\raggedleft\small
\textit{Superest ut ex iisdem principiis doceamus}\\
\textit{constitutionem Systematis Viarum Brevissimarum}\\
\textit{in Retibus Spatialibus.}\footnotemark[1]\\[2pt]
--- with apologies to I.~Newton,
\textit{Principia}~\cite{newton1726principia}, Book~III}
\end{flushright}
\footnotetext[1]{It remains that, from the same principles, we
demonstrate the frame of the system of shortest paths in spatial
networks.}

\begin{figure}[htbp]
\centering
\includegraphics[width=0.8\linewidth]
{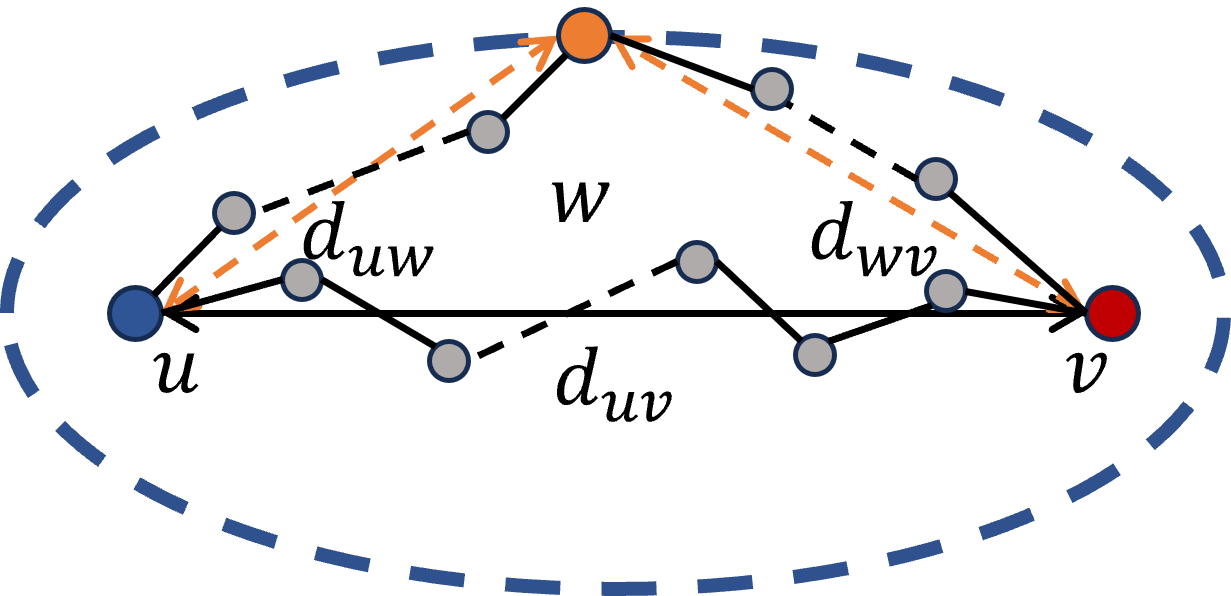}
\caption{\textbf{Nodes on one ellipse give detours of equal length.}
The detour from the source \(u\) to the destination \(v\) through the node
\(w\) (orange dashed lines) is longer than the distance \(d_{uv}\) by
\(\rho_{uvw}d_{uv}=d_{uw}+d_{wv}-d_{uv}\). The lower path is a shortest
path from \(u\) to \(v\) with \(H_{uv}\) hops. The upper path consists of
a shortest path from \(u\) to \(w\) and a shortest path from \(w\) to
\(v\), with \(H_{uw}+H_{wv}\) hops. The difference between the two
hopcounts is the hopcount excess \(Z_{uvw}\), whose mean is about
\(\rho_{uvw}\mu(H_{uv})\). Black dashed lines stand for omitted hops. All
positions on the blue dashed ellipse with foci \(x_u\) and \(x_v\) have
the same relative geometric deviation \(\rho_{uvw}\).}
\label{fig:detour}
\end{figure}

\subsection{Gaussian estimate}
\label{sec:gaussian}

Figure~\ref{fig:detour} shows a path from the source \(u\) to the
destination \(v\) through the node \(w\). The path runs close to the
detour from \(x_u\) through \(x_w\) to \(x_v\), which is longer than the
distance \(d_{uv}\) by \(d_{uw}+d_{wv}-d_{uv}\). The relative geometric
deviation of the node \(w\) is
\begin{equation}
\rho_{uvw}=\frac{d_{uw}+d_{wv}-d_{uv}}{d_{uv}}.
\label{eq:relative_deviation}
\end{equation}
The relative geometric deviation \(\rho_{uvw}\) is zero on the geodesic
from \(x_u\) to \(x_v\) and grows as the node \(w\) moves away from the
geodesic. In Euclidean
space the curves of constant relative geometric deviation
\(\rho_{uvw}\) are ellipses with foci \(x_u\) and \(x_v\)
(Fig.~\ref{fig:detour}). By the linear law~\eqref{eq:linear_law}, the
product \(\rho_{uvw}\mu(H_{uv})\) approximates the
mean number of extra hops of the shortest path from \(u\) to \(v\)
through the node \(w\), that is, the mean of the hopcount excess
\(Z_{uvw}\), whose variance without correlations is
\begin{equation}
V_{uvw}=\sigma^2(H_{uv})+\sigma^2(H_{uw})+\sigma^2(H_{wv}).
\label{eq:independent_variance}
\end{equation}

\begin{proposition}[Gaussian estimate]
\label{prop:gaussian}
Assume that (i) each of the hopcounts \(H_{uv}\), \(H_{uw}\) and
\(H_{wv}\) has a discrete Gaussian distribution with the mean of the
linear law~\eqref{eq:linear_law} and the variance of the variance
law~\eqref{eq:variance_law}, (ii) the three hopcounts are independent
and (iii) the standard deviations of the hopcounts are large compared
with one hop. Then the shortest-path node probability is
\begin{equation}
\pi_{uv}(w)\approx\frac{1}{\sqrt{2\pi V_{uvw}}}
\exp\!\left(-\frac{[\rho_{uvw}\,\mu(H_{uv})]^2}{2V_{uvw}}\right).
\label{eq:gaussian_membership}
\end{equation}
\end{proposition}

\begin{proof}
By assumption (i), the probability that the hopcount \(H_{uv}\) between
two nodes \(u\) and \(v\) equals the integer \(k\) is
\begin{equation}
\begin{split}
\Pr[H_{uv}=k]\approx{}&\frac{1}{\sqrt{2\pi\sigma^2(H_{uv})}}\\
&\times\exp\!\left(-\frac{[k-\mu(H_{uv})]^2}{2\sigma^2(H_{uv})}\right).
\end{split}
\label{eq:discrete_gaussian}
\end{equation}
By assumption (ii), ignoring the correlations between the hopcounts
\(H_{uv}\), \(H_{uw}\) and \(H_{wv}\), the distribution of the sum \(H_{uw}+H_{wv}\) is the
convolution
\begin{equation}
\begin{split}
\Pr[H_{uw}+H_{wv}=k]={}&\sum_{j=0}^{k}\Pr[H_{uw}=j]\\
&\times\Pr[H_{wv}=k-j],
\end{split}
\label{eq:convolution}
\end{equation}
which is close to a discrete Gaussian
with the mean \(\mu(H_{uw})+\mu(H_{wv})\) and the variance
\(\sigma^2(H_{uw})+\sigma^2(H_{wv})\). The shortest-path node
probability \(\pi_{uv}(w)\) in Eq.~\eqref{eq:membership_probability} is
then the inner product of the distribution of the sum \(H_{uw}+H_{wv}\)
and the distribution of the hopcount \(H_{uv}\),
\begin{equation}
\pi_{uv}(w)=\sum_{k=0}^{\infty}\Pr[H_{uw}+H_{wv}=k]\,\Pr[H_{uv}=k].
\label{eq:inner_product}
\end{equation}
By assumption (iii), the sum over \(k\) is close to an integral over
\(k\) from \(-\infty\) to \(\infty\). The inner product of two Gaussian
functions is a Gaussian function of the difference
\(\mu(H_{uw})+\mu(H_{wv})-\mu(H_{uv})\) between their means, with the
variance \(V_{uvw}\) of Eq.~\eqref{eq:independent_variance}. The
difference between the means follows from the geometry. By
assumption (i) and Eq.~\eqref{eq:relative_deviation}, the difference
between the means is \((d_{uw}+d_{wv}-d_{uv})/\nu=\rho_{uvw}\,\mu(H_{uv})\),
which gives the Gaussian estimate~\eqref{eq:gaussian_membership}.
\end{proof}

The advantage of the Gaussian estimate~\eqref{eq:gaussian_membership} is
that the estimate needs only the positions of the three nodes, the
linear law~\eqref{eq:linear_law} and the variance
law~\eqref{eq:variance_law}, but no shortest path of the pair
\((u,v)\). The limitation is that the estimate
ignores the correlations between the hopcounts \(H_{uv}\), \(H_{uw}\)
and \(H_{wv}\), which Section~\ref{sec:correlations} measures. In
addition, the fitted linear law~\eqref{eq:app_linear_fit} has an
intercept (Appendix~\ref{app:mean_excess}) and the hopcount distribution
is not exactly a discrete Gaussian. We therefore use the Gaussian
estimate~\eqref{eq:gaussian_membership} only as a tool to analyze how
the shortest-path node probability \(\pi_{uv}(w)\) changes with the
positions of the nodes and with the hopcount statistics.
Appendix~\ref{app:accuracy} analyzes the performance of the Gaussian
estimate~\eqref{eq:gaussian_membership} as a model for the values of the
shortest-path node probability \(\pi_{uv}(w)\).

The three variables, the relative geometric deviation
\(\rho_{uvw}\), the mean hopcount \(\mu(H_{uv})\) and the variance
\(V_{uvw}\), enter the exponent of the Gaussian
estimate~\eqref{eq:gaussian_membership} only through the standardized
detour
\begin{equation}
X_{uvw}=\frac{\rho_{uvw}\,\mu(H_{uv})}{\sqrt{V_{uvw}}},
\label{eq:standardized_mismatch}
\end{equation}
the mean number of extra hops \(\rho_{uvw}\mu(H_{uv})\) divided by the
standard deviation \(\sqrt{V_{uvw}}\) of the hopcount excess
\(Z_{uvw}\). With the standardized detour \(X_{uvw}\), the Gaussian
estimate~\eqref{eq:gaussian_membership} reads
\begin{equation}
\pi_{uv}(w)\approx\frac{\exp\!\left(-X_{uvw}^2/2\right)}{\sqrt{2\pi V_{uvw}}}.
\label{eq:gaussian_X}
\end{equation}
The node \(w\) lies on a shortest path when a fluctuation of the
hopcount excess \(Z_{uvw}\) cancels the mean number of extra hops
\(\rho_{uvw}\mu(H_{uv})\). Ignoring the correlations between the
hopcounts, the larger the standardized detour \(X_{uvw}\), the smaller
the shortest-path node probability \(\pi_{uv}(w)\) at the same variance
\(V_{uvw}\).

\subsection{Comparison with simulations}
\label{sec:mismatch}

\begin{figure}[htbp]
\centering
\includegraphics[width=1\linewidth]{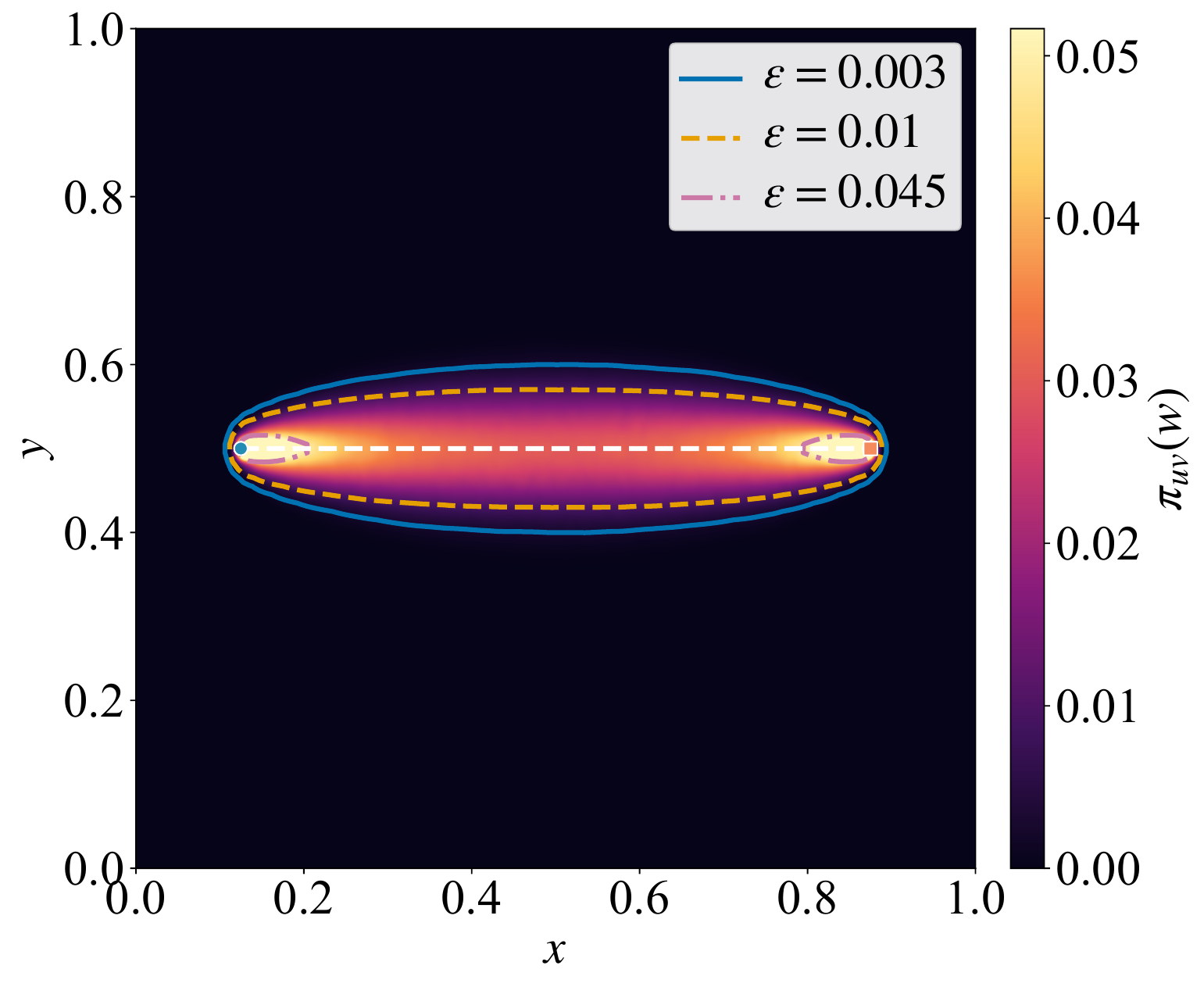}
\caption{%
\textbf{Simulated shortest-path node probability \(\pi_{uv}(w)\) of one
pair.} The source \(u\) (circle) and the destination \(v\) (square) lie
at the distance \(d_{uv}=0.75\) in a soft random geometric graph with
\(N=10^{5}\) nodes in the unit square, mean degree \(14\) and steepness
\(\beta=4\). Contours mark the levels
\(\varepsilon=0.003\), \(0.01\) and \(0.045\) of \(\pi_{uv}(w)\). The
contour of the highest level \(\varepsilon=0.045\) splits into two
separate lobes, one near the source \(u\) and one near the destination
\(v\). The dashed line is the geodesic from \(x_u\) to \(x_v\).}
\label{fig:probability}
\end{figure}

Figure~\ref{fig:probability} shows the heatmap of the shortest-path
node probability \(\pi_{uv}(w)\) from our simulations
(Appendix~\ref{app:simulations}).

\begin{figure*}[htbp]
\centering
\subfloat[Relative geometric deviation \(\rho_{uvw}\).\label{fig:result1_geometric_deviation}]{%
\includegraphics[width=0.32\linewidth]{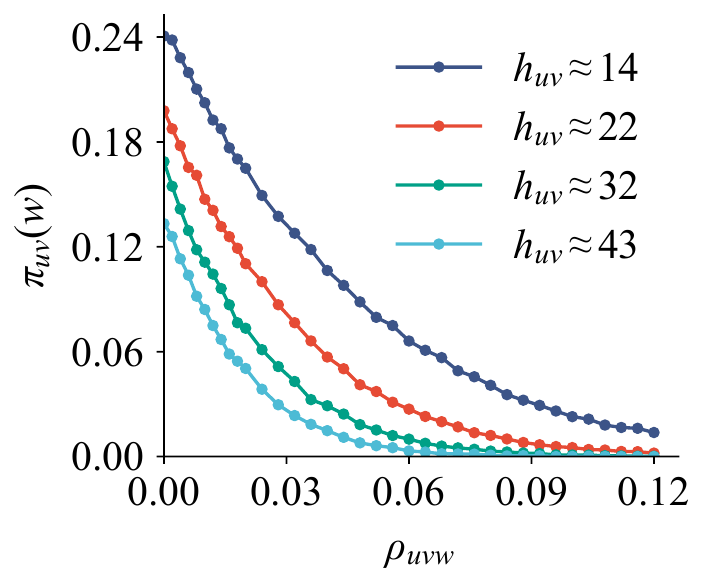}}
\hfill
\subfloat[Mean hopcount \(\mu(H_{uv})\).\label{fig:result1_path_scale}]{%
\includegraphics[width=0.32\linewidth]{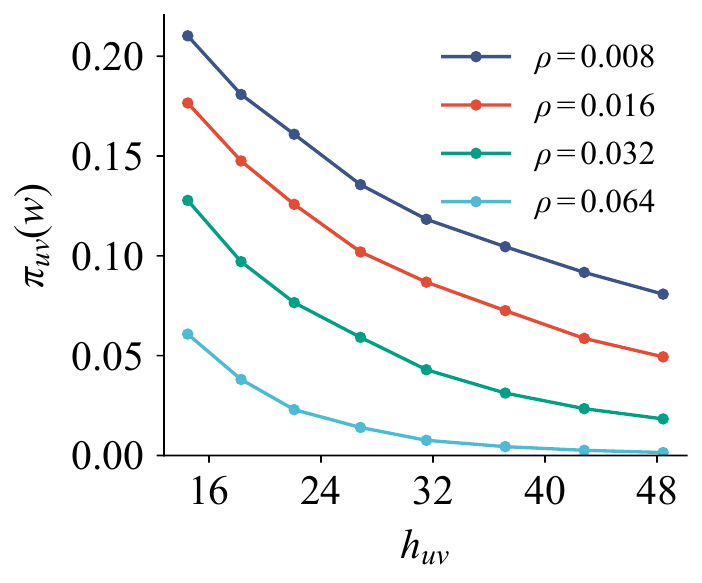}}
\hfill
\subfloat[Hopcount variance \(\sigma_0^2\).\label{fig:result1_hop_fluctuations}]{%
\includegraphics[width=0.32\linewidth]{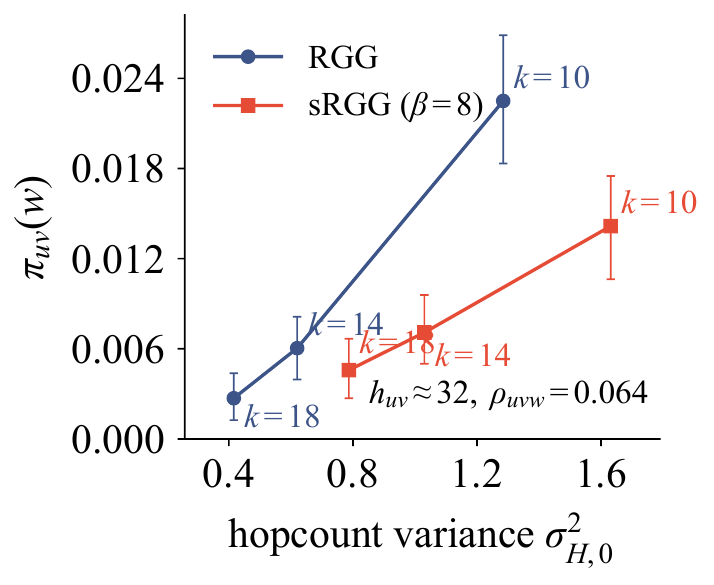}}
\caption{%
\textbf{Three variables affect the shortest-path node probability
\(\pi_{uv}(w)\).}
(a) The simulated shortest-path node probability \(\pi_{uv}(w)\)
decreases with the relative geometric deviation \(\rho_{uvw}\). Curves are labeled by
the mean hopcount \(\mu(H_{uv})\).
(b) At a fixed relative geometric deviation \(\rho_{uvw}\), the
simulated shortest-path node probability \(\pi_{uv}(w)\) decreases with the mean
hopcount \(\mu(H_{uv})\). Curves are labeled by the relative geometric
deviation \(\rho_{uvw}\).
Panels (a) and (b) show a soft random geometric graph of \(30{,}000\)
nodes with mean degree \(14\) and steepness \(\beta=8\) in the square
\([0,4]^2\).
(c) Simulated shortest-path node probability \(\pi_{uv}(w)\) at the
relative geometric deviation \(\rho_{uvw}=0.064\) and the mean hopcount
\(\mu(H_{uv})\approx32\) in hard random geometric graphs (RGG) and soft
random geometric graphs (sRGG, \(\beta=8\)) in Euclidean space. Within
each connection rule, the simulated shortest-path node probability
\(\pi_{uv}(w)\) is higher in networks with a larger hopcount variance
\(\sigma_0^2\), measured at the mean hopcount \(\mu_0=32\). Points are
labeled by mean degree. Error bars are 95\% confidence intervals from a
bootstrap over network realizations.
Test nodes lie at equal distance from the source and the destination
(Appendix~\ref{app:simulations}).}
\label{fig:result1_controls}
\end{figure*}

\textit{(i) Relative geometric deviation \(\rho_{uvw}\).} In
Fig.~\ref{fig:probability}, the contours of equal probability are close
to ellipses with foci \(x_u\) and \(x_v\). In the Euclidean plane, the
positions with the same relative geometric deviation \(\rho_{uvw}\) form
the ellipse \(d_{uw}+d_{wv}=(1+\rho_{uvw})\,d_{uv}\)
(Fig.~\ref{fig:detour}). In the Gaussian
estimate~\eqref{eq:gaussian_membership}, which ignores the correlations
between the hopcounts, the position of
the node \(w\) enters only through the mean number of extra hops
\(\rho_{uvw}\mu(H_{uv})\) and the variance \(V_{uvw}\). On the ellipse,
the mean number of extra hops \(\rho_{uvw}\mu(H_{uv})\) is constant and
the variance \(V_{uvw}=O(\sigma^2(H_{uv}))\) changes more slowly with the
position of the node \(w\) than the relative geometric deviation
\(\rho_{uvw}\). The Gaussian
estimate~\eqref{eq:gaussian_membership} is thus nearly constant on each
ellipse, which gives the elliptical contours. Because the term
\([\rho_{uvw}\mu(H_{uv})]^2/(2V_{uvw})\) in the exponent of
Eq.~\eqref{eq:gaussian_membership} grows with the relative geometric
deviation \(\rho_{uvw}\), the Gaussian
estimate~\eqref{eq:gaussian_membership} decreases with the relative
geometric deviation \(\rho_{uvw}\), as the shortest-path node
probability \(\pi_{uv}(w)\) does in our simulations at every mean
hopcount \(\mu(H_{uv})\) [Fig.~\ref{fig:result1_controls}(a)].

\textit{(ii) Mean hopcount \(\mu(H_{uv})\).} At a fixed relative
geometric deviation \(\rho_{uvw}\) and fixed ratios \(d_{uw}/d_{uv}\) and
\(d_{wv}/d_{uv}\), the variance law~\eqref{eq:variance_law} gives
\(V_{uvw}\propto\mu(H_{uv})^{\gamma}\). As the mean hopcount
\(\mu(H_{uv})\) grows, the term
\([\rho_{uvw}\mu(H_{uv})]^2/(2V_{uvw})\propto\mu(H_{uv})^{2-\gamma}\) in
the exponent of Eq.~\eqref{eq:gaussian_membership} grows and the
prefactor \(1/\sqrt{2\pi V_{uvw}}\propto\mu(H_{uv})^{-\gamma/2}\) falls,
so that the Gaussian estimate~\eqref{eq:gaussian_membership} decreases
with the mean hopcount \(\mu(H_{uv})\) (Appendix~\ref{app:hopcount}).
In our simulations at a fixed relative
geometric deviation \(\rho_{uvw}\), the shortest-path node probability
\(\pi_{uv}(w)\) decreases with the mean hopcount \(\mu(H_{uv})\)
[Fig.~\ref{fig:result1_controls}(b)], as the Gaussian estimate predicts.

\textit{(iii) Variance \(V_{uvw}\).} The variance \(V_{uvw}\) changes the
Gaussian estimate~\eqref{eq:gaussian_membership} in a nonmonotonic way.
At a fixed relative geometric deviation \(\rho_{uvw}\) and a fixed mean
hopcount \(\mu(H_{uv})\), the derivative
\(\partial\ln\pi_{uv}(w)/\partial V_{uvw}\) of
Eq.~\eqref{eq:gaussian_membership} has the sign of
\([\rho_{uvw}\mu(H_{uv})]^2-V_{uvw}\) [Eq.~\eqref{eq:app_dlnpi_dV}].
Near the geodesic, the mean number of extra hops
\(\rho_{uvw}\mu(H_{uv})\) is smaller than the standard deviation
\(\sqrt{V_{uvw}}\) and a larger variance \(V_{uvw}\) lowers the Gaussian
estimate. Far from the geodesic, the mean number of extra hops
\(\rho_{uvw}\mu(H_{uv})\) exceeds the standard deviation
\(\sqrt{V_{uvw}}\) and a larger variance \(V_{uvw}\) raises the Gaussian
estimate. We discuss only the monotonic part far from the geodesic.
Figure~\ref{fig:result1_controls}(c) varies the variance \(V_{uvw}\)
through the mean degree, because a larger mean degree generally gives a
smaller hopcount variance at the same mean hopcount
(Appendix~\ref{app:degree}). In our simulations at the relative
geometric deviation \(\rho_{uvw}=0.064\) and the mean hopcount
\(\mu(H_{uv})\approx32\), the mean number of extra hops
\(\rho_{uvw}\mu(H_{uv})\approx2\) exceeds the standard deviation
\(\sqrt{V_{uvw}}\) in all six networks (Table~\ref{tab:fig3c}). Within
each connection rule, the simulated networks with a larger hopcount
variance \(\sigma_0^2\) at the mean hopcount \(\mu_0=32\) have a higher
shortest-path node probability \(\pi_{uv}(w)\)
[Fig.~\ref{fig:result1_controls}(c)], as the Gaussian estimate predicts.

\begin{figure}[htbp]
\centering
\subfloat[Before standardization.\label{fig:result2_before_scaling}]{%
\includegraphics[width=0.49\linewidth]{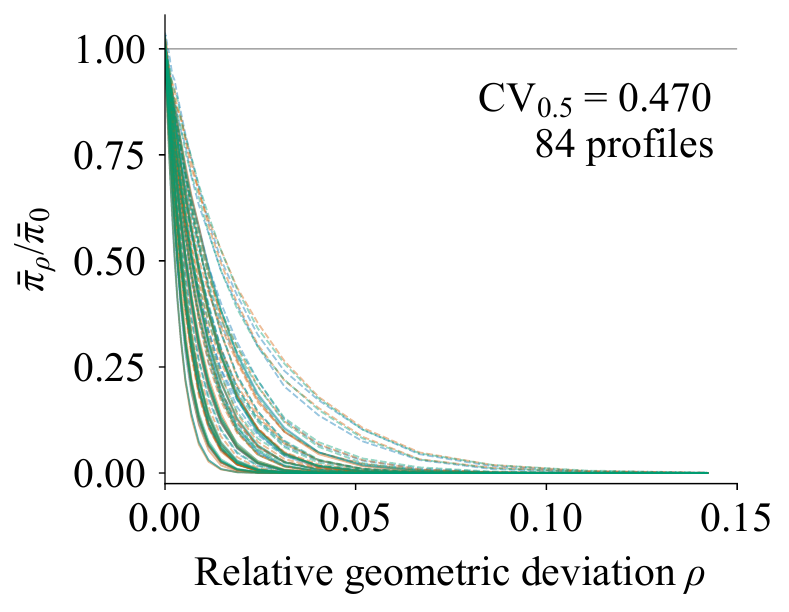}}
\hfill
\subfloat[After standardization.\label{fig:result2_after_scaling}]{%
\includegraphics[width=0.49\linewidth]{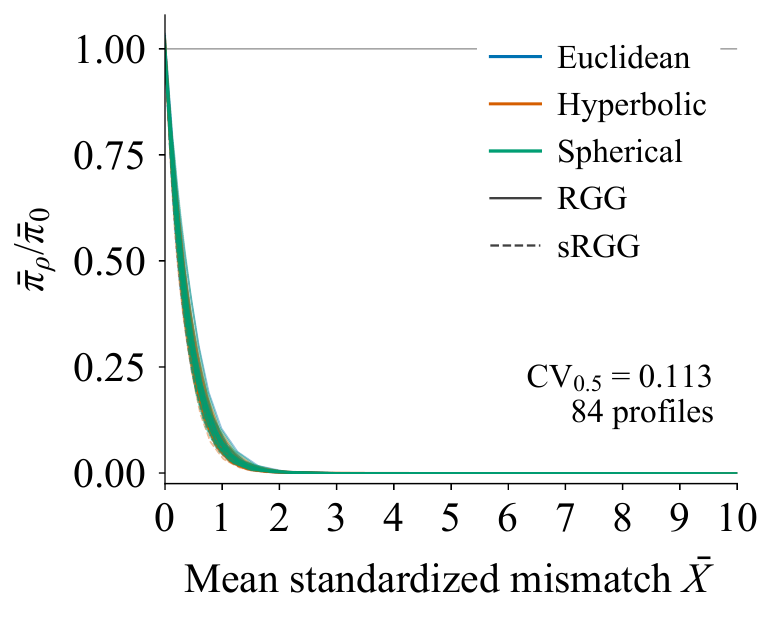}}
 
\vspace{1.5ex}
 
\subfloat[Shuffle control.\label{fig:result2_shuffle}]{%
\includegraphics[width=0.49\linewidth]{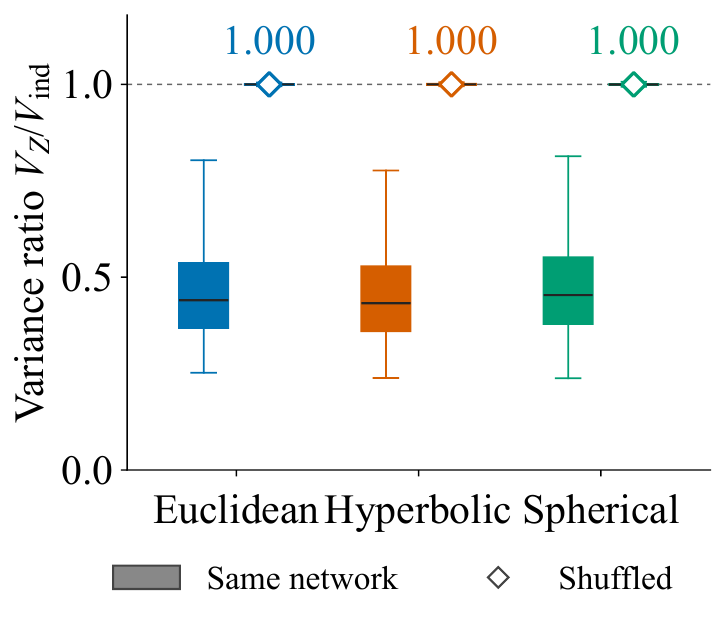}}
\hfill
\subfloat[Held-out covariance prediction.\label{fig:result2_covariance}]{%
\includegraphics[width=0.49\linewidth]{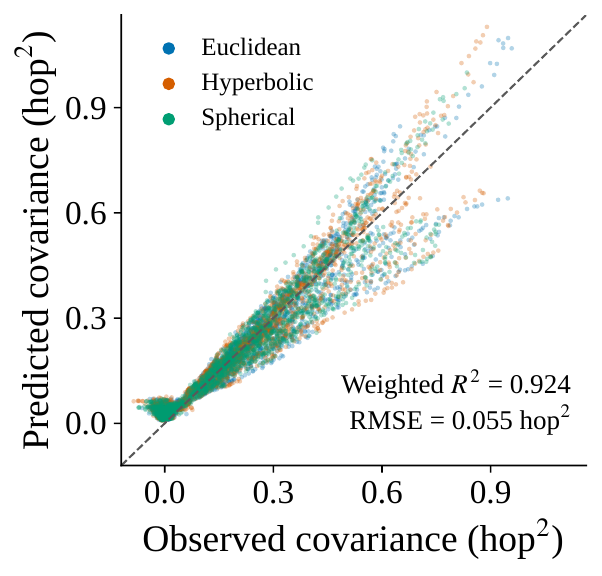}}
 
\caption{%
\textbf{Hopcount statistics align the simulated shortest-path node
probability \(\pi_{uv}(w)\) across network models and spaces.}
(a), (b) The same 84 normalized profiles \(\widetilde\pi\) against the
relative geometric deviation \(\rho\) (a) and the mean standardized
detour \(\overline X\) (b). Profiles are averages over curves of equal
\(\rho\), divided by their value at \(\rho=0\) (Appendix~\ref{app:profiles}). Colors mark the space.
Solid and dashed lines mark hard and soft random geometric graphs
(legend in (b)).
(c), (d) The three hopcounts measured together in our simulations at the mean hopcounts
\(\mu(H_{uv})=14\), 22, 32 and 43 (Appendix~\ref{app:correlations}).
(c) Ratio of the measured variance \(V_Z\) of the hopcount excess
\(Z_{uvw}\) to the independent value \(V_{uvw}\), the sum of the three measured hopcount variances, for test nodes at equal distance from the source and the destination with \(0.004\le\rho_{uvw}\le0.064\). Boxes show quartiles
and medians across conditions. Whiskers span the full range. Open
diamonds mark the medians after shuffling the hopcount records across
realizations.
(d) Predicted and simulated covariances for \(6{,}912\) held-out terms.
Each term is the covariance of two hopcounts at one position of the
node \(w\) in one network condition. The prediction is the linear function~\eqref{eq:app_linear_corr} of the normalized overlap \(E[S]/\sqrt{\mu(H_{uw})\mu(H_{uv})}\), where the overlap \(E[S]\) is the mean number of initial links that the two paths share, fitted on training realizations with one set
of parameters for the three spaces. The overlaps \(E[S]\) and the hopcount variances \(\sigma^2(H_{uw})\) and \(\sigma^2(H_{uv})\) of the test terms are measured in new simulated realizations. The dashed line is
\(y=x\). The weighted coefficient of determination is \(R^2=0.924\).}
\label{fig:result2_statistical_control}
\end{figure}

\textit{(iv) Standardized detour \(X_{uvw}\).} The shortest-path nodes
concentrate in a narrow region around the geodesic
(Fig.~\ref{fig:probability}). By Eq.~\eqref{eq:gaussian_X}, the
shortest-path node probability \(\pi_{uv}(w)\) decays as
\(\exp(-X_{uvw}^2/2)\). The standardized detour
\(X_{uvw}=\rho_{uvw}\mu(H_{uv})/\sqrt{V_{uvw}}\) grows rapidly with the
relative geometric deviation \(\rho_{uvw}\), so that the shortest-path
nodes concentrate in a narrow region around the geodesic. Let \(\overline\pi_\rho\) be
the average of the shortest-path node probability \(\pi_{uv}(w)\) over
the curve of equal relative geometric deviation \(\rho\). In particular,
\(\overline\pi_0\) is the average of the shortest-path node probability
\(\pi_{uv}(w)\) on the geodesic, where \(\rho=0\). Dividing by the
average \(\overline\pi_0\) removes most of the prefactor
\(1/\sqrt{2\pi V_{uvw}}\) in Eq.~\eqref{eq:gaussian_X}, so that the
normalized profile \(\widetilde\pi=\overline\pi_\rho/\overline\pi_0\)
depends approximately only on the mean standardized detour
\(\overline X\) on the same curve (Appendix~\ref{app:profiles}). In our simulations of 21 network
conditions in three spaces at four distances \(d_{uv}\), the 84
normalized profiles \(\widetilde\pi\) are widely separated against the
relative geometric deviation \(\rho\)
[Fig.~\ref{fig:result2_statistical_control}(a)] but approximately follow
one curve against the mean standardized detour \(\overline X\)
[Fig.~\ref{fig:result2_statistical_control}(b)]. In our simulations,
the spread of the value at which the normalized profile
\(\widetilde\pi\) drops below 1/2, the standard deviation divided by the
mean, falls from 0.470 for the relative geometric deviation \(\rho\) to
0.113 for the mean standardized detour \(\overline X\)
(Appendix~\ref{app:profiles}). Every quantity in the standardized detour
\(X_{uvw}\) is calibrated on training realizations, so the agreement is
a prediction, not a fit.

\subsection{Hopcount correlations}
\label{sec:correlations}

The Gaussian estimate~\eqref{eq:gaussian_membership} is nearly constant
on each ellipse with foci \(x_u\) and \(x_v\) (Sec.~\ref{sec:mismatch})
and thus does not produce the two lobes of Fig.~\ref{fig:probability}
near the end nodes. Proposition~\ref{prop:gaussian} treats the hopcounts
\(H_{uv}\), \(H_{uw}\) and \(H_{wv}\) as independent [assumption (ii)].
With correlations, the variance of the hopcount excess
\(Z_{uvw}\) is
\begin{equation}
\begin{split}
V_Z={}&V_{uvw}-2\,\mathrm{Cov}[H_{uw},H_{uv}]\\
&-2\,\mathrm{Cov}[H_{wv},H_{uv}]+2\,\mathrm{Cov}[H_{uw},H_{wv}].
\end{split}
\label{eq:excess_variance}
\end{equation}
Under this condition, with the mean \(m_Z\) of the hopcount excess
\(Z_{uvw}\), the Gaussian estimate~\eqref{eq:gaussian_membership} becomes
\begin{equation}
\pi_{uv}(w)=\Pr[Z_{uvw}=0]\approx\frac{1}{\sqrt{2\pi V_Z}}
\exp\!\left(-\frac{m_Z^2}{2V_Z}\right).
\label{eq:correlated_estimate}
\end{equation}
In Eq.~\eqref{eq:correlated_estimate}, the position of the node \(w\)
enters through the mean \(m_Z\) and the variance \(V_Z\).

Shortest paths with a common end node often leave the end node along the
same links~\cite{tang2008effect}. By Eq.~\eqref{eq:covariance_split}, the
variance \(\sigma^2(S)\) of the number \(S\) of shared links adds to the
covariance of the two hopcounts, so that the hopcounts \(H_{uv}\),
\(H_{uw}\) and \(H_{wv}\) are positively correlated
(Appendix~\ref{app:overlap}). A similar principle has been discussed in
Refs.~\cite{pinto2012locating,chatterjee2014superconcentration,ahlberg2023chaos}.
In our simulations, the overlap of the shortest paths predicts the
covariances of the hopcounts
[Fig.~\ref{fig:result2_statistical_control}(d)]. In the same
simulations, the ratio \(V_Z/V_{uvw}\) lies below one and returns to one
after the hopcounts are shuffled across realizations
[Fig.~\ref{fig:result2_statistical_control}(c)].

The longer the shared part of two shortest paths, the larger the
covariance of their hopcounts. When the node \(w\) lies near the
destination \(v\), the shortest paths from \(u\) to \(w\) and from \(u\)
to \(v\) share most of their links. The same holds for the shortest paths
from \(w\) to \(v\) and from \(u\) to \(v\) when the node \(w\) lies near
the source \(u\). At equal relative geometric deviation \(\rho_{uvw}\)
near the geodesic, the correlations are therefore strongest near the end
nodes, where the correlations raise the shortest-path node probability
\(\pi_{uv}(w)\). In our end-node experiment, the rise of the
shortest-path node probability \(\pi_{uv}(w)\) toward the end nodes is
weaker after the hopcounts are shuffled across realizations
(Appendix~\ref{app:end_nodes}). The higher probability near the end
nodes gives the two lobes of Fig.~\ref{fig:probability}
(Appendix~\ref{app:lobes}).

\section{Quantized distribution of shortest-path nodes}
\label{sec:quantization}
 
The Gaussian estimate~\eqref{eq:gaussian_membership} treats the hopcount \(H_{uv}\) as a continuous variable, which is accurate when the standard deviation \(\sigma(H_{uv})\) of the hopcount is large compared with one hop
(Sec.~\ref{sec:gaussian}). We now take the opposite limit, in which the
hopcount fluctuations vanish. In our simulations, the hopcount variance \(\sigma^2(H_{uv})\) often falls as the mean degree grows (Table~\ref{tab:fig3c}).
 
With the velocity \(\nu\) of the linear law~\eqref{eq:linear_law}, the
predicted hopcount
\begin{equation}
k^\star(d)=\left\lceil d/\nu\right\rceil
\label{eq:predicted_hopcount}
\end{equation}
is the smallest number of hops that covers the distance \(d\) along the
geodesic when every hop advances the distance \(\nu\).

\begin{proposition}[Quantization rule]
\label{prop:quantization}
Assume that (i) the hopcount \(H_{uv}\) between two nodes \(u\) and \(v\)
is a normal random variable with the mean \(d_{uv}/\nu\) of the linear
law~\eqref{eq:linear_law} and the variance \(s_{uv}^2\), rounded up to an
integer and (ii) the variance \(s_{uv}^2\) of every pair of nodes tends
to zero. Then the shortest-path node
probability~\eqref{eq:membership_probability} becomes
\begin{equation}
\begin{split}
\pi_{uv}(w)
&\to\begin{cases}
1 & \text{if } k^\star(d_{uw})+k^\star(d_{wv})=k^\star(d_{uv}),\\
0 & \text{otherwise}.
\end{cases}
\end{split}
\label{eq:quantization_rule}
\end{equation}
\end{proposition}

\begin{proof}
By assumption (i), the hopcount \(H_{uv}\) equals the integer \(k\) when
the normal random variable lies in the interval \((k-1,k]\), which
happens with the probability
\begin{equation}
\begin{split}
\Pr[H_{uv}=k]={}&\int_{k-1}^{k}\frac{1}{\sqrt{2\pi s_{uv}^2}}\\
&\times\exp\!\left(-\frac{(y-d_{uv}/\nu)^2}{2s_{uv}^2}\right)dy.
\end{split}
\label{eq:unit_bin}
\end{equation}
Because rounding up a real number \(y\) gives the integer
\(\lceil y\rceil\) with \(0\le\lceil y\rceil-y<1\), the mean hopcount
\(\mu(H_{uv})\) lies between \(d_{uv}/\nu\) and \(d_{uv}/\nu+1\).

By assumption (ii), the normal density
\begin{equation}
\frac{1}{\sqrt{2\pi s_{uv}^2}}\exp\!\left(-\frac{(y-d_{uv}/\nu)^2}{2s_{uv}^2}\right)
\to\delta(y-d_{uv}/\nu),
\label{eq:density_limit}
\end{equation}
because the Dirac delta function is the limit of a normal density whose
standard deviation tends to zero~\cite[Sec.~7.1]{vanmieghem2023graph}.
Apart from the distances \(d_{uv}\) that are multiples of the velocity
\(\nu\), which mark the transitions between neighboring hopcounts and
have zero measure, the integral~\eqref{eq:unit_bin} then tends to one if
\(k-1<d_{uv}/\nu<k\) and to zero otherwise. The distribution of the
hopcount \(H_{uv}\) thus degenerates into a Kronecker delta at the
predicted hopcount \(k^\star(d_{uv})\) of Eq.~\eqref{eq:predicted_hopcount},
\begin{equation}
\Pr[H_{uv}=k]\to\delta_{k,k^\star(d_{uv})}.
\label{eq:delta_limit}
\end{equation}
The Kronecker delta \(\delta_{k,k^\star(d_{uv})}\) equals one for
\(k=k^\star(d_{uv})\) and zero otherwise. The mean hopcount \(\mu(H_{uv})\) then tends to the predicted hopcount
\(k^\star(d_{uv})\) and the hopcount variance \(\sigma^2(H_{uv})\) tends
to zero.

By the convolution~\eqref{eq:convolution}, the limits~\eqref{eq:delta_limit}
give \(\Pr[H_{uw}+H_{wv}=k]\to\delta_{k,k^\star(d_{uw})+k^\star(d_{wv})}\).
Substituting the limits into the inner product~\eqref{eq:inner_product}
gives
\begin{equation}
\begin{split}
\pi_{uv}(w)&\to\sum_{k=0}^{\infty}\delta_{k,k^\star(d_{uw})+k^\star(d_{wv})}\,
\delta_{k,k^\star(d_{uv})}\\
&=\delta_{k^\star(d_{uw})+k^\star(d_{wv}),\,k^\star(d_{uv})},
\end{split}
\label{eq:delta_product}
\end{equation}
which is the quantization rule~\eqref{eq:quantization_rule}.
\end{proof}

The quantization rule~\eqref{eq:quantization_rule} does not need the
independence behind the inner product~\eqref{eq:inner_product}. Each of
the three hopcounts \(H_{uw}\), \(H_{wv}\) and \(H_{uv}\) takes the
predicted value with a probability that tends to one, so that the
definition~\eqref{eq:membership_probability} gives the quantization
rule~\eqref{eq:quantization_rule} directly.
 
\begin{figure}[htbp]
\centering
\subfloat[\(d_{uv}=0.900\), \(k^{\star}(d_{uv})=31\).\label{fig:result3_euclidean_0900}]{%
\includegraphics[width=0.4\textwidth]{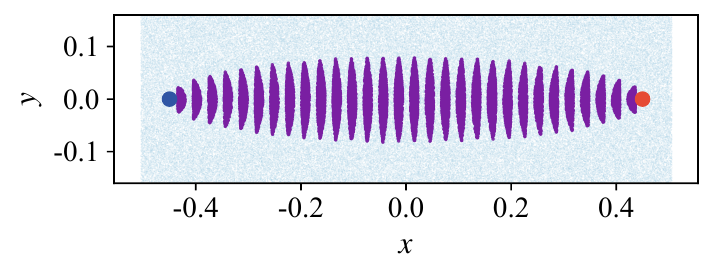}}
\hfill
\subfloat[\(d_{uv}=0.910\), \(k^{\star}(d_{uv})=31\).\label{fig:result3_euclidean_0910}]{%
\includegraphics[width=0.4\textwidth]{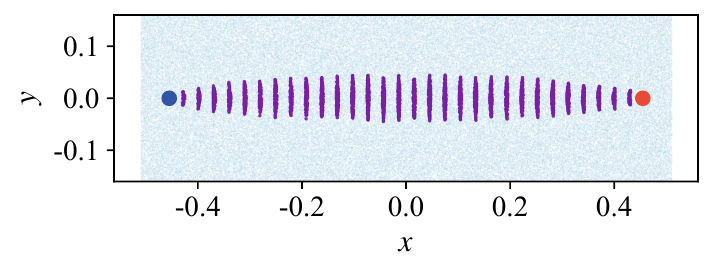}}
\hfill
\subfloat[\(d_{uv}=0.920\), \(k^{\star}(d_{uv})=32\).\label{fig:result3_euclidean_0920}]{%
\includegraphics[width=0.4\textwidth]{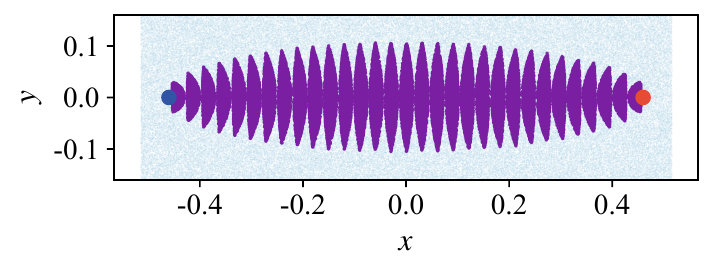}}
\hfill
\subfloat[Sawtooth slack \(W(d_{uv})\).\label{fig:result3_sawtooth}]{%
\includegraphics[width=0.4\textwidth]{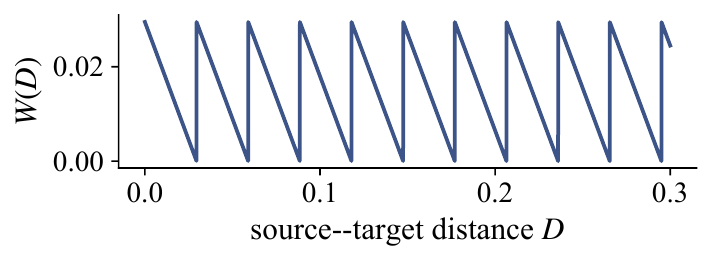}}
\caption{%
\textbf{Quantized distribution of shortest-path nodes in the Euclidean
plane.} Positions selected by
the quantization rule~\eqref{eq:quantization_rule} among uniformly
placed positions (purple) form petals between the source \(u\) (blue)
and the destination \(v\) (red). The velocity \(\nu=0.0295\) is
estimated in our simulations of a hard random geometric graph with
\(N=2\times10^5\) nodes and the connection radius \(r=0.03\). From
\(d_{uv}=0.900\) (a) to \(d_{uv}=0.910\) (b), the predicted hopcount
stays at \(k^\star(d_{uv})=31\) and the petals narrow. At
\(d_{uv}=0.920\) (c), the predicted hopcount is \(k^\star(d_{uv})=32\)
and the petals are wider. (d) The slack
\(W(d_{uv})=\lceil d_{uv}/\nu\rceil\nu-d_{uv}\) decreases with slope
\(-1\) and jumps up by \(\nu\) at every multiple of the velocity
\(\nu\).}
\label{fig:quant_euclid}
\end{figure}

The \(k^\star(d)\) hops of length \(\nu\) cover the length
\(k^\star(d)\,\nu\ge d\). We define the slack of the distance \(d\) as the
amount by which the covered length exceeds the distance \(d\),
\begin{equation}
W(d)=k^\star(d)\,\nu-d.
\label{eq:slack}
\end{equation}
Because the predicted hopcount \(k^\star(d)\) is the smallest number of
hops that covers the distance \(d\), the slack lies in the range
\(0\le W(d)<\nu\). The slack \(W(d_{uv})\) sets the width of the petals
in Fig.~\ref{fig:quant_euclid}, which the quantization
rule~\eqref{eq:quantization_rule} selects along the geodesic
(Appendix~\ref{app:quantization}). As the distance \(d_{uv}\) grows, the
slack \(W(d_{uv})\) decreases linearly until the slack reaches zero at a
multiple of the velocity \(\nu\). Beyond that distance, the predicted
hopcount \(k^\star(d_{uv})\) increases by one and the slack \(W(d_{uv})\)
jumps up by \(\nu\) [Fig.~\ref{fig:quant_euclid}(d)]. From
\(d_{uv}=0.900\) to \(d_{uv}=0.910\), the predicted hopcount stays at 31,
the slack \(W(d_{uv})\) falls from 0.0145 to 0.0045 and the petals
narrow. At \(d_{uv}=0.920>31\nu=0.9145\), the predicted hopcount jumps to
32, the slack \(W(d_{uv})\) jumps to 0.024 and the petals widen. In one simulated realization of the network of
Fig.~\ref{fig:quant_euclid}, the shortest paths at these three distances
have 31, 31 and 32 hops.

\begin{figure}[tb]
\centering
\subfloat[Hyperbolic plane.\label{fig:quant_hyp_disk}]{%
\includegraphics[width=0.49\linewidth]{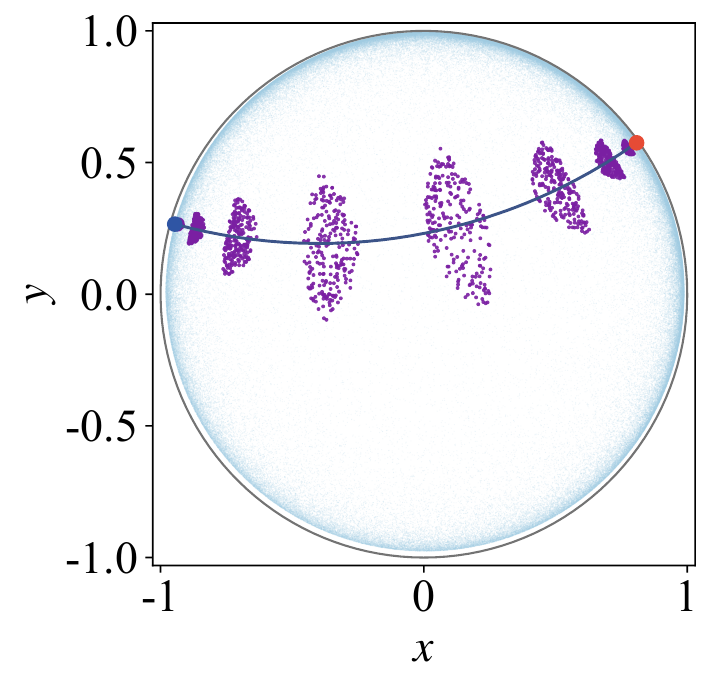}}
\hfill
\subfloat[Sphere.\label{fig:quant_sph_view}]{%
\includegraphics[width=0.49\linewidth]{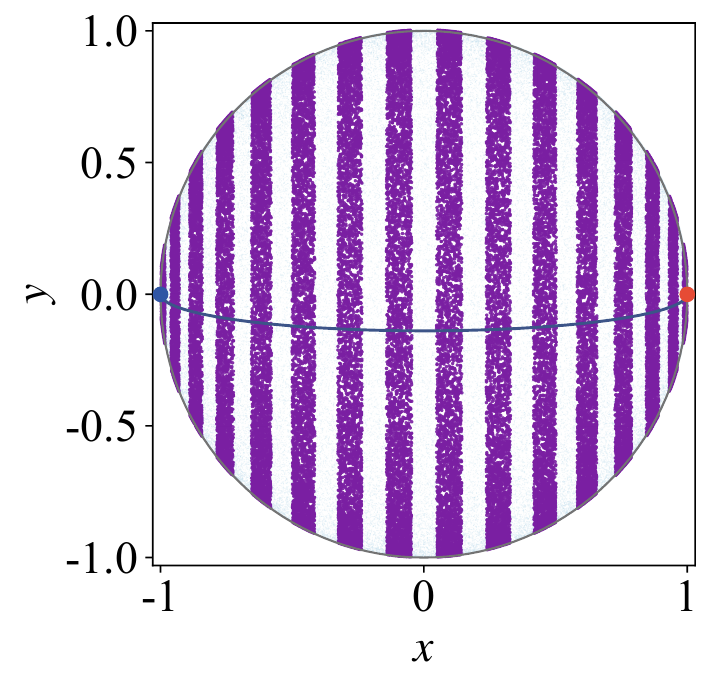}}
 
\vspace{1ex}
 
\subfloat[Hyperbolic plane, along the arc.\label{fig:quant_hyp_arc}]{%
\includegraphics[width=0.49\linewidth]{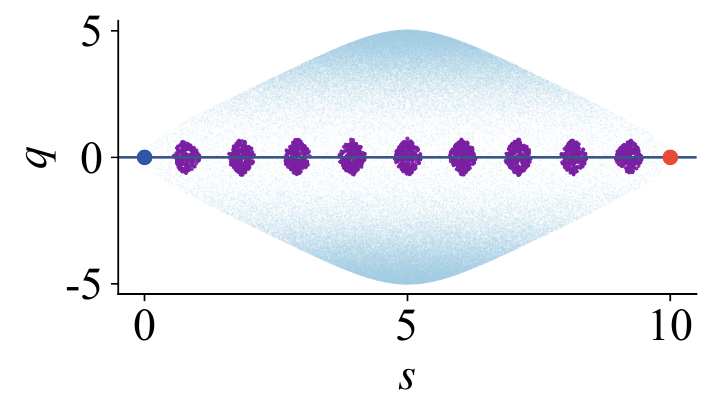}}
\hfill
\subfloat[Sphere, along the arc.\label{fig:quant_sph_arc}]{%
\includegraphics[width=0.49\linewidth]{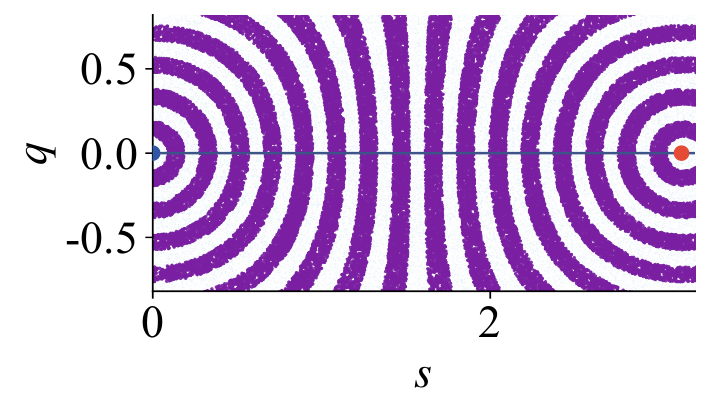}}
\caption{%
\textbf{Quantized distribution of shortest-path nodes in curved
space.} Positions selected by
the quantization rule~\eqref{eq:quantization_rule} among uniformly placed positions
(purple). (a), (c) Hyperbolic plane with the end nodes at the distance
\(d_{uv}=10\) and the velocity \(\nu=1.05\), in the Poincar\'e disk (a)
and against the arc length \(s\) along the geodesic arc and the signed
distance \(q\) from the arc (c). The rule selects nine separate petals.
(b), (d) Unit sphere with the end nodes at the distance
\(d_{uv}=3.133\) and the velocity \(\nu=0.19\), in an orthographic view
(b) and against the arc length \(s\) and the signed distance \(q\) (d).
The selected regions close into bands around the axis through the end
nodes. The two values of the velocity \(\nu\) are chosen for
illustration.}
\label{fig:quant_curved}
\end{figure}

The quantization rule~\eqref{eq:quantization_rule} depends only on the
distances between the nodes, so that the separate regions also appear in
curved space (Fig.~\ref{fig:quant_curved} and
Appendix~\ref{app:quantization}). In the hyperbolic plane, as in the
Euclidean plane, the extra distance of a detour grows without bound away
from the geodesic, so that the rule selects nine separate petals along
the geodesic [Figs.~\ref{fig:quant_curved}(a),(c)]. On the unit sphere,
the extra distance is bounded. The end nodes in
Figs.~\ref{fig:quant_curved}(b),(d) lie at the distance
\(d_{uv}=3.133\), close to the antipodal distance \(\pi\). Every great
circle through \(x_u\) passes through the antipode of \(x_u\), which lies
at the distance \(\pi-d_{uv}=0.009\) from \(x_v\). A detour from \(x_u\) to
\(x_v\) through a position in any direction from \(x_u\) is therefore at
most \(2(\pi-d_{uv})=0.017\) longer than the distance \(d_{uv}\), much
less than the slack \(W(d_{uv})=0.097\). A position is selected when
the extra distance of the detour and the slack \(W(d_{uw})\) together do not
exceed the slack \(W(d_{uv})\) (Appendix~\ref{app:quantization}).
Every position with \(W(d_{uw})\le0.080\) is thus selected in every
direction from \(x_u\), so that the selected regions close into bands
around the axis through the end nodes.
 
\section{Spatially inhomogeneous networks}
\label{sec:heterogeneous}
 
The previous sections analyze shortest paths in homogeneous RSNs, in which the node density \(\lambda(x)\) and the link rule are the same at every position \(x\).
In real networks, the node density \(\lambda(x)\) and the link rule vary with the position \(x\). We call such networks spatially inhomogeneous. In this section, we build a model based on a velocity field \(\nu(x)\) and the travel time \(\tau(u,v)\) to locate the shortest paths in spatially
inhomogeneous networks.
 
\subsection{Velocity field}
\label{sec:velocity_field}
 
The velocity field \(\nu(x)\) is the distance that a shortest path near
the position \(x\) advances per hop. In a homogeneous RSN, the velocity
field \(\nu(x)\) is the constant velocity \(\nu\) of the linear
law~\eqref{eq:linear_law}. We estimate the velocity from local
information, without hopcount data. For a node \(u\), let \(L_{\max}\)
be the length of the longest link of the node \(u\) and \(P_{\max}\) the
largest projection of the links of the node \(u\) on a unit direction
\(e\). Under two assumptions, the means of the two quantities bound the
velocity, \(E[P_{\max}]\le\nu\le E[L_{\max}]\)
(Appendix~\ref{app:travel}). Appendix~\ref{app:travel} also computes
both means from the node density and the link rule. In all \(45\)
conditions of our simulations of homogeneous networks, the velocity
\(\nu\) obtained from the mean hopcount \(\mu(H_{uv})\) lies between the
means \(E[P_{\max}]\) and \(E[L_{\max}]\), with the ratio
\(\nu/E[L_{\max}]\) between 0.78 and 0.87 and the ratio
\(\nu/E[P_{\max}]\) between 1.04 and 1.41. In our simulations, the
midpoint \(\widehat\nu=(E[P_{\max}]+E[L_{\max}])/2\) exceeds the
velocity \(\nu\) by 5\% to 10\% in every condition of hard random
geometric graphs and differs from the velocity \(\nu\) by at most 7.1\%
in every condition of soft random geometric graphs
(Table~\ref{tab:velocity}).

In an inhomogeneous network, we assume that the node density
\(\lambda(x)\) and the link rule vary slowly on the scale of one link.
Near the position \(x\), the network then looks like a homogeneous RSN
with the node density \(\lambda(x)\) and the link rule at \(x\). The
bounds then hold locally,
\(E[P_{\max}(x)]\le\nu(x)\le E[L_{\max}(x)]\)
(Appendix~\ref{app:travel}). We estimate the velocity field by the
midpoint
\begin{equation}
\widehat\nu(x)=\tfrac12\bigl(E[P_{\max}(x)]+E[L_{\max}(x)]\bigr).
\label{eq:velocity_field}
\end{equation}
In a model network, the two means follow from the model fields of the node density \(\lambda(x)\) and the link rule. In an empirical network, we measure the
two quantities at the nodes (Sec.~\ref{sec:empirical}). No shortest-path
data enter the velocity field \(\widehat\nu(x)\).
 
\subsection{Travel time}
\label{sec:travel_time}

In optics, Fermat's principle states that light follows a curve
\(\Gamma\) along which the optical path length \(\int_\Gamma n(x)\,ds\)
is stationary, where \(n(x)\) is the refractive
index~\cite{born1999principles}. Locating the shortest paths of an
inhomogeneous network is an optimization problem of the same kind. We
assume that the velocity \(\nu(x)\) is the same in every direction. A
piece of length \(ds\) of a curve \(\Gamma\) near the position \(x\)
then takes \(ds/\nu(x)\) hops. The travel time \(\tau(u,v)\) is the
smallest number of hops along a curve from \(x_u\) to \(x_v\),
\begin{equation}
\tau(u,v)=\inf_{\Gamma:\,x_u\to x_v}\int_\Gamma\frac{ds}{\nu(x)},
\label{eq:travel_time}
\end{equation}
where \(\Gamma\) runs over the curves from \(x_u\) to \(x_v\). The
inverse velocity \(1/\nu(x)\) plays the role of the refractive index
\(n(x)\). A shortest path follows the curve with the fewest hops, so
that the mean hopcount \(\mu(H_{uv})\) between the nodes \(u\) and
\(v\) is approximately the travel time \(\tau(u,v)\). Unlike the
hopcount \(H_{uv}\), the travel time \(\tau(u,v)\) is not restricted to
integer values. For a constant velocity \(\nu\), the travel time is
\(\tau(u,v)=d_{uv}/\nu\) and the approximation
\(\mu(H_{uv})\approx\tau(u,v)\) reduces to the linear
law~\eqref{eq:linear_law}.

We call the curve of the shortest travel time \(\tau(u,v)\) from
\(x_u\) to \(x_v\) the alignment of the shortest paths from \(u\) to
\(v\). For a constant velocity \(\nu\), the alignment is the geodesic
from \(x_u\) to \(x_v\), around which the shortest-path nodes of a
homogeneous network concentrate (Fig.~\ref{fig:probability}). The
fastest curve from \(x_u\) to \(x_v\) through the position \(x_w\)
takes the travel time \(\tau(u,w)+\tau(w,v)\), which exceeds the travel
time \(\tau(u,v)\) by the excess travel time
\begin{equation}
t_{uvw}=\tau(u,w)+\tau(w,v)-\tau(u,v)\ge0.
\label{eq:excess_travel_time}
\end{equation}
The excess travel time \(t_{uvw}\) is zero on the alignment and grows
as the position \(x_w\) moves away from the alignment. With
\(\mu(H_{uv})\approx\tau(u,v)\) for every pair of nodes, the excess
travel time \(t_{uvw}\) approximates the mean number of extra hops
\(\mu(H_{uw})+\mu(H_{wv})-\mu(H_{uv})\) of the shortest path from \(u\)
to \(v\) through the node \(w\). For a constant velocity \(\nu\), the
excess travel time
\(t_{uvw}=\rho_{uvw}d_{uv}/\nu=\rho_{uvw}\,\mu(H_{uv})\) is the mean
number of extra hops in the Gaussian
estimate~\eqref{eq:gaussian_membership}.

Local link statistics give the estimated velocity field
\(\widehat\nu(x)\) but no estimate of the variance \(V_{uvw}\)
(Appendix~\ref{app:degree}), so that we cannot use the Gaussian
estimate~\eqref{eq:gaussian_membership} in an inhomogeneous network.
As in the Gaussian estimate~\eqref{eq:gaussian_membership} at a fixed
variance \(V_{uvw}\), however, the shortest-path node probability
\(\pi_{uv}(w)\) grows as the excess travel time \(t_{uvw}\) decreases.
We therefore use the excess travel time \(t_{uvw}\) to determine the
region in which the shortest-path nodes are likely to lie. We compare
the travel-time region \(C^\tau\), where the excess travel time
\(t_{uvw}\) is at most a threshold \(t_0\), with the near-geodesic
region \(C^E\), where the relative geometric deviation \(\rho_{uvw}\)
is at most a threshold \(\rho_0\),
\begin{equation}
C^\tau=\{x_w:t_{uvw}\le t_0\},\qquad
C^E=\{x_w:\rho_{uvw}\le\rho_0\}.
\label{eq:regions}
\end{equation}
The area budget \(A\) is the fraction of the area of the network that a
region covers. The thresholds \(t_0\) and \(\rho_0\) are chosen so that
the two regions have the same area budget \(A\)
(Appendixes~\ref{app:hetero} and~\ref{app:empirical}). The
near-geodesic region \(C^E\) is an ellipse with foci \(x_u\) and
\(x_v\) (Fig.~\ref{fig:detour}). The travel-time region \(C^\tau\)
follows the alignment and bends toward positions of larger velocity
\(\nu(x)\).

Let \(\mathcal S_{uv}(G)\) be the set of the shortest-path nodes of the
pair \((u,v)\) in a realization \(G\), the nodes \(w\) that satisfy
Eq.~\eqref{eq:membership_event}. The set \(\mathcal S_{uv}(G)\)
contains the end nodes \(u\) and \(v\), which lie in both regions
because the relative geometric deviation \(\rho_{uvw}\) and the excess
travel time \(t_{uvw}\) are zero for \(w=u\) and \(w=v\). The coverage
\(\kappa(C)\) of a region \(C\) is the fraction of the shortest-path
nodes in the set \(\mathcal S_{uv}(G)\) that lie inside the region
\(C\),
\begin{equation}
\kappa(C)=\frac{\bigl|\{w\in\mathcal S_{uv}(G):x_w\in C\}\bigr|}
{\bigl|\mathcal S_{uv}(G)\bigr|}.
\label{eq:coverage}
\end{equation}
The paired gain \(\Delta\kappa=\kappa(C^\tau)-\kappa(C^E)\) is the
difference between the coverages of the travel-time region \(C^\tau\)
and the near-geodesic region \(C^E\) of the same pair \((u,v)\) in the
same realization \(G\). A positive paired gain \(\Delta\kappa\) means
that the travel-time region \(C^\tau\) contains more shortest-path
nodes than the near-geodesic region \(C^E\) of the same area.

\subsection{Inhomogeneous model network}
\label{sec:hetero_model}
 
We test the regions first in an inhomogeneous model network in the unit
square (Appendix~\ref{app:hetero}). The node positions form a Poisson
point process whose intensity \(\lambda(x)\) has 14 Gaussian hubs on a
uniform background. The connection radius \(r(x)\) grows with the node
density \(\lambda(x)\). The velocity field \(\widehat\nu(x)\) in
Eq.~\eqref{eq:velocity_field} follows from the two model fields
\(\lambda(x)\) and \(r(x)\), not from the realizations.

\begin{figure}[tb]
\centering
\includegraphics[width=\linewidth]{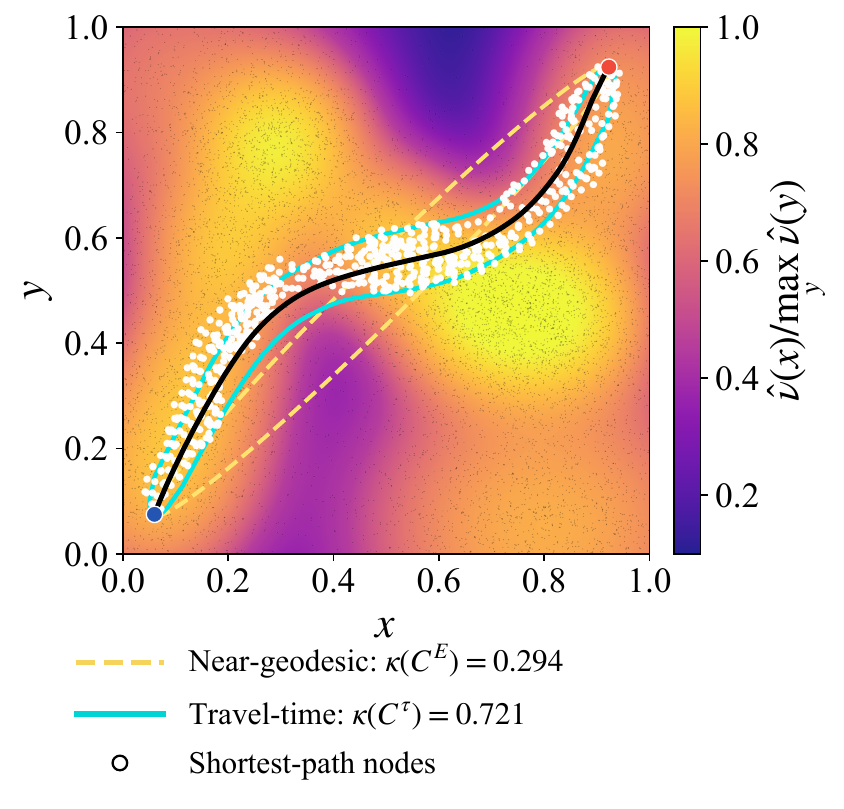}
\caption{%
\textbf{For one pair in our simulations of an inhomogeneous model network,
the travel-time region \(C^\tau\) contains more shortest-path nodes than
the near-geodesic region \(C^E\) of the same area.} The area budget is
\(A=0.10\). The background shows the velocity field \(\widehat\nu(x)\)
normalized by its maximum. White points are shortest-path nodes. The
dashed curve bounds the near-geodesic region \(C^E\) and the cyan curve the
travel-time region \(C^\tau\), which contain 0.29 and 0.72 of the
shortest-path nodes. The black curve is the alignment, which has the
shortest travel time \(\tau(u,v)\).}
\label{fig:hetero_pair}
\end{figure}

\begin{figure*}[tb]
\centering
\subfloat[Paired gain \(\Delta\kappa\) against the area budget \(A\).\label{fig:hetero_gain}]{%
\includegraphics[width=0.32\linewidth]{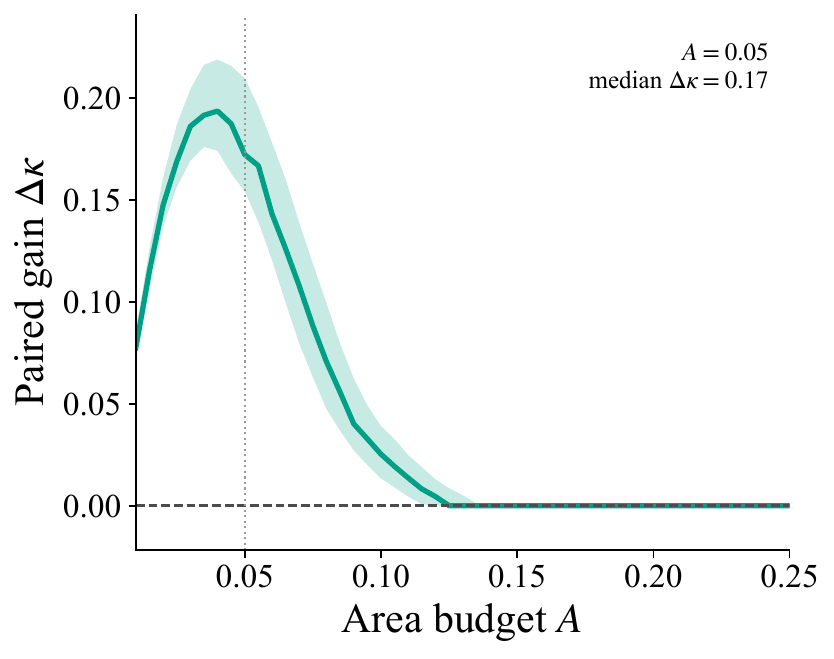}}
\hfill
\subfloat[Paired coverage \(\kappa\) at \(A=0.05\).\label{fig:hetero_paired}]{%
\includegraphics[width=0.32\linewidth]{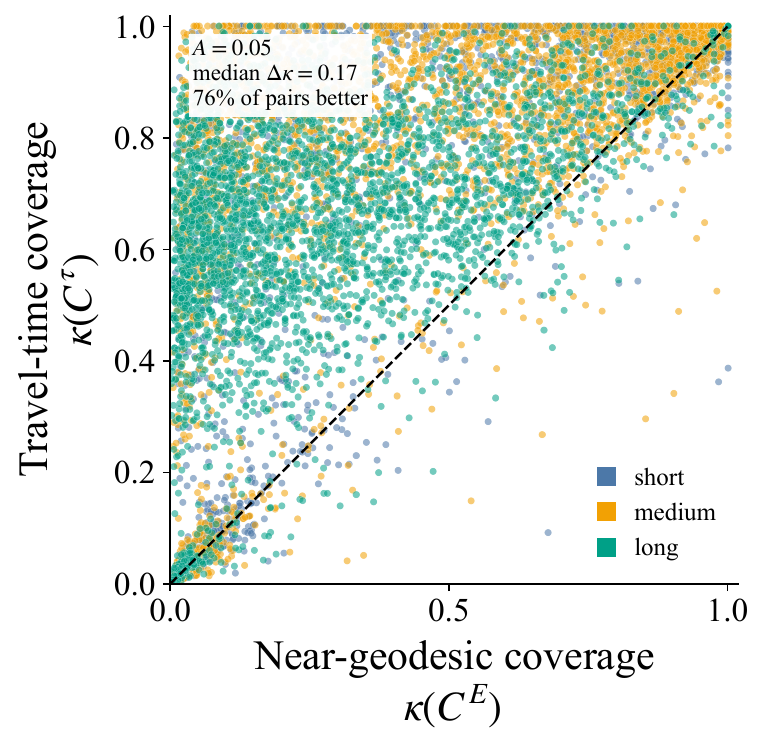}}
\hfill
\subfloat[Ratio \(A_{90}(C^\tau)/A_{90}(C^E)\) by distance \(d_{uv}\).\label{fig:hetero_A90}]{%
\includegraphics[width=0.32\linewidth]{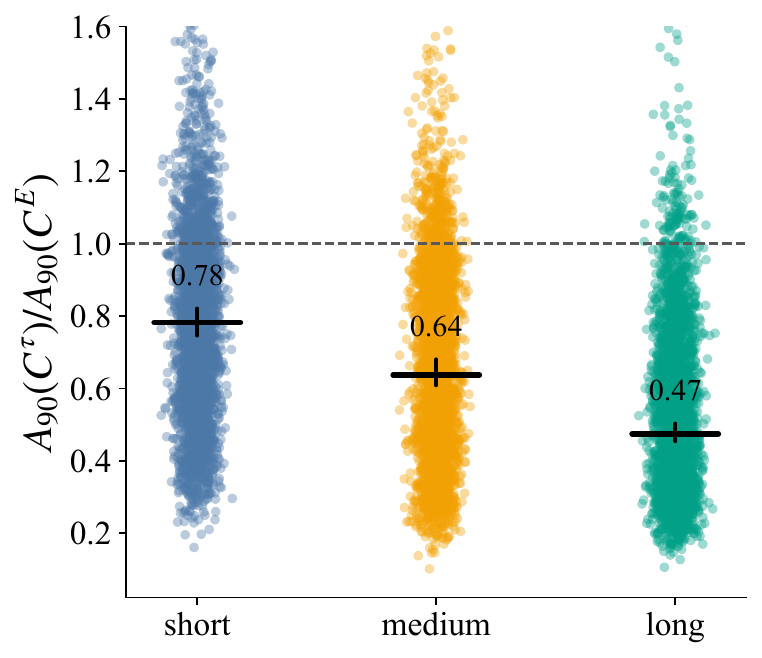}}
\caption{%
\textbf{In our simulations of an inhomogeneous model network, travel-time
regions \(C^\tau\) contain more shortest-path nodes than near-geodesic
regions \(C^E\) of the same area.} (a) Median paired gain
\(\Delta\kappa\) against the area budget \(A\). The band is the 95\%
interval of a hierarchical bootstrap over realizations and anchors.
(b) Coverage \(\kappa(C^\tau)\) against \(\kappa(C^E)\) at \(A=0.05\)
for 10{,}176 pairs in 20 realizations, colored by the third of the
distance \(d_{uv}\). Points above the diagonal are pairs for which the
travel-time region \(C^\tau\) contains more shortest-path nodes.
(c) Ratio \(A_{90}(C^\tau)/A_{90}(C^E)\) for the shortest, the middle
and the longest third of the pairs, with the median and its 95\%
interval. The dashed line marks equal areas.}
\label{fig:hetero_model}
\end{figure*}

Figure~\ref{fig:hetero_pair} shows one pair at the area budget
\(A=0.10\). In this simulated realization, the near-geodesic region
\(C^E\) contains 0.29 and the travel-time region \(C^\tau\) 0.72 of the
shortest-path nodes. For the statistics, we take 10{,}176 pairs of
anchor nodes spread over the square in 20 realizations
(Appendix~\ref{app:hetero}). In our simulations at the area budget
\(A=0.05\), the median paired gain is \(\Delta\kappa=0.17\)
[Fig.~\ref{fig:hetero_model}(a)]. At the same area budget \(A\), the
travel-time region \(C^\tau\) contains more shortest-path nodes in
76\% of the simulated pairs, the same number in 16\% and fewer in 8\%
[Fig.~\ref{fig:hetero_model}(b)].

In our simulations, the longer the distance \(d_{uv}\), the larger the
advantage of the travel-time region \(C^\tau\) around the alignment over
the near-geodesic region \(C^E\) around the geodesic. Let \(A_{90}\) be
the smallest area budget at which a region contains 90\% of the
shortest-path nodes of a pair. The travel-time region \(C^\tau\) needs
less area than the near-geodesic region \(C^E\) when the ratio
\(A_{90}(C^\tau)/A_{90}(C^E)\) is below one. In our simulations, the
median ratio \(A_{90}(C^\tau)/A_{90}(C^E)\) falls from 0.78 for the
shortest third of the pairs to 0.47 for the longest third
[Fig.~\ref{fig:hetero_model}(c)].
Differences in the velocity \(\widehat\nu(x)\) add up along a path, so
that the near-geodesic region \(C^E\) misses more shortest-path nodes on
long paths.

\subsection{Empirical networks}
\label{sec:empirical}
 
We test the travel time \(\tau(u,v)\) in 22 empirical networks: 15 urban street
networks, three rural and two regional road networks, the European power
grid and the OpenFlights airline network (Appendix~\ref{app:empirical}).
The airline network is an exploratory case, because its long links do
not describe local transport. Every link counts as one hop, whatever its
length. An empirical network has no model fields. For every node \(i\),
we measure the length \(L_{\max}\) of its longest link and the largest
projections \(P_{\max}\) of its links on eight equally spaced
directions. The midpoint of the length \(L_{\max}\) and the mean of the
eight projections, averaged over the nearest nodes of the node \(i\),
gives the estimated velocity \(\widehat\nu_i\)
(Appendix~\ref{app:empirical}). The travel time \(\widehat\tau(u,v)\) is
measured along the links. The link between the nodes \(i\) and \(j\)
takes the time \(2d_{ij}/(\widehat\nu_i+\widehat\nu_j)\). The travel
time \(\widehat\tau(u,v)\) is the smallest total time of a path from
\(u\) to \(v\). The travel-time region \(C^\tau\) in Eq.~\eqref{eq:regions} then uses the
travel time \(\widehat\tau\) in Eq.~\eqref{eq:excess_travel_time}.
 
We compare the travel-time region \(C^\tau\) and the near-geodesic
region \(C^E\) with the measures of the model network
(Sec.~\ref{sec:hetero_model}) in 1{,}000 sampled pairs per network
(Appendix~\ref{app:empirical}). Here the area budget \(A\) is a fraction
of the spatial cells that contain nodes. A median over pairs is the
median over the pairs of each network, followed by the median over the
22 networks.

\begin{figure}[!tbp]
\centering
\subfloat[Coverage \(\kappa\) of single pairs at \(A=0.05\).\label{fig:empirical_coverage}]{%
\includegraphics[width=0.88\linewidth]{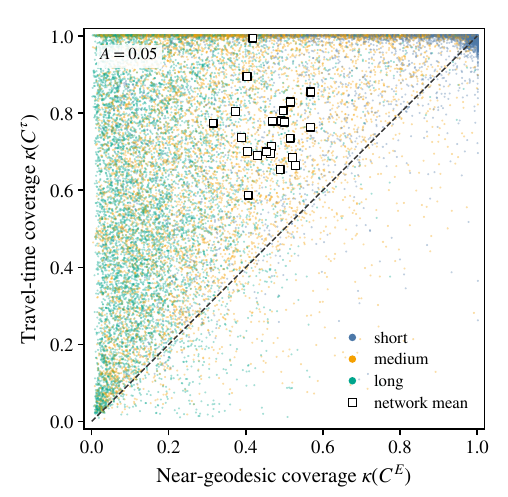}}

\subfloat[Ratio \(A_{90}(C^\tau)/A_{90}(C^E)\) by distance \(d_{uv}\).\label{fig:empirical_A90}]{%
\includegraphics[width=0.88\linewidth]{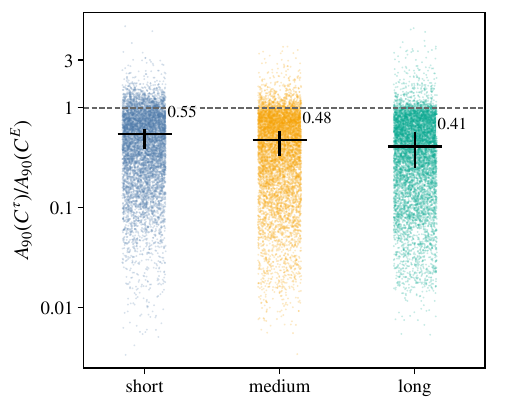}}
\caption{%
\textbf{In our tests on 22 empirical networks, travel-time regions
\(C^\tau\) contain more shortest-path nodes than near-geodesic regions
\(C^E\) of the same area.} (a) Coverage \(\kappa(C^\tau)\) of the
travel-time region against the coverage \(\kappa(C^E)\) of the
near-geodesic region at the area budget \(A=0.05\) for 22{,}000 pairs,
1{,}000 in each network, colored by the third of the distance \(d_{uv}\)
within the network. Each coverage is shifted at random by less than
\(1/(2|\mathcal S_{uv}(G)|)\), half the coverage of one node, so that
equal fractions of different pairs do not line up. Squares are the
network means. The coverages count the end nodes. Points above the dashed diagonal are pairs for which the
travel-time region \(C^\tau\) contains more shortest-path nodes.
(b) Ratio \(A_{90}(C^\tau)/A_{90}(C^E)\) of the same pairs for the
shortest, the middle and the longest third of the distance \(d_{uv}\)
in each network, on a logarithmic axis. Horizontal bars give the median
and vertical bars the interquartile range over the 22 networks of the
median in each network. The dashed line marks equal areas.}
\label{fig:empirical}
\end{figure}

In our tests at the area budget \(A=0.05\), the median paired gain is
\(\Delta\kappa=0.23\). In our tests at the same area budget \(A\), the
travel-time region \(C^\tau\) contains more shortest-path nodes in 74\%
of the sampled pairs, the same number in 21\% and fewer in 5\%
[Fig.~\ref{fig:empirical}(a)]. In our tests, the mean coverage
\(\kappa\) over the pairs of a network is larger for the travel-time
region \(C^\tau\) than for the near-geodesic region \(C^E\) in all 22
networks. As in the model network, the median ratio
\(A_{90}(C^\tau)/A_{90}(C^E)\) in our tests falls with the distance
\(d_{uv}\), from 0.55 for the shortest third of the pairs of each
network to 0.41 for the longest third [Fig.~\ref{fig:empirical}(b)].
Appendix~\ref{app:empirical} also compares the hopcounts predicted by
the travel time \(\widehat\tau\) and the profiles of the shortest-path
frequencies.
 
\section{Conclusion}
\label{sec_conclusions}

To answer the question of where shortest-path nodes lie in spatial
networks, we have proposed a Gaussian estimate of the shortest-path node
probability \(\pi_{uv}(w)\) that is based on two statistics of the
hopcount \(H_{uv}\), the mean hopcount \(\mu(H_{uv})\) and the hopcount
variance \(\sigma^2(H_{uv})\). Without the correlations between the
hopcounts \(H_{uw}\), \(H_{wv}\) and \(H_{uv}\), the mathematical form of
the Gaussian estimate predicts that three variables affect the
shortest-path node probability \(\pi_{uv}(w)\): the relative geometric
deviation \(\rho_{uvw}\), the mean hopcount \(\mu(H_{uv})\) and the
variance \(V_{uvw}\). In our simulations, each variable changes the
shortest-path node probability \(\pi_{uv}(w)\) in the direction that the
Gaussian estimate predicts, which supports the validity of the Gaussian
estimate. We have further combined the three variables into the
standardized detour \(X_{uvw}=\rho_{uvw}\mu(H_{uv})/\sqrt{V_{uvw}}\),
the only variable through which the three variables enter the exponent of
the Gaussian estimate. In our simulations of hard and soft random
geometric graphs in the Euclidean plane, the hyperbolic plane and the
sphere, 84 normalized profiles \(\widetilde\pi\) of the shortest-path
node probability \(\pi_{uv}(w)\) are widely separated against the
relative geometric deviation \(\rho\) but approximately follow one curve
against the mean standardized detour \(\overline X\). Because the mean
number of extra hops \(\rho_{uvw}\mu(H_{uv})\) is the same at every
position of an ellipse with foci \(x_u\) and \(x_v\), the Gaussian
estimate gives nearly the same shortest-path node probability
\(\pi_{uv}(w)\) along each ellipse. Shortest paths with a common end
node, however, share links, so that the hopcounts \(H_{uw}\), \(H_{wv}\)
and \(H_{uv}\) are correlated. In our simulations near the geodesic, the
correlations make the shortest-path node probability \(\pi_{uv}(w)\)
higher close to the end nodes \(u\) and \(v\) than elsewhere on the same
ellipse, which helps explain the two lobes near the end nodes.

When the hopcount variance \(\sigma^2(H_{uv})\) is very small, which in
our simulations happens at a large mean degree, the hopcount \(H_{uv}\)
at a given distance \(d_{uv}\) is strongly concentrated and the field of
the shortest-path node probability \(\pi_{uv}(w)\) collapses. The
continuous field turns into separate regions. Inside the regions, the
shortest-path node probability \(\pi_{uv}(w)\) tends to one. Outside the
regions, the shortest-path node probability \(\pi_{uv}(w)\) tends to
zero. The transition between the inside and the outside is narrow. We
call this phenomenon the quantized distribution of shortest-path nodes.
The width of the regions is set by the slack \(W(d_{uv})\), the amount
by which the length covered by the predicted hops of length \(\nu\)
exceeds the distance \(d_{uv}\) between the two nodes. The width
shrinks as the distance \(d_{uv}\) grows and jumps up at each multiple of
the velocity \(\nu\). In the Euclidean plane and the hyperbolic plane,
the regions are petals along the geodesic. On the sphere, the regions
can close into bands.

To address shortest paths in real spatial networks, we have studied
spatially inhomogeneous networks, in which the node density \(\lambda(x)\)
and the link rule vary with the position \(x\). We have proposed a
method that estimates the velocity field \(\nu(x)\) from local
information alone, the mean length \(E[L_{\max}]\) of the longest link
and the mean largest link projection \(E[P_{\max}]\) near the position
\(x\), without shortest-path data. In the 45 conditions of our
simulations of homogeneous networks, the estimated velocity
\(\widehat\nu\) differs from the velocity \(\nu\) by at most 10\%. The
travel time \(\tau(u,v)\), which follows from the velocity field
\(\nu(x)\), takes the place of the distance \(d_{uv}\). The travel-time
region \(C^\tau\) then takes the place of the near-geodesic region
\(C^E\). In our simulations of an
inhomogeneous model network at the area budget \(A=0.05\), the median
paired gain \(\Delta\kappa\) of the travel-time region \(C^\tau\) over
the near-geodesic region \(C^E\) of the same area is 0.17. In our tests on 22 empirical networks, the
travel-time region \(C^\tau\) contains more shortest-path nodes than the
near-geodesic region \(C^E\) of the same area in every network. At the area
budget \(A=0.05\), the median coverage \(\kappa\) in our tests rises from
0.47 for the near-geodesic region \(C^E\) to 0.75 for the travel-time region
\(C^\tau\).

\begin{acknowledgments}
P. Van Mieghem is supported by the European Research Council under the European Union’s Horizon 2020 research and innovation program (Grant Agreement 101019718).
\end{acknowledgments}

\newpage

\newpage

\appendix

\section{Simulations}
\label{app:simulations}

\subsection{Shortest-path nodes in a realization}

In each realization \(G\) of our simulations, the source \(u\) and the
destination \(v\) are added to the \(N\) nodes and are linked by the
same rule as the other nodes. A breadth-first search from the source
\(u\) and one from the destination \(v\) give the hopcounts \(h_{uw}\)
and \(h_{wv}\) of every node \(w\) and the hopcount \(h_{uv}\). A node
\(w\) is a shortest-path node when Eq.~\eqref{eq:membership_event}
holds. The test needs no list of shortest paths and counts a
shortest-path node once, however many shortest paths contain the node.

We use the realizations in which the source \(u\) and the destination
\(v\) are connected. A node outside the component that contains the
source \(u\) and the destination \(v\) is not a shortest-path node. To
estimate a probability field, we divide the space into cells. The
estimate of the shortest-path node probability \(\pi_{uv}(w)\) in a
cell is the number of shortest-path nodes in the cell divided by the
number of nodes in the cell, both summed over the realizations. The
source \(u\) and the destination \(v\) are not counted.

Several experiments place a test node \(w\) at a prescribed position.
The test node \(w\) is added to the realization at the prescribed
position and is linked by the same rule as the other nodes. Only one
test node is added to a graph at a time. The test node \(w\) can shorten
the shortest path from the source \(u\) to the destination \(v\), so
the hopcount \(h_{uv}\) is computed after the test node \(w\) is added.

\subsection{Experiments}

Table~\ref{tab:experiments} lists the seven simulation experiments. In the
profile experiment, the geodesic reference experiment and the
joint-hopcount experiment, lengths are measured in units of the
connection radius \(r\) of the hard RGG. In these three experiments, the
node density \(\lambda\) of a network is the mean degree of the network
divided by the area of a ball of radius \(r\) in the space. A soft RGG then has the
same node density \(\lambda\) as the hard RGG with the same mean degree.
The scale \(\alpha\) of the link probability
\(p(d)=1/[1+(\alpha d)^\beta]\) is chosen so that the integral of the
link probability \(p(d)\) over the space equals the area of a ball of
radius \(r\). The link probability \(p(d)\) is zero beyond the cutoff
distance \(d_c\) at which the link probability \(p(d_c)\) equals the
cutoff probability \(p_c\). The cutoff probability is \(p_c=10^{-4}\) in
the profile experiment and in the geodesic reference experiment and
\(p_c=10^{-3}\) in the joint-hopcount experiment.

\begin{table*}[tb]
\caption{Simulation experiments. The mean hopcount \(\mu(H_{uv})\) of
the source--destination pairs is given where the distances
\(d_{uv}\) are chosen by the mean hopcount.}
\label{tab:experiments}
\begin{ruledtabular}
\footnotesize
\begin{tabular}{p{2.0cm}p{3.0cm}p{3.4cm}p{1.9cm}p{3.0cm}p{2.6cm}}
Experiment & Used in & Networks & Space & Nodes & Pairs \((u,v)\)\\
\hline
Unit square & Fig.~\ref{fig:probability} & Soft RGG, mean degree 14, \(\beta=4\) & Euclidean plane & \(N=10^5\) in the unit square & \(d_{uv}=0.75\)\\
Deviation & Figs.~\ref{fig:result1_controls}(a), \ref{fig:result1_controls}(b) & Soft RGG, mean degree 14, \(\beta=8\) & Euclidean plane & \(N=30{,}000\) in the square \([0,4]^2\) & \(\mu(H_{uv})\) from 14 to 48\\
Profile & Figs.~\ref{fig:result2_statistical_control}(a), \ref{fig:result2_statistical_control}(b), Tables~\ref{tab:velocity}, \ref{tab:regions}, \ref{tab:geodesic_positions}, \ref{tab:equal_detour} & 21 conditions (Table~\ref{tab:profile_conditions}) & Three spaces & 126{,}947 to 321{,}288 & \(d_{uv}=60\), 100, 140, 180\\
Geodesic reference & Figs.~\ref{fig:result2_statistical_control}(a), \ref{fig:result2_statistical_control}(b) & As in the profile experiment & Three spaces & As in the profile experiment & As in the profile experiment\\
Joint hopcount & Figs.~\ref{fig:result1_controls}(c), \ref{fig:result2_statistical_control}(c), \ref{fig:result2_statistical_control}(d), Table~\ref{tab:velocity} & 24 conditions & Three spaces & 14{,}942 to 37{,}286 & \(\mu(H_{uv})\approx14\), 22, 32, 43\\
End node & Appendix~\ref{app:correlations} & 6 networks & Euclidean plane & \(N=20{,}000\) & \(d_{uv}=0.7\) to 2.5\\
Cross-space & Appendix~\ref{app:accuracy}, Fig.~\ref{fig:computation}, Table~\ref{tab:computation} & 189 networks, 567 conditions & Three spaces & \(N_b=4{,}000\) to 64{,}000 in a domain of radius 1 & \(d_{uv}=0.25\) to 1.7\\
\end{tabular}
\end{ruledtabular}
\end{table*}

\paragraph*{Unit-square experiment.}
The probability field of Fig.~\ref{fig:probability} comes from a soft
RGG with \(N=10^5\) nodes in the unit square, the mean degree 14 and the
steepness \(\beta=4\). The source \(u\) and the destination \(v\) are at
the distance \(d_{uv}=0.75\). The probability field is estimated on a
square grid of cells.

\paragraph*{Deviation experiment.}
Figures~\ref{fig:result1_controls}(a) and \ref{fig:result1_controls}(b)
use a soft RGG with \(N=30{,}000\) nodes in the square \([0,4]^2\), the
mean degree 14 and the steepness \(\beta=8\). The midpoint \(x_m\) of
the positions \(x_u\) and \(x_v\) is the center of the square. For a
relative geometric deviation \(\rho_{uvw}\), two test nodes are placed at
the positions
\begin{equation}
x_w=x_m\pm\tfrac12\,d_{uv}\sqrt{(1+\rho_{uvw})^2-1}\;n,
\label{eq:app_test_positions}
\end{equation}
where \(n\) is a unit vector perpendicular to the geodesic from \(x_u\)
to \(x_v\). Both positions have \(d_{uw}=d_{wv}=(1+\rho_{uvw})d_{uv}/2\),
so that the test nodes lie at equal distance from the source \(u\) and
the destination \(v\). In Fig.~\ref{fig:result1_controls}(a), the
relative geometric deviation \(\rho_{uvw}\) runs from 0 to 0.12 at four
distances \(d_{uv}\) with the mean hopcounts \(\mu(H_{uv})\approx14\),
22, 32 and 43. In Fig.~\ref{fig:result1_controls}(b), the relative
geometric deviation \(\rho_{uvw}\) takes the values 0.008, 0.016, 0.032
and 0.064 at eight distances \(d_{uv}\) with the mean hopcount
\(\mu(H_{uv})\) from 14 to 48.

\paragraph*{Profile experiment.}
Table~\ref{tab:profile_conditions} lists the 21 conditions. The domain
\(\Omega\) is a disk of radius \(R_\Omega=130\) in the Euclidean plane,
a disk of radius \(R_\Omega=130\) in the hyperbolic plane of curvature
\(-1/L^2\) and a cap of radius \(R_\Omega=130\) on a sphere of radius
\(L\), with \(L=150\). One hard RGG with the mean degree 10 in each space
uses the domain radius \(R_\Omega=170\) to test the effect of the
boundary. The nodes are uniform in the area of the space. The number of
nodes \(N\) is the node density \(\lambda\) times the area
\(|\Omega|\) of the domain, rounded to an integer. The source \(u\) and
the destination \(v\) lie on a geodesic through the center of the
domain, at equal distance from the center, at the distances
\(d_{uv}=60\), 100, 140 and 180. The direction of the pair is uniformly
random in each realization. We use geodesic disks because a disk is
defined in the same way in all three spaces and keeps the boundary at
the same distance in every direction of the pair. The four distances use the same nodes, but
each distance is evaluated in a separate graph that contains one pair
\((u,v)\). In our simulations of the profile experiment, the source
\(u\) and the destination \(v\) are connected in at least 99.6\% of the
realizations of every condition and distance. In our simulations of the
profile experiment, the mean hopcount \(\mu(H_{uv})\) of the test
realizations lies between 47.7 and 245.9. The 92{,}416 realizations
are split by the index of the realization into training (40\%), validation (20\%) and
test (40\%) realizations. The training realizations fix the linear law,
the variance law and the merging of grid cells (Appendixes~\ref{app:hopcount}
and~\ref{app:profiles}). The profiles of
Fig.~\ref{fig:result2_statistical_control} use the test realizations.

\begin{table}[tb]
\caption{Conditions of the profile experiment. The hyperbolic plane has
the curvature \(-1/L^2\) and the sphere has the radius \(L\), with
\(L=150\). Lengths are in units of the connection radius \(r\) of the
hard RGG. Degree is the mean degree. The steepness \(\beta\) applies to
soft RGGs.}
\label{tab:profile_conditions}
\begin{ruledtabular}
\footnotesize
\begin{tabular}{lcccrrrr}
Space & Rule & Degree & \(\beta\) & \(N\) & \multicolumn{3}{c}{Realizations}\\
 & & & & & Training & Validation & Test\\
\hline
Euclidean & Hard & 8 & -- & 135{,}200 & 5{,}324 & 2{,}660 & 5{,}320\\
Euclidean & Hard & 10 & -- & 169{,}000 & 3{,}816 & 1{,}904 & 3{,}808\\
Euclidean & Hard & 14 & -- & 236{,}600 & 928 & 460 & 920\\
Euclidean & Soft & 8 & 6 & 135{,}200 & 1{,}096 & 544 & 1{,}088\\
Euclidean & Soft & 10 & 6 & 169{,}000 & 736 & 364 & 728\\
Euclidean & Soft & 10 & 12 & 169{,}000 & 1{,}448 & 724 & 1{,}448\\
Euclidean\(^\dagger\) & Hard & 10 & -- & 289{,}000 & 844 & 420 & 840\\
Hyperbolic & Hard & 8 & -- & 143{,}877 & 2{,}568 & 1{,}284 & 2{,}560\\
Hyperbolic & Hard & 10 & -- & 179{,}846 & 1{,}880 & 940 & 1{,}880\\
Hyperbolic & Hard & 14 & -- & 251{,}784 & 464 & 232 & 456\\
Hyperbolic & Soft & 8 & 6 & 143{,}877 & 968 & 484 & 964\\
Hyperbolic & Soft & 10 & 6 & 179{,}846 & 648 & 324 & 648\\
Hyperbolic & Soft & 10 & 12 & 179{,}846 & 1{,}320 & 656 & 1{,}312\\
Hyperbolic\(^\dagger\) & Hard & 10 & -- & 321{,}288 & 400 & 200 & 392\\
Spherical & Hard & 8 & -- & 126{,}947 & 5{,}600 & 2{,}796 & 5{,}592\\
Spherical & Hard & 10 & -- & 158{,}684 & 3{,}888 & 1{,}944 & 3{,}880\\
Spherical & Hard & 14 & -- & 222{,}157 & 936 & 468 & 932\\
Spherical & Soft & 8 & 6 & 126{,}947 & 1{,}096 & 544 & 1{,}088\\
Spherical & Soft & 10 & 6 & 158{,}684 & 708 & 352 & 704\\
Spherical & Soft & 10 & 12 & 158{,}684 & 1{,}464 & 732 & 1{,}456\\
Spherical\(^\dagger\) & Hard & 10 & -- & 259{,}362 & 896 & 448 & 892\\
\end{tabular}
\end{ruledtabular}
\begin{flushleft}
\footnotesize \(^\dagger\)Domain radius \(R_\Omega=170\). All other
conditions have \(R_\Omega=130\).
\end{flushleft}
\end{table}

\paragraph*{Geodesic reference experiment.}
The geodesic reference experiment uses the 21 conditions and the four
distances \(d_{uv}\) of the profile experiment and places test nodes on
the geodesic from \(x_u\) to \(x_v\) (Appendix~\ref{app:profiles}).

\paragraph*{Joint-hopcount experiment.}
The joint-hopcount experiment records the three hopcounts \(H_{uw}\),
\(H_{wv}\) and \(H_{uv}\) of the same realization. The experiment
contains eight networks in each of the three spaces: hard RGGs with the
mean degrees 10, 14 and 18, soft RGGs with the steepness \(\beta=8\) and
the mean degrees 10, 14 and 18 and soft RGGs with the mean degree 14 and
the steepness \(\beta=6\) or \(\beta=12\). The domain \(\Omega\) is a
disk of radius 42 in the Euclidean plane, a disk of radius 42 in the
hyperbolic plane of curvature \(-1/L^2\) and a cap of radius 42 on a
sphere of radius \(L\), with \(L=30\). The networks have between
14{,}942 and 37{,}286 nodes. Each condition has 1{,}200 realizations,
split by the index of the realization into 480 training, 240 validation and 480 test
realizations. In each realization, two pairs \((u,v)\) with independent
random directions are placed at each of four distances \(d_{uv}\). The
four distances \(d_{uv}\) give the mean hopcounts
\(\mu(H_{uv})\approx14\), 22, 32 and 43. Most test nodes lie at equal
distance from the source \(u\) and the destination \(v\), at 24 values of
the relative geometric deviation \(\rho_{uvw}\) between 0 and 0.12. Four
further test nodes lie on the geodesic. In the Euclidean plane, the
positions at equal distance follow Eq.~\eqref{eq:app_test_positions}. In the hyperbolic plane
and on the sphere, the positions are constructed with the intrinsic
distances. Figure~\ref{fig:result1_controls}(c) uses the six Euclidean
networks with the hard rule or with the steepness \(\beta=8\), at the
mean hopcount \(\mu(H_{uv})\approx32\) and the relative geometric
deviation \(\rho_{uvw}=0.064\).

\paragraph*{End-node experiment.}
The end-node experiment places \(N=20{,}000\) nodes uniformly in the
Euclidean plane at the node density 1{,}875. The experiment contains
hard RGGs with the mean degrees 10, 14 and 18 and soft RGGs with the
mean degree 14 and the steepness \(\beta=6\), 8 or 12. Each network is
simulated at the five distances \(d_{uv}=0.7\), 1.0, 1.4, 2.0 and 2.5,
which gives 30 conditions. Test nodes lie on the geodesic at 0.2, 0.5
and 0.8 of the way from the source \(u\) to the destination \(v\). In
our simulations of the end-node experiment, the mean hopcount
\(\mu(H_{uv})\) lies between 12 and 80. In our simulations of the
end-node experiment, every condition and position has at least
23{,}813 recorded triples \((h_{uw},h_{wv},h_{uv})\).

\paragraph*{Cross-space experiment.}
The cross-space experiment places \(N_b=4{,}000\), 16{,}000 or 64{,}000
nodes uniformly in the area of a disk of radius 1 in the Euclidean
plane, a disk of radius 1 in the hyperbolic plane of curvature
\(-1/L^2\) or a cap of radius 1 on a sphere of radius \(L\). The
curvature radius is \(L=2/3\) in the hyperbolic plane and \(L=5/6\) on
the sphere, with the further values \(L=1/3\) and \(L=2\) in the
hyperbolic plane and \(L=1/\pi\) and \(L=2\) on the sphere. For
\(L=1/\pi\), the cap covers the whole sphere. The networks are hard RGGs
and soft RGGs with the steepness \(\beta=3\), 4, 6, 8, 12 or 16 and the
cutoff probability \(p_c=10^{-4}\). The connection radius \(r\) of a
hard RGG or the scale \(\alpha\) of a soft RGG sets the expected number
6, 8, 14, 24 or 40 of neighbors of a node at the center of the domain.
The experiment combines the curvature radii, the rules, the numbers of
nodes and the expected numbers of neighbors in 189 networks. The source \(u\) and
the destination \(v\) lie on a geodesic through the center of the
domain, at equal distance from the center, at the distance
\(d_{uv}=0.5\), 1.1 or 1.7. On the whole sphere, the distance is
\(d_{uv}=0.25\), 0.55 or 0.85. The 189 networks and three distances give
567 conditions. Test nodes lie at 13 positions along the geodesic, from
0.01 to 0.99 of the way from the source \(u\) to the destination \(v\).
At each position, the relative geometric deviation is
\(\rho_{uvw}=0\), 0.0005, 0.001, 0.002, 0.005, 0.01, 0.02, 0.04, 0.08,
0.12, 0.2 or 0.35 on both sides of the geodesic. Test positions outside the domain are left out, which
leaves 247 to 299 test positions per condition and 167{,}685 test
positions in total. The realizations are split by the index of the
realization into training (30\%), validation (20\%) and test (50\%)
realizations. In our simulations of the cross-space experiment, each
condition has 1{,}073 to 1{,}219 realizations, of which 324 to 366 are
training realizations and 535 to 609 are test realizations. The shortest-path node
probability \(\pi_{uv}(w)\) is the frequency of \(Z_{uvw}=0\) in the
test realizations in which the three hopcounts \(H_{uw}\), \(H_{wv}\)
and \(H_{uv}\) are finite. In our simulations of the cross-space
experiment, the source \(u\) and the destination \(v\) are connected in
at least 96.6\% of the realizations of every condition. In the same
simulations, the mean hopcount \(\mu(H_{uv})\) lies between 3.6 and 220.

\section{Hopcount statistics}
\label{app:hopcount}

\subsection{Distribution of the hopcount}

In our simulations of the joint-hopcount experiment, we inspect the
hopcount \(H_{uv}\) of the two pairs \((u,v)\) in the first 150
realizations of every condition, which gives 300 values at each of the
four distances \(d_{uv}\). We ignore the hopcount values that occur in
at most 1\% of the pairs, all of which lie in the tails. In our
simulations, the histogram of the hopcount \(H_{uv}\) then has a single
peak in all 96 combinations of condition and distance: the counts do not
increase away from the most frequent value. In our simulations, the
skewness of the hopcount \(H_{uv}\) lies between \(-0.40\) and 1.04.

\subsection{Linear law and variance law}

For each condition, we fit the linear law~\eqref{eq:linear_law} with an
intercept \(c\),
\begin{equation}
\mu(H_{uv})=\frac{d_{uv}}{\nu}+c,
\label{eq:app_linear_fit}
\end{equation}
by weighted least squares. Because the distance \(d_{uv}\) is bounded on
the sphere and in the finite domains of our simulations, the velocity
\(\nu\) is the inverse of the fitted slope (Table~\ref{tab:velocity}),
not a limit for \(d_{uv}\to\infty\). We fit the variance
law~\eqref{eq:variance_law} as a straight line of the logarithm of the
hopcount variance \(\sigma^2(H_{uv})\) against the logarithm of the mean
hopcount \(\mu(H_{uv})\). In the profile experiment, both fits use the
training realizations at the ten distances \(d_{uv}=24\), 40, 60, 80,
100, 120, 140, 160, 180 and 200, weighted by the number of connected
pairs. In the joint-hopcount experiment, both fits use 100 distances
\(d_{uv}\) per condition.

In our simulations, the coefficient of determination of the linear
law~\eqref{eq:app_linear_fit} is at least 0.9999 in all 45 conditions.
In our simulations, the intercept \(c\) lies between 0.56 and 3.30
hops. In our simulations of the profile experiment, the mean hopcount at
the distance \(d_{uv}=12\), below the fitted distances and with the mean
hopcount between 11 and 18, exceeds the term
\(d_{uv}/\nu\) by 0.66 to 0.74 times the intercept \(c\). In our
simulations, the fitted exponent \(\gamma\) of the
variance law~\eqref{eq:variance_law} lies between 0.31 and 0.75 in the
45 conditions. In our simulations, the coefficient of determination of
the logarithmic variance fit lies between 0.80 and 0.99.

\subsection{Effect of the variance}

The Gaussian estimate~\eqref{eq:gaussian_membership} depends on the
mean number of extra hops \(m=\rho_{uvw}\mu(H_{uv})\) and on the
variance \(V_{uvw}\). At a fixed mean number of extra hops \(m\),
\begin{equation}
\frac{\partial\ln\pi_{uv}(w)}{\partial V_{uvw}}
=\frac{m^2-V_{uvw}}{2V_{uvw}^2}.
\label{eq:app_dlnpi_dV}
\end{equation}
A larger variance \(V_{uvw}\) thus raises the Gaussian estimate where
the mean number of extra hops \(m\) exceeds the standard deviation
\(\sqrt{V_{uvw}}\) and lowers the Gaussian estimate where the mean
number of extra hops \(m\) is smaller than the standard deviation
\(\sqrt{V_{uvw}}\).

To vary the second variable, the mean hopcount \(\mu(H_{uv})\), we fix
the relative geometric deviation \(\rho_{uvw}\) and the ratios
\(d_{uw}/d_{uv}\) and \(d_{wv}/d_{uv}\). By
the variance law~\eqref{eq:variance_law}, the variance \(V_{uvw}\) is
then proportional to \(\mu(H_{uv})^\gamma\). With the standardized
detour \(X_{uvw}\) of Eq.~\eqref{eq:standardized_mismatch},
\begin{equation}
\frac{d\ln\pi_{uv}(w)}{d\ln\mu(H_{uv})}
=-\frac{\gamma}{2}-\frac{2-\gamma}{2}\,X_{uvw}^2,
\label{eq:app_dlnpi_dmu}
\end{equation}
which is negative for \(\gamma<2\). A longer pair therefore has a smaller
Gaussian estimate at the same relative geometric deviation
\(\rho_{uvw}\).

Near the geodesic, the two effects of the variance \(V_{uvw}\) nearly
balance in our simulations. In our simulations of the six Euclidean
networks of Fig.~\ref{fig:result1_controls}(c), the shortest-path node
probability \(\pi_{uv}(w)\) at the midpoint of the geodesic, where
\(\rho_{uvw}=0\), lies between 0.36 and 0.39 in the three hard RGGs and
between 0.15 and 0.17 in the three soft RGGs. In our simulations at the
relative geometric deviation \(\rho_{uvw}=0.008\), the shortest-path
node probability \(\pi_{uv}(w)\) of the three hard RGGs lies between
0.264 and 0.267. In our simulations at the relative geometric deviation
\(\rho_{uvw}=0.064\), the shortest-path node probability
\(\pi_{uv}(w)\) instead changes by a factor of 8 in the hard RGGs (from
0.0027 to 0.0225) and by a factor of 3 in the soft RGGs (from 0.0046 to
0.0142).

\subsection{Networks of Fig.~\ref{fig:result1_controls}(c)}

Table~\ref{tab:fig3c} lists the six networks of
Fig.~\ref{fig:result1_controls}(c). The hopcount variance
\(\sigma_0^2\) is the sample variance of the hopcount \(H_{uv}\) of the
2{,}400 pairs \((u,v)\) of a network. The variance \(V_{uvw}\) is the sum
of the measured hopcount variances at the mean hopcounts of the three
pairs \((u,v)\), \((u,w)\) and \((w,v)\), interpolated linearly in the
measured mean hopcount. Each network has 1{,}200 realizations with two
pairs and two test positions each, which gives 4{,}800 tests. The 95\%
intervals come from 2{,}000 bootstrap samples of whole realizations. In
our simulations, the mean number of extra hops
\(\rho_{uvw}\mu(H_{uv})\) exceeds the standard deviation
\(\sqrt{V_{uvw}}\) by at least 0.068 hops in all six networks. In our
simulations, the shortest-path node probability \(\pi_{uv}(w)\) of the
soft RGG with the mean degree 14 exceeds the value of the soft RGG with
the mean degree 18 by 0.0025, with the 95\% interval from \(-0.0004\) to
0.0054. In our simulations, the other three differences between
networks with neighboring mean degrees are positive at the 95\% level.

\begin{table*}[tb]
\caption{The six Euclidean networks of
Fig.~\ref{fig:result1_controls}(c) at the relative geometric deviation
\(\rho_{uvw}=0.064\). The interval is the 95\% bootstrap interval of the
shortest-path node probability \(\pi_{uv}(w)\). Degree is the mean degree.}
\label{tab:fig3c}
\begin{ruledtabular}
\begin{tabular}{lcrrrrcrr}
Rule & Degree & \(d_{uv}\) & \(\mu(H_{uv})\) & \(\sigma_0^2\) & \(\pi_{uv}(w)\) & Interval & \(\rho_{uvw}\mu(H_{uv})\) & \(\sqrt{V_{uvw}}\)\\
\hline
Hard & 10 & 23.82 & 31.93 & 1.28 & 0.0225 & [0.0183, 0.0269] & 2.04 & 1.81\\
Hard & 14 & 25.71 & 32.26 & 0.62 & 0.0060 & [0.0040, 0.0081] & 2.06 & 1.24\\
Hard & 18 & 26.72 & 32.30 & 0.42 & 0.0027 & [0.0013, 0.0044] & 2.07 & 1.00\\
Soft & 10 & 30.30 & 32.02 & 1.63 & 0.0142 & [0.0106, 0.0175] & 2.05 & 1.98\\
Soft & 14 & 33.68 & 32.27 & 1.03 & 0.0071 & [0.0050, 0.0096] & 2.07 & 1.58\\
Soft & 18 & 35.63 & 32.10 & 0.79 & 0.0046 & [0.0027, 0.0067] & 2.05 & 1.38\\
\end{tabular}
\end{ruledtabular}
\end{table*}

\section{Velocity}
\label{app:travel}

\subsection{Local link statistics}

For a node \(u\) at the position \(x_u\), let \(L_{\max}\) be the length
of the longest link of the node \(u\). For a unit direction \(e\), let
\begin{equation}
P_{\max}=\max\Bigl(0,\ \max_{v}\,(x_v-x_u)\cdot e\Bigr)
\label{eq:app_Pmax}
\end{equation}
be the largest projection of the links of the node \(u\) on the
direction \(e\), where \(v\) runs over the neighbors of the node \(u\).
The largest projection \(P_{\max}\) is zero when no link of the node
\(u\) has a positive projection on the direction \(e\). The length
\(L_{\max}\) and the projection \(P_{\max}\) are zero for an isolated
node.

\subsection{Means for a Poisson point process}

We make two assumptions. First, the nodes other than the node \(u\) form
a Poisson point process with the node density \(\lambda\) in the
Euclidean plane. Second, a node at the distance \(d\) is linked to the
node \(u\) independently with the link probability \(p(d)\). The
neighbors of the node \(u\) then form a Poisson point process with the
density \(\lambda p(d)\). The length \(L_{\max}\) is at most \(t\) when
no neighbor lies farther than \(t\), so that
\begin{equation}
F_L(t)=\Pr[L_{\max}\le t]
=\exp\!\left[-2\pi\lambda\int_t^\infty p(d)\,d\,\mathrm{d}d\right].
\label{eq:app_FL}
\end{equation}
The projection \(P_{\max}\) is at most \(t\ge0\) when no neighbor lies
beyond the line \(x\cdot e=t\). A circle of radius \(d>t\) crosses this
line on an arc of the angle \(2\arccos(t/d)\), so that
\begin{equation}
\begin{split}
F_P(t)&=\Pr[P_{\max}\le t]\\
&=\exp\!\left[-2\lambda\int_t^\infty p(d)\,d\,\arccos(t/d)\,\mathrm{d}d\right].
\end{split}
\label{eq:app_FP}
\end{equation}
Both quantities are nonnegative, so that
\begin{equation}
\begin{split}
E[L_{\max}]&=\int_0^\infty[1-F_L(t)]\,\mathrm{d}t,\\
E[P_{\max}]&=\int_0^\infty[1-F_P(t)]\,\mathrm{d}t.
\end{split}
\label{eq:app_means}
\end{equation}
For the hard RGG, the link probability is \(p(d)=1\) for \(d\le r\) and
\(p(d)=0\) otherwise. The mean degree is then \(\lambda\pi r^2\).
Equations~\eqref{eq:app_FL} and~\eqref{eq:app_means} give
\begin{equation}
E[L_{\max}]=r\left[1-\frac{\sqrt{\pi}\,e^{-\lambda\pi r^2}}{2\sqrt{\lambda\pi r^2}}\,
\operatorname{erfi}\bigl(\sqrt{\lambda\pi r^2}\bigr)\right],
\label{eq:app_hard_EL}
\end{equation}
where \(\operatorname{erfi}(z)=2\pi^{-1/2}\int_0^z e^{y^2}\mathrm{d}y\).
The mean \(E[P_{\max}]\) follows by numerical integration. For the soft
RGG, Eqs.~\eqref{eq:app_FL}--\eqref{eq:app_means} use the link
probability \(p(d)\) with the cutoff distance \(d_c\) of the experiment
(Appendix~\ref{app:simulations}). Because the links are short compared
with the curvature radius \(L\), we use the planar formulas in all three
spaces.

\subsection{Bounds on the velocity}

Let a shortest path from the source \(u\) to the destination \(v\) visit
the nodes at the positions \(X_0=x_u,X_1,\ldots,X_{H}=x_v\), where
\(H=H_{uv}\). Each link of the path is at most as long as the longest
link of the node at the start of the link, so that
\begin{equation}
d_{uv}\le\sum_{j=0}^{H-1}d(X_j,X_{j+1})
\le\sum_{j=0}^{H-1}L_{\max}(X_j).
\label{eq:app_path_bound}
\end{equation}
The first assumption is that the nodes of shortest paths have the same
mean longest link \(E[L_{\max}]\) as a typical node. Taking the mean of
Eq.~\eqref{eq:app_path_bound} then gives
\(d_{uv}\le\mu(H_{uv})\,E[L_{\max}]\). With the linear
law~\eqref{eq:linear_law}, the velocity obeys \(\nu\le E[L_{\max}]\).

For the lower bound, consider a greedy path that at each hop moves to the
neighbor with the largest projection on the direction to the destination
\(v\). The second assumption is that the greedy path advances on average the
mean projection \(E[P_{\max}]\) per hop and reaches the destination
\(v\) after about \(d_{uv}/E[P_{\max}]\) hops. A shortest path has at
most as many hops as the greedy path, so that
\(\mu(H_{uv})\le d_{uv}/E[P_{\max}]\) and the velocity obeys
\(\nu\ge E[P_{\max}]\). Neither assumption holds exactly. Shortest paths
can favor nodes with long links. Successive steps of the greedy path see
overlapping neighborhoods. Table~\ref{tab:velocity} therefore tests the
bounds in our simulations.

\begin{table*}[tb]
\caption{Velocity \(\nu\) in our simulations and the local means
\(E[P_{\max}]\) and \(E[L_{\max}]\) of Eqs.~\eqref{eq:app_FL}--\eqref{eq:app_means}.
The velocity \(\nu\) is the inverse of the fitted slope of the linear
law~\eqref{eq:app_linear_fit} in the Euclidean plane (E), the hyperbolic
plane (H) and on the sphere (S). The midpoint is
\(\widehat\nu=(E[P_{\max}]+E[L_{\max}])/2\). The cutoff probability is
\(p_c=10^{-4}\) in the profile experiment and \(p_c=10^{-3}\) in the
joint-hopcount experiment. Lengths are in units of the connection
radius \(r\) of the hard RGG. Degree is the mean degree.}
\label{tab:velocity}
\begin{ruledtabular}
\begin{tabular}{llccccccccc}
Experiment & Rule & Degree & \(\beta\) & \(R_\Omega\) & \(E[P_{\max}]\) & \(E[L_{\max}]\) & \(\nu\) (E) & \(\nu\) (H) & \(\nu\) (S) & \(\widehat\nu\)\\
\hline
Profile & Hard & 8 & -- & 130 & 0.668 & 0.932 & 0.740 & 0.739 & 0.740 & 0.800\\
Profile & Hard & 10 & -- & 130 & 0.714 & 0.947 & 0.779 & 0.778 & 0.780 & 0.831\\
Profile & Hard & 14 & -- & 130 & 0.774 & 0.963 & 0.826 & 0.825 & 0.826 & 0.868\\
Profile & Hard & 10 & -- & 170 & 0.714 & 0.947 & 0.779 & 0.779 & 0.779 & 0.831\\
Profile & Soft & 8 & 6 & 130 & 0.865 & 1.449 & 1.216 & 1.212 & 1.216 & 1.157\\
Profile & Soft & 10 & 6 & 130 & 0.945 & 1.535 & 1.332 & 1.332 & 1.335 & 1.240\\
Profile & Soft & 10 & 12 & 130 & 0.762 & 1.087 & 0.879 & 0.877 & 0.879 & 0.925\\
\hline
Joint hopcount & Hard & 10 & -- & 42 & 0.714 & 0.947 & 0.759 & 0.757 & 0.760 & 0.831\\
Joint hopcount & Hard & 14 & -- & 42 & 0.774 & 0.963 & 0.809 & 0.809 & 0.811 & 0.868\\
Joint hopcount & Hard & 18 & -- & 42 & 0.810 & 0.971 & 0.839 & 0.840 & 0.840 & 0.891\\
Joint hopcount & Soft & 10 & 8 & 42 & 0.825 & 1.241 & 0.973 & 0.972 & 0.978 & 1.033\\
Joint hopcount & Soft & 14 & 6 & 42 & 1.038 & 1.594 & 1.324 & 1.321 & 1.328 & 1.316\\
Joint hopcount & Soft & 14 & 8 & 42 & 0.911 & 1.315 & 1.076 & 1.073 & 1.077 & 1.113\\
Joint hopcount & Soft & 14 & 12 & 42 & 0.832 & 1.129 & 0.921 & 0.920 & 0.924 & 0.980\\
Joint hopcount & Soft & 18 & 8 & 42 & 0.971 & 1.370 & 1.143 & 1.142 & 1.146 & 1.170\\
\end{tabular}
\end{ruledtabular}
\end{table*}

\subsection{Mean degree and hopcount variance}
\label{app:degree}

By the bounds on the velocity \(\nu\), the hopcount \(H_{uv}\) lies
roughly between \(d_{uv}/E[L_{\max}]\) and \(d_{uv}/E[P_{\max}]\) for
typical values of the longest link \(L_{\max}\) and the largest
projection \(P_{\max}\). At a fixed shape of the link probability
\(p(d)\), the node density \(\lambda\) grows in proportion to the mean
degree. As the mean degree grows, the distribution functions
\(F_L(t)\) and \(F_P(t)\) of Eqs.~\eqref{eq:app_FL}
and~\eqref{eq:app_FP} tend to zero for every \(t\) below the range of
the links, the connection radius \(r\) in the hard RGG and the cutoff
distance \(d_c\) in the soft RGG. The mean longest link \(E[L_{\max}]\)
and the mean projection \(E[P_{\max}]\) then both tend to the range,
which narrows the window. In the hard RGGs of
Table~\ref{tab:velocity}, the relative width
\(E[L_{\max}]/E[P_{\max}]-1\) of the window falls from 0.33 at the mean
degree 10 to 0.24 at the mean degree 14 and to 0.20 at the mean degree
18. A narrower window leaves less room for the hopcount \(H_{uv}\) to
fluctuate, so that a larger mean degree generally gives a smaller
hopcount variance at the same mean hopcount, as for the hopcount
variance \(\sigma_0^2\) in Table~\ref{tab:fig3c}. At the mean hopcount
32 in the hard RGG with the mean degree 10, the window spans about eight
hops, whereas in our simulations the standard deviation of the hopcount
is about one hop. The argument therefore explains the direction in which
the hopcount variance changes with the mean degree, not the size of the
change.

\subsection{Local velocity field and travel time}

In a spatially inhomogeneous network, we assume that the node density
\(\lambda(x)\) and the link rule change little over the length of a
link. Equations~\eqref{eq:app_FL}--\eqref{eq:app_means} with the local
node density \(\lambda(x)\) and the local link rule then give the means
\(E[P_{\max}(x)]\) and \(E[L_{\max}(x)]\) and the velocity
field~\eqref{eq:velocity_field}. A segment of the length \(ds\) of a
curve near the position \(x\) needs about \(ds/\nu(x)\) hops. Adding the
hops along a curve and minimizing over the curves gives the travel
time~\eqref{eq:travel_time}. For a smooth velocity field \(\nu(x)\), the
travel time \(T(x)=\tau(u,x)\) from the source \(u\) obeys the eikonal
equation
\begin{equation}
|\nabla T(x)|=\frac{1}{\nu(x)},\qquad T(x_u)=0,
\label{eq:app_eikonal}
\end{equation}
which describes the propagation of a
front~\cite{sethian1996fast,sethian1999level} with the local speed
\(\nu(x)\).

\section{Profiles}
\label{app:profiles}

\subsection{Average over a curve of equal relative geometric deviation}

In the profile experiment, the probability field is recorded on a grid
in the coordinate \(s\) along the geodesic and the coordinate \(t\)
across the geodesic, both measured as intrinsic distances. The midpoint
of the geodesic is at \(s=0\) and the end nodes are at
\(s=\pm d_{uv}/2\). The grid has 160 cells of equal length between the
end nodes and 24 cells beyond each end node. Across the geodesic, the
cells are 0.01 wide at the geodesic and grow up to the distance
\(0.38\,d_{uv}\) from the geodesic. Neighboring cells along the
geodesic are merged by a factor that is fixed with the training
realizations.

For a relative geometric deviation \(\rho>0\), let \(\Gamma_\rho\) be
the closed curve of the positions \(x\) with
\(d(x_u,x)+d(x,x_v)=(1+\rho)d_{uv}\) and let \(\ell_\rho\) be the length
of the curve \(\Gamma_\rho\). The average of the shortest-path node
probability \(\pi_{uv}(w)\) over the curve is
\begin{equation}
\overline\pi_\rho=\frac{1}{\ell_\rho}\oint_{\Gamma_\rho}\pi_{uv}(x)\,\mathrm{d}\ell,
\label{eq:app_contour_average}
\end{equation}
where \(\pi_{uv}(x)\) is the estimate in the cell that contains the
position \(x\). Each cell thus contributes in proportion to the length
of the curve \(\Gamma_\rho\) inside the cell. The average covers the
whole curve \(\Gamma_\rho\), including the parts near the end nodes.
Each plotted point of a profile is the mean of the average
\(\overline\pi_\rho\) over a narrow interval of the relative geometric
deviation \(\rho\). The 33 intervals have the edges zero and 33 values
of the relative geometric deviation \(\rho\) spaced geometrically
between \(5\times10^{-5}\) and 0.16.

\subsection{Reference on the geodesic}

The reference \(\overline\pi_0\) comes from the geodesic reference
experiment. The geodesic from \(x_u\) to \(x_v\) is divided into 40
segments of equal length. In each realization, one test node is placed
at a uniformly random position in each segment, each test node in a
separate graph with \(N-1\) other nodes, so that the test node \(w\) is
one of \(N\) nodes. The reference \(\overline\pi_0\) is the mean over
the 40 segments of the fraction of test nodes that are shortest-path
nodes. Each condition and distance has 20{,}000 realizations. In our
simulations, the half-width of the 95\% interval of the reference
\(\overline\pi_0\) lies between 0.7\% and 1.2\% of the reference
\(\overline\pi_0\).

\subsection{Mean standardized detour}

The standardized detour \(X_{uvw}\) uses the linear
law~\eqref{eq:app_linear_fit} and the variance law~\eqref{eq:variance_law}
fitted on the training realizations of the condition. The mean hopcount
\(\mu(H_{uv})\) is the fitted value \(d_{uv}/\nu+c\). The mean hopcounts
\(\mu(H_{uw})\) and \(\mu(H_{wv})\) of the pairs \((u,w)\) and \((w,v)\) are taken
in proportion to the distances \(d_{uw}\) and \(d_{wv}\), as in the
linear law~\eqref{eq:linear_law}. The variance~\eqref{eq:independent_variance}
then becomes
\begin{equation}
V_{uvw}=\hat\sigma^2(H_{uv})\left[1+\left(\frac{d_{uw}}{d_{uv}}\right)^{\gamma}
+\left(\frac{d_{wv}}{d_{uv}}\right)^{\gamma}\right].
\label{eq:app_profile_variance}
\end{equation}
The mean standardized detour \(\overline X\) of a plotted point is
the mean of the standardized detour \(X_{uvw}\) over the same curves
and intervals as the profile. No parameter is fitted to the test
realizations.

\subsection{Spread of the profiles}

The normalized profile is \(\widetilde\pi=\overline\pi_\rho/\overline\pi_0\).
For a level \(p\), we locate the first point at which the normalized
profile \(\widetilde\pi\) falls below the level \(p\) and interpolate
linearly between the two neighboring points, both in the relative
geometric deviation \(\rho\) and in the mean standardized detour
\(\overline X\). The spread is the standard deviation divided by the
mean over the 84 profiles. In our simulations, the spread falls from
0.470 for the relative geometric deviation \(\rho\) to 0.113 for the
mean standardized detour \(\overline X\) at the level \(p=0.5\). In
our simulations, the spread falls from 0.506 to 0.081 at the level
\(p=0.2\).

\section{Correlations}
\label{app:correlations}

\subsection{Variance of the hopcount excess}

When the three hopcounts \(H_{uw}\), \(H_{wv}\) and \(H_{uv}\) are
measured in the same realization, the variance \(V_Z\) of the hopcount
excess \(Z_{uvw}=H_{uw}+H_{wv}-H_{uv}\) follows
Eq.~\eqref{eq:excess_variance}.
Without correlations, the variance \(V_Z\) equals the sum of the three
hopcount variances, the independent value \(V_{uvw}\). In
Fig.~\ref{fig:result2_statistical_control}(c), the independent value
\(V_{uvw}\) is the sum of the three measured hopcount variances. A
positive covariance between the hopcount \(H_{uv}\) and the hopcount
\(H_{uw}\) or \(H_{wv}\) lowers the variance \(V_Z\). A positive covariance
between the hopcounts \(H_{uw}\) and \(H_{wv}\) raises the variance
\(V_Z\).

\subsection{Shuffle control}
\label{app:shuffle}

For one condition, distance and position of the test node, the shuffle
permutes the realizations independently for each of the three hopcounts
\(H_{uw}\), \(H_{wv}\) and \(H_{uv}\). The shuffle keeps the
distribution of each hopcount and removes the correlations, so that the
expected variance of the shuffled hopcount excess equals the
independent value \(V_{uvw}\). We average the ratio \(V_Z/V_{uvw}\)
over independent shuffles. In our simulations of the joint-hopcount
experiment, the shuffled ratio \(V_Z/V_{uvw}\) has the mean 1.000 and
the standard deviation 0.002 over the 1{,}536 combinations of
condition, distance and test position with
\(0.004\le\rho_{uvw}\le0.064\). In our simulations, the ratio
\(V_Z/V_{uvw}\) without the shuffle lies between 0.3 and 0.6 at the same
test positions [Fig.~\ref{fig:result2_statistical_control}(c)]. The
test positions lie off the geodesic, where the ratio \(V_Z/V_{uvw}\) is
larger than on the geodesic (Table~\ref{tab:equal_detour}).

\subsection{Overlap and covariance}
\label{app:overlap}

In Eq.~\eqref{eq:excess_variance}, each covariance belongs to two pairs
with a common end node. For two pairs \((u,w)\) and \((u,v)\) with the
common end node \(u\), we take one shortest path per pair from the
breadth-first search tree rooted at the end node \(u\). Let \(S\) be the number of links that the two paths share from the end
node \(u\) before the paths first separate. With the remaining hopcounts
\(R_{uw}=H_{uw}-S\) and \(R_{uv}=H_{uv}-S\),
\begin{equation}
\begin{split}
\mathrm{Cov}[H_{uw},H_{uv}]={}&\sigma^2(S)+\mathrm{Cov}[S,R_{uw}+R_{uv}]\\
&+\mathrm{Cov}[R_{uw},R_{uv}],
\end{split}
\label{eq:covariance_split}
\end{equation}
so that the shared links add their variance \(\sigma^2(S)\) to the
covariance. In the model of Ref.~\cite{pinto2012locating}, the delay
along a path is the sum of independent delays on the links of the path,
so that the covariance of the delays along two paths is proportional to
the number of shared links. The overlap \(E[S]\) is the mean of the
number \(S\) of shared links. The normalized overlap is \(\omega=E[S]/\sqrt{\mu(H_{uw})\mu(H_{uv})}\). In
our simulations, the correlation coefficient of the hopcounts \(H_{uw}\)
and \(H_{uv}\) grows approximately linearly with the normalized overlap
\(\omega\). We fit four forms of the covariance
\(\mathrm{Cov}[H_{uw},H_{uv}]\) on the training realizations of the
three spaces together (Table~\ref{tab:cov_models}).
Figure~\ref{fig:result2_statistical_control}(d) uses the linear form of
the correlation coefficient,
\begin{equation}
\frac{\mathrm{Cov}[H_{uw},H_{uv}]}{\sqrt{\sigma^2(H_{uw})\,\sigma^2(H_{uv})}}
=a_0+a_1\omega,
\label{eq:app_linear_corr}
\end{equation}
with the parameters \(a_0=0.045\) and \(a_1=1.494\) fitted in our
simulations. The predicted correlation coefficient is limited to the
interval from \(-0.95\) to 0.95. In our simulations, no test term
reaches the limits. In our simulations, the linear
form~\eqref{eq:app_linear_corr} predicts the 6{,}912 covariances of the
test realizations with the coefficient of determination \(R^2=0.924\),
weighted by the number of records of each covariance
[Fig.~\ref{fig:result2_statistical_control}(d)]. In our
simulations, the coefficient of determination of the test covariances in
each space lies between 0.922 and 0.926. In our simulations, the
saturating form of Table~\ref{tab:cov_models} fits the validation
realizations better and predicts the test covariances with the
coefficient of determination \(R^2=0.961\), at the cost of a third
parameter. In the test,
the overlaps \(E[S]\) and the hopcount variances are measured in the
test realizations. The prediction therefore uses the measured overlap
\(E[S]\), not the node positions alone.

For each of the 24 conditions, we average the ratio \(V_Z/V_{uvw}\) and
the overlap \(E[S]\) of the shortest paths of the pairs \((u,v)\) and
\((u,w)\) over the
four distances and over the test positions at equal distance from the
end nodes with \(0.004\le\rho_{uvw}\le0.064\). In our simulations, the
Pearson correlation of the two averages over the 24 conditions is
\(-0.97\).

\begin{table*}[tb]
\caption{Forms of the covariance \(\mathrm{Cov}[H_{uw},H_{uv}]\) fitted
in our simulations of the joint-hopcount experiment. Here
\(\sigma_{uw}\sigma_{uv}=\sqrt{\sigma^2(H_{uw})\sigma^2(H_{uv})}\) and
\(\bar\gamma=0.545\) is the median fitted exponent \(\gamma\) of the 24
conditions. The coefficient of determination \(R^2\) and the
root-mean-square error refer to the validation realizations.
Figure~\ref{fig:result2_statistical_control}(d) uses the linear
correlation form.}
\label{tab:cov_models}
\begin{ruledtabular}
\begin{tabular}{lllcc}
Form & Model & Parameters & RMSE & \(R^2\)\\
\hline
Linear covariance & \(a_0+a_1E[S]\) & \(a_0=0.016\), \(a_1=0.066\) & 0.106 & 0.712\\
Power covariance & \(a_0+a_1E[S]^{\bar\gamma}\) & \(a_0=-0.038\), \(a_1=0.164\) & 0.106 & 0.713\\
Linear correlation & \(\sigma_{uw}\sigma_{uv}(a_0+a_1\omega)\) & \(a_0=0.045\), \(a_1=1.494\) & 0.054 & 0.924\\
Saturating correlation & \(\sigma_{uw}\sigma_{uv}[\eta_0+\eta_1\omega/(1+\eta_2\omega)]\) & \(\eta_0=-0.006\), \(\eta_1=3.51\), \(\eta_2=3.89\) & 0.041 & 0.957\\
\end{tabular}
\end{ruledtabular}
\end{table*}

\subsection{Position of the node \(w\)}
\label{app:position}

By the linear law~\eqref{eq:linear_law} and the variance
law~\eqref{eq:variance_law}, the independent value \(V_{uvw}\) of a pair
\((u,v)\) is proportional to the bracket in
Eq.~\eqref{eq:app_profile_variance}. On the geodesic, where
\(d_{uw}+d_{wv}=d_{uv}\), the bracket grows from 2 at the end nodes to
\(1+2^{1-\gamma}\) at the midpoint. For \(0<\gamma<1\), the independent
value \(V_{uvw}\) at the midpoint thus exceeds the value at the end nodes
by the factor \((1+2^{1-\gamma})/2\), which lies between 1 and 1.5. In
our simulations, the fitted exponents \(\gamma\) of
Appendix~\ref{app:hopcount} give the factor 1.09 to 1.31.

Near the geodesic, the directions from the node \(w\) to the source
\(u\) and to the destination \(v\) enclose a large angle. The shortest
paths from \(w\) to \(u\) and from \(w\) to \(v\) then share almost no
links, so that \(\mathrm{Cov}[H_{uw},H_{wv}]\approx0\). For the pairs
\((w,v)\) and \((u,v)\), which share the end node \(v\), the number \(S\)
of shared links cannot exceed the hopcount \(H_{wv}\), so that
\(E[S]\le\mu(H_{wv})\). Dividing by \(\sqrt{\mu(H_{wv})\mu(H_{uv})}\)
bounds the normalized overlap as
\(E[S]/\sqrt{\mu(H_{wv})\mu(H_{uv})}\le\sqrt{\mu(H_{wv})/\mu(H_{uv})}\).
By the linear law~\eqref{eq:linear_law}, the bound is \(1/\sqrt2\) at
the midpoint of the geodesic and tends to one near the source \(u\). The
same holds near the destination \(v\) for the pairs \((u,w)\) and
\((u,v)\). The covariances in Eq.~\eqref{eq:excess_variance} thus lower
the variance \(V_Z\) most near the end nodes.

In our simulations of the profile experiment (Appendix~\ref{app:off_geodesic}),
the ratio \(V_Z/V_{uvw}\) grows with the mean number of extra hops
\(\rho_{uvw}\mu(H_{uv})\) in every condition and is smaller in the end
regions than in the middle region in every group of every condition
(Table~\ref{tab:equal_detour}). In the same simulations, at the mean
number of extra hops \(\rho_{uvw}\mu(H_{uv})\) between 2 and 3, the ratio
\(V_Z/V_{uvw}\) is 0.23 to 0.44 within \(0.1\,d_{uv}\) of an end node and
0.37 to 0.64 within \(0.05\,d_{uv}\) of the midpoint
(Table~\ref{tab:equal_detour}).

\subsection{Mean of the hopcount excess}
\label{app:mean_excess}

By the fitted linear law~\eqref{eq:app_linear_fit}, the mean hopcount of
each of the pairs \((u,w)\), \((w,v)\) and \((u,v)\) is the distance
divided by the velocity \(\nu\) plus the intercept \(c\). In the mean
\(m_Z\) of the hopcount excess~\eqref{eq:hopcount_excess}, the intercepts
\(c\) of the pairs \((u,w)\) and \((w,v)\) enter with a plus sign and the
intercept \(c\) of the pair \((u,v)\) enters with a minus sign. With the
relative geometric deviation~\eqref{eq:relative_deviation},
\begin{equation}
\begin{split}
m_Z&=\mu(H_{uw})+\mu(H_{wv})-\mu(H_{uv})\\
&\approx\rho_{uvw}\,\mu(H_{uv})+(1-\rho_{uvw})\,c\\
&\approx\rho_{uvw}\,\mu(H_{uv})+c,
\end{split}
\label{eq:mean_excess}
\end{equation}
where the last step uses \(\rho_{uvw}\ll1\).
On the geodesic, where \(\rho_{uvw}=0\), Eq.~\eqref{eq:mean_excess}
gives \(m_Z\approx c\), whereas the mean \(\rho_{uvw}\mu(H_{uv})\) of
Proposition~\ref{prop:gaussian} is zero. Near the destination \(v\), the
pair \((w,v)\) is short. By Appendix~\ref{app:hopcount}, the mean
hopcount \(\mu(H_{wv})\) of a short pair exceeds the term \(d_{wv}/\nu\)
by less than the intercept \(c\), so that the mean \(m_Z\) falls below
the intercept \(c\). The same holds near the source \(u\). Away from the
geodesic, the mean number of extra hops \(\rho_{uvw}\mu(H_{uv})\) adds to
the intercept \(c\) and dominates the mean \(m_Z\) far from the
geodesic.

\subsection{End nodes and midpoint}
\label{app:end_nodes}

The end-node experiment tests whether the rise of the shortest-path node
probability \(\pi_{uv}(w)\) toward the end nodes comes from the
correlations. In the end-node experiment, the rise of a condition is the mean of the
shortest-path node probability \(\pi_{uv}(w)\) at 0.2 and 0.8 of the
way from the source \(u\) to the destination \(v\), divided by the
shortest-path node probability \(\pi_{uv}(w)\) at the midpoint. The
shuffle permutes the realizations independently for each hopcount, as in
the joint-hopcount experiment. In our simulations, the median rise over
the 30 conditions is 1.20 with the correlations and 1.08 after the
shuffle. In our simulations, the rise with the correlations is the
larger one in 29 of the 30 conditions. In our simulations, the mean of
the shortest-path node probability \(\pi_{uv}(w)\) at 0.2 and 0.8 is
larger with the correlations than after the shuffle in all 15
conditions of hard RGGs and in 7 of the 15 conditions of soft RGGs. In
our simulations at these positions, the median of the measured ratio
\(m_Z^2/V_Z\) over 15 conditions is 1.13 for hard RGGs and 2.08 for
soft RGGs.

\subsection{Regions near the midpoint and near the end nodes}
\label{app:regions}

In the profile experiment at the largest distance \(d_{uv}=180\), in
units of the connection radius \(r\), we compare a
middle region, \(|s|<0.05\,d_{uv}\), with two end regions,
\(0.40\,d_{uv}<|s|<0.50\,d_{uv}\), all in the band \(|t|<0.02\) around
the geodesic. The regions contain the nodes of the realizations, not
test nodes. The end regions do not contain the end nodes. In our
simulations, the mean hopcount \(\mu(H_{uv})\) at the distance
\(d_{uv}=180\) lies between 138 and 246. For each
region, we pool the nodes of all realizations and compute the measured
mean \(m_Z\) and the measured variance \(V_Z\) of the hopcount excess
\(Z_{uvw}\) and the fraction \(\pi\) of shortest-path nodes
(Table~\ref{tab:regions}). In our simulations, the measured mean
\(m_Z\) is 1.64 to 1.90 times smaller and the measured variance \(V_Z\)
is 1.57 to 2.14 times smaller in the end regions than in the middle
region. In our simulations, the correlation coefficient of the sum
\(H_{uw}+H_{wv}\) and the hopcount \(H_{uv}\) over the pooled nodes is
0.77 to 0.83 in the middle region and 0.86 to 0.91 in the end regions.
In our simulations, Eq.~\eqref{eq:correlated_estimate} with the
measured mean \(m_Z\) and the measured variance \(V_Z\) gives a ratio of
1.99 to 2.59 between the end regions and the middle region, against a
measured ratio of 2.31 to 3.83.

\begin{table*}[tb]
\caption{Measured mean \(m_Z\) and variance \(V_Z\) of the hopcount
excess and fraction \(\pi\) of shortest-path nodes in the middle region
(m) and in the end regions (e) at the distance \(d_{uv}=180\), with the
mean hopcount \(\mu(H_{uv})\) between 138 and 246, in our simulations of
the profile experiment. Degree is the mean degree.}
\label{tab:regions}
\begin{ruledtabular}
\begin{tabular}{llcccrrrrrr}
Space & Rule & Degree & \(\beta\) & \(R_\Omega\) & \(m_Z\) (m) & \(m_Z\) (e) & \(V_Z\) (m) & \(V_Z\) (e) & \(\pi\) (m) & \(\pi\) (e)\\
\hline
Euclidean & Hard & 8 & -- & 130 & 2.66 & 1.44 & 3.88 & 2.28 & 0.105 & 0.280\\
Euclidean & Hard & 10 & -- & 130 & 1.89 & 1.04 & 1.70 & 0.94 & 0.125 & 0.313\\
Euclidean & Hard & 10 & -- & 170 & 1.90 & 1.02 & 1.67 & 0.90 & 0.114 & 0.323\\
Euclidean & Hard & 14 & -- & 130 & 1.42 & 0.80 & 0.89 & 0.51 & 0.151 & 0.357\\
Euclidean & Soft & 8 & 6 & 130 & 3.48 & 2.00 & 3.88 & 1.95 & 0.036 & 0.127\\
Euclidean & Soft & 10 & 6 & 130 & 3.09 & 1.82 & 2.77 & 1.46 & 0.034 & 0.120\\
Euclidean & Soft & 10 & 12 & 130 & 2.25 & 1.23 & 2.08 & 1.02 & 0.089 & 0.239\\
Hyperbolic & Hard & 8 & -- & 130 & 2.54 & 1.42 & 3.54 & 2.26 & 0.109 & 0.287\\
Hyperbolic & Hard & 10 & -- & 130 & 1.81 & 1.01 & 1.60 & 0.92 & 0.136 & 0.326\\
Hyperbolic & Hard & 10 & -- & 170 & 1.90 & 1.03 & 1.78 & 0.91 & 0.129 & 0.313\\
Hyperbolic & Hard & 14 & -- & 130 & 1.38 & 0.79 & 0.84 & 0.49 & 0.155 & 0.358\\
Hyperbolic & Soft & 8 & 6 & 130 & 3.33 & 1.97 & 3.78 & 1.96 & 0.038 & 0.134\\
Hyperbolic & Soft & 10 & 6 & 130 & 3.04 & 1.85 & 2.67 & 1.46 & 0.034 & 0.114\\
Hyperbolic & Soft & 10 & 12 & 130 & 2.11 & 1.18 & 1.87 & 0.99 & 0.094 & 0.257\\
Spherical & Hard & 8 & -- & 130 & 2.77 & 1.46 & 4.05 & 2.28 & 0.098 & 0.276\\
Spherical & Hard & 10 & -- & 130 & 1.95 & 1.04 & 1.79 & 0.94 & 0.122 & 0.313\\
Spherical & Hard & 10 & -- & 170 & 1.97 & 1.04 & 1.81 & 0.97 & 0.117 & 0.323\\
Spherical & Hard & 14 & -- & 130 & 1.46 & 0.81 & 0.93 & 0.50 & 0.148 & 0.351\\
Spherical & Soft & 8 & 6 & 130 & 3.67 & 2.01 & 4.21 & 1.97 & 0.033 & 0.126\\
Spherical & Soft & 10 & 6 & 130 & 3.25 & 1.85 & 3.06 & 1.45 & 0.030 & 0.112\\
Spherical & Soft & 10 & 12 & 130 & 2.32 & 1.26 & 2.20 & 1.06 & 0.084 & 0.232\\
\end{tabular}
\end{ruledtabular}
\end{table*}

\subsection{Positions on the geodesic}
\label{app:on_geodesic}

We use the cells of the probability field of the profile experiment at
the distance \(d_{uv}=180\) (Appendix~\ref{app:profiles}), with the
realizations of all splits. Only cells between the end nodes with more
than 200 nodes of finite hopcount enter. For each cell, the nodes in the
cell give the measured mean \(m_Z\), the measured variance \(V_Z\) and
the fraction of shortest-path nodes. The mean number of extra hops of a
cell is \(\rho_{uvw}\mu(H_{uv})\) at the center of the cell, with the
intrinsic distances and the mean hopcount \(\mu(H_{uv})\) of the fitted
linear law~\eqref{eq:app_linear_fit}. The independent value
\(V_{uvw}\) follows Eq.~\eqref{eq:app_profile_variance}.

On the geodesic, we use the band \(|t|<0.02\) of the middle region and
the end regions and group the cells of the band by the distance from the
nearer end node. As in the middle region and the end regions, we pool
the nodes of each group. The independent value \(V_{uvw}\) of a group is
taken at the mean distance of the nodes from the nearer end node.
Table~\ref{tab:geodesic_positions} lists the groups. In our
simulations, the measured mean \(m_Z\) of the middle region is 1.04 to
1.17 times the intercept \(c\) of the condition. In our simulations, the
measured mean \(m_Z\) of the middle region is 2.46 to 3.19 times the
value within \(0.025\,d_{uv}\) of an end node, whereas the measured
standard deviation \(\sqrt{V_Z}\) of the middle region is 1.57 to 1.96
times the value within \(0.025\,d_{uv}\) of an end node. In our
simulations, the factor of the mean \(m_Z\) exceeds the factor of the
standard deviation \(\sqrt{V_Z}\) in every condition. In the same
simulations, the shortest-path node probability \(\pi_{uv}(w)\) on the
geodesic within \(0.025\,d_{uv}\) of an end node is 3.2 to 7.3 times the
value within \(0.05\,d_{uv}\) of the midpoint, against 3.0 to 4.6 times
for Eq.~\eqref{eq:correlated_estimate} with the measured mean \(m_Z\)
and the measured variance \(V_Z\).

\subsection{Positions off the geodesic}
\label{app:off_geodesic}

We group the cells of Appendix~\ref{app:on_geodesic} that lie in the
middle region or in the end regions by the mean number of extra hops,
without the restriction to the band \(|t|<0.02\). The mean \(m_Z\), the variance \(V_Z\),
the ratio \(V_Z/V_{uvw}\) and the estimate~\eqref{eq:correlated_estimate}
of a group are averages of the cell values weighted by the number of
nodes. Table~\ref{tab:equal_detour} compares the end regions with the
middle region. In our simulations, the difference
\(m_Z-\rho_{uvw}\mu(H_{uv})\) in the middle region changes by at most
0.11 hops between 0 and 6 extra hops, as Eq.~\eqref{eq:mean_excess}
predicts. In our simulations at the same mean number of extra hops
\(\rho_{uvw}\mu(H_{uv})\), the shortest-path node probability
\(\pi_{uv}(w)\) in the end regions is 2.0 to 3.0 times the value in the
middle region below 0.1 extra hops and smaller than the value in the
middle region at 4 to 6 extra hops in all 21 conditions. In the same
simulations, Eq.~\eqref{eq:correlated_estimate} with the measured mean
\(m_Z\) and the measured variance \(V_Z\) gives the same trend.

\begin{table*}[tb]
\caption{Groups on the geodesic in our simulations of the profile
experiment at the distance \(d_{uv}=180\), with the mean hopcount
\(\mu(H_{uv})\) between 138 and 246. The distance from the nearer
end node is in units of the distance \(d_{uv}\). The shortest-path node
probability \(\pi_{uv}(w)\), the estimate~\eqref{eq:correlated_estimate}
with the measured mean \(m_Z\) and the measured variance \(V_Z\), the
mean \(m_Z\) and the variance \(V_Z\) are divided by the values of the
middle region. Entries give the range over the 21 conditions.}
\label{tab:geodesic_positions}
\begin{ruledtabular}
\begin{tabular}{lccccc}
Distance from the nearer end node & \(\pi_{uv}(w)\) & Eq.~\eqref{eq:correlated_estimate} & \(m_Z\) & \(V_Z\) & \(V_Z/V_{uvw}\)\\
\hline
0 to 0.025 & 3.15--7.32 & 3.01--4.58 & 0.31--0.41 & 0.26--0.41 & 0.060--0.102\\
0.025 to 0.05 & 2.32--3.52 & 2.07--2.47 & 0.50--0.60 & 0.38--0.58 & 0.082--0.135\\
0.05 to 0.1 & 1.79--2.36 & 1.60--1.85 & 0.64--0.72 & 0.51--0.71 & 0.103--0.166\\
0.1 to 0.2 & 1.34--1.73 & 1.22--1.41 & 0.78--0.84 & 0.71--0.82 & 0.131--0.207\\
0.2 to 0.3 & 1.12--1.29 & 1.10--1.20 & 0.89--0.93 & 0.82--0.94 & 0.155--0.239\\
0.3 to 0.45 & 0.97--1.16 & 0.98--1.08 & 0.95--1.00 & 0.92--1.03 & 0.169--0.252\\
0.45 to 0.5 (middle region) & 1 & 1 & 1 & 1 & 0.172--0.249\\
\end{tabular}
\end{ruledtabular}
\end{table*}

\begin{table*}[tb]
\caption{End regions and middle region at the same mean number of extra
hops \(\rho_{uvw}\mu(H_{uv})\) in our simulations of the profile
experiment at the distance \(d_{uv}=180\), with the mean hopcount
\(\mu(H_{uv})\) between 138 and 246. The second and third columns give
the shortest-path node probability \(\pi_{uv}(w)\) in the end regions
divided by the value in the middle region, measured and from
Eq.~\eqref{eq:correlated_estimate} with the measured mean \(m_Z\) and
the measured variance \(V_Z\). The last two columns give the ratio
\(V_Z/V_{uvw}\) in the end regions (e) and in the middle region (m).
Entries give the range over the 21 conditions.}
\label{tab:equal_detour}
\begin{ruledtabular}
\begin{tabular}{lcccc}
Extra hops & \(\pi_{uv}(w)\) & Eq.~\eqref{eq:correlated_estimate} & \(V_Z/V_{uvw}\) (e) & \(V_Z/V_{uvw}\) (m)\\
\hline
0 to 0.1 & 1.96--3.02 & 1.88--2.19 & 0.09--0.16 & 0.18--0.26\\
0.1 to 0.5 & 1.80--2.83 & 1.71--2.02 & 0.11--0.21 & 0.20--0.33\\
0.5 to 1 & 1.55--2.43 & 1.44--1.83 & 0.14--0.28 & 0.25--0.43\\
1 to 2 & 1.10--1.95 & 1.02--1.60 & 0.19--0.36 & 0.31--0.53\\
2 to 3 & 0.59--1.53 & 0.62--1.49 & 0.23--0.44 & 0.37--0.64\\
3 to 4 & 0.28--1.26 & 0.32--1.31 & 0.27--0.49 & 0.42--0.71\\
4 to 6 & 0.04--0.86 & 0.18--1.08 & 0.32--0.56 & 0.48--0.80\\
\end{tabular}
\end{ruledtabular}
\end{table*}

\subsection{Two lobes near the end nodes}
\label{app:lobes}

The correlations lower the variance \(V_Z\) most near the end nodes.
Along the geodesic, the independent value \(V_{uvw}\) changes by less
than the factor 1.5 (Appendix~\ref{app:position}), so that the variance
\(V_Z\) changes mainly through the covariances in
Eq.~\eqref{eq:excess_variance}. Near the geodesic, the shortest paths
from \(w\) to \(u\) and from \(w\) to \(v\) leave the node \(w\) in
opposite directions, so that \(\mathrm{Cov}[H_{uw},H_{wv}]\approx0\).
Near the destination \(v\), the pair \((u,w)\) nearly coincides with the
pair \((u,v)\), so that the covariance \(\mathrm{Cov}[H_{uw},H_{uv}]\)
can cancel most of the independent value \(V_{uvw}\). The same holds for
the covariance \(\mathrm{Cov}[H_{wv},H_{uv}]\) near the source \(u\).
Away from the geodesic, the shortest paths from \(w\) share fewer links
with the shortest path from \(u\) to \(v\), so that the variance \(V_Z\)
grows toward the independent value \(V_{uvw}\).

Near the geodesic, the shortest-path node probability \(\pi_{uv}(w)\) is
then largest near the end nodes. On the geodesic, the variance \(V_Z\)
grows from the end nodes to the midpoint. The mean \(m_Z\) also grows,
from below the intercept \(c\) of the fitted linear
law~\eqref{eq:app_linear_fit} near the end nodes to about the intercept
\(c\) near the midpoint (Appendix~\ref{app:mean_excess}). In our
simulations, the mean \(m_Z\) grows by a larger factor than the standard
deviation \(\sqrt{V_Z}\), so that both the exponent \(m_Z^2/(2V_Z)\) and
the prefactor \(1/\sqrt{2\pi V_Z}\) in Eq.~\eqref{eq:correlated_estimate}
lower the shortest-path node probability \(\pi_{uv}(w)\) toward the
midpoint (Appendix~\ref{app:on_geodesic}). In the same simulations,
Eq.~\eqref{eq:correlated_estimate} with the measured mean \(m_Z\) and the
measured variance \(V_Z\) gives the rise of the shortest-path node
probability \(\pi_{uv}(w)\) toward the end nodes
(Appendices~\ref{app:regions} and~\ref{app:on_geodesic}). In our
end-node experiment, the rise is weaker after the hopcounts are shuffled
across realizations (Appendix~\ref{app:end_nodes}).

Far from the geodesic, the smaller variance \(V_Z\) near the end nodes
instead lowers the shortest-path node probability \(\pi_{uv}(w)\). The
mean number of extra hops \(\rho_{uvw}\mu(H_{uv})\) then dominates the
mean \(m_Z\) (Appendix~\ref{app:mean_excess}), so that the mean \(m_Z\)
changes little along a curve of equal relative geometric deviation
\(\rho_{uvw}\). At a fixed mean \(m_Z\), the derivative of
\(\ln\pi_{uv}(w)\) in Eq.~\eqref{eq:correlated_estimate} with respect to
the variance \(V_Z\) has the sign of \(m_Z^2-V_Z\), as in
Eq.~\eqref{eq:app_dlnpi_dV}. Where \(m_Z^2>V_Z\), the smaller variance
\(V_Z\) near the end nodes thus lowers the shortest-path node probability
\(\pi_{uv}(w)\). In our simulations, the shortest-path node probability
\(\pi_{uv}(w)\) far from the geodesic is smaller near the end nodes than
near the midpoint at equal relative geometric deviation \(\rho_{uvw}\)
(Appendix~\ref{app:off_geodesic}).

Near the end nodes, the shortest-path node probability
\(\pi_{uv}(w)\) is thus higher close to the geodesic and lower far from
the geodesic than near the midpoint, which gives the two lobes of
Fig.~\ref{fig:probability}.

\section{Accuracy of the Gaussian form}
\label{app:accuracy}

This appendix tests how accurately a Gaussian distribution of the
hopcount excess \(Z_{uvw}\) gives the values of the shortest-path node
probability \(\pi_{uv}(w)\) in the cross-space experiment
(Appendix~\ref{app:simulations}). We compare three choices of the mean
\(m_Z\) and the variance \(V_Z\) of the hopcount excess \(Z_{uvw}\). The
moments of the Gaussian estimate~\eqref{eq:gaussian_membership} are the
mean \(\rho_{uvw}\mu(H_{uv})\) and the variance \(V_{uvw}\). The measured
moments are the mean \(m_Z\) and the variance \(V_Z\) in the training
realizations. The predicted moments follow from the hopcounts of node
pairs without the hopcounts of the test node \(w\).

\subsection{Interval form}

Because the hopcount excess \(Z_{uvw}\) takes integer values, we use the
Gaussian probability of the interval from \(-1/2\) to \(1/2\),
\begin{equation}
\pi_{uv}(w)\approx\Phi\!\left(\frac{1/2-m_Z}{\sqrt{V_Z}}\right)
-\Phi\!\left(\frac{-1/2-m_Z}{\sqrt{V_Z}}\right),
\label{eq:interval_estimate}
\end{equation}
where \(\Phi\) is the standard normal distribution function. For a large
variance \(V_Z\), Eq.~\eqref{eq:interval_estimate} reduces to
Eq.~\eqref{eq:correlated_estimate}. Unlike
Eq.~\eqref{eq:correlated_estimate}, Eq.~\eqref{eq:interval_estimate}
never exceeds one.

\subsection{Predicted moments}

The mean \(m_Z\) follows from the mean hopcount of node pairs. By
Eq.~\eqref{eq:hopcount_excess}, the mean
\(m_Z=\mu(H_{uw})+\mu(H_{wv})-\mu(H_{uv})\) holds exactly. In a
homogeneous network, the mean hopcount \(\mu(H_{uw})\) of a pair depends
only on the distance \(d_{uw}\) of the pair, so that the mean \(m_Z\)
needs only the mean hopcount of node pairs as a function of the
distance. The Gaussian estimate~\eqref{eq:gaussian_membership} replaces
the mean hopcount by the linear law~\eqref{eq:linear_law}, which omits
the intercept \(c\) of the fitted linear law~\eqref{eq:app_linear_fit}
and the deviation of short pairs from the linear law
(Appendix~\ref{app:mean_excess}). In our simulations of the cross-space
experiment, the mean \(\rho_{uvw}\mu(H_{uv})\) of the Gaussian
estimate~\eqref{eq:gaussian_membership} lies on average 1.21 hops below
the measured mean \(m_Z\).

For each condition of the cross-space experiment, we predict the mean
\(m_Z\) and the variance \(V_Z\) of the hopcount excess without the
hopcounts of the test node \(w\) in that condition. The mean hopcount
and the hopcount variance of a pair as functions of the distance come
from the training realizations of the same network at the two other
distances \(d_{uv}\). We average the mean hopcount and the hopcount
variance of the pairs \((u,w)\) and \((w,v)\) of those realizations in
60 bins of the distance and add the pairs \((u,v)\) of the three
distances. Linear interpolation between the bins gives the mean hopcount
\(\mu(H_{uw})\) and the hopcount variance \(\sigma^2(H_{uw})\) at the
distance \(d_{uw}\). The same holds for the pair \((w,v)\). The mean
hopcount is made nondecreasing in the distance. Beyond the largest bin,
the mean hopcount is extrapolated linearly and the hopcount variance as
a power of the distance. For the pair \((u,v)\), we use the training
values of the mean hopcount \(\mu(H_{uv})\) and the hopcount variance
\(\sigma^2(H_{uv})\), which the Gaussian
estimate~\eqref{eq:gaussian_membership} also uses. The predicted
variance is \(V_Z=\zeta_{uvw}V_{uvw}\), with the predicted independent
value \(V_{uvw}\) and the correlation factor \(\zeta_{uvw}\) of
Eq.~\eqref{eq:correlation_factor}.

In our simulations, the predicted mean \(m_Z\) differs from the measured
mean \(m_Z\) by 0.084 hops on average. In the same simulations, the
linear law~\eqref{eq:app_linear_fit} with a constant intercept \(c\),
fitted on the three pairs \((u,v)\) of a network, gives the mean \(m_Z\)
with the mean absolute difference 0.43 hops. In our simulations, the
predicted independent value \(V_{uvw}\) has the median ratio 1.00 to the
measured independent value \(V_{uvw}\).

\subsection{Correlation factor}

The correlations lower the variance \(V_Z\) by a factor that depends
mainly on the position of the node \(w\). Let
\(\zeta_{uvw}=V_Z/V_{uvw}\) be the correlation factor and
\(f=\min(d_{uw},d_{wv})/(d_{uw}+d_{wv})\) the relative distance of the
node \(w\) to the nearer end node. In our simulations of the cross-space
experiment, the correlation factor follows
\begin{equation}
\begin{split}
\zeta_{uvw}&\approx\zeta_0+\frac{1-\zeta_0}{1+(0.12/\rho_{uvw})^{0.86}},\\
\zeta_0&=0.33\,[4f(1-f)]^{0.26}\,[\mu(H_{uv})/30]^{-0.42}
\end{split}
\label{eq:correlation_factor}
\end{equation}
[Fig.~\ref{fig:computation}(f)]. The correlation factor
\(\zeta_{uvw}\) is smallest on the geodesic next to the end nodes, where
the shared links cancel most of the independent value \(V_{uvw}\)
(Sec.~\ref{sec:correlations}). Far from the geodesic, the correlation
factor \(\zeta_{uvw}\) approaches one. In our simulations, the
correlation factor \(\zeta_{uvw}\) also decreases with the mean hopcount
\(\mu(H_{uv})\).

We fit Eq.~\eqref{eq:correlation_factor} to the logarithm of the
measured ratio \(V_Z/V_{uvw}\) at all test positions by least squares
with a soft \(\ell_1\) loss, which limits the influence of outlying
ratios. In our simulations, the fit on all three spaces gives the five
parameters 0.33, 0.26, 0.12, 0.86 and \(-0.42\) of
Eq.~\eqref{eq:correlation_factor}. In our simulations, the fits on two
spaces give the five parameters between 0.31 and 0.36, 0.25 and 0.27,
0.10 and 0.14, 0.83 and 0.90 and \(-0.45\) and \(-0.38\).
The predicted moments in Fig.~\ref{fig:computation} and
Table~\ref{tab:computation} use the parameters fitted on the two other
spaces. In our simulations, Eq.~\eqref{eq:correlation_factor} with the
parameters fitted on the two other spaces gives the coefficient of
determination 0.68 for the measured ratio \(V_Z/V_{uvw}\).

\subsection{Errors}

\begin{figure*}[tb]
\centering
\includegraphics[width=\linewidth]{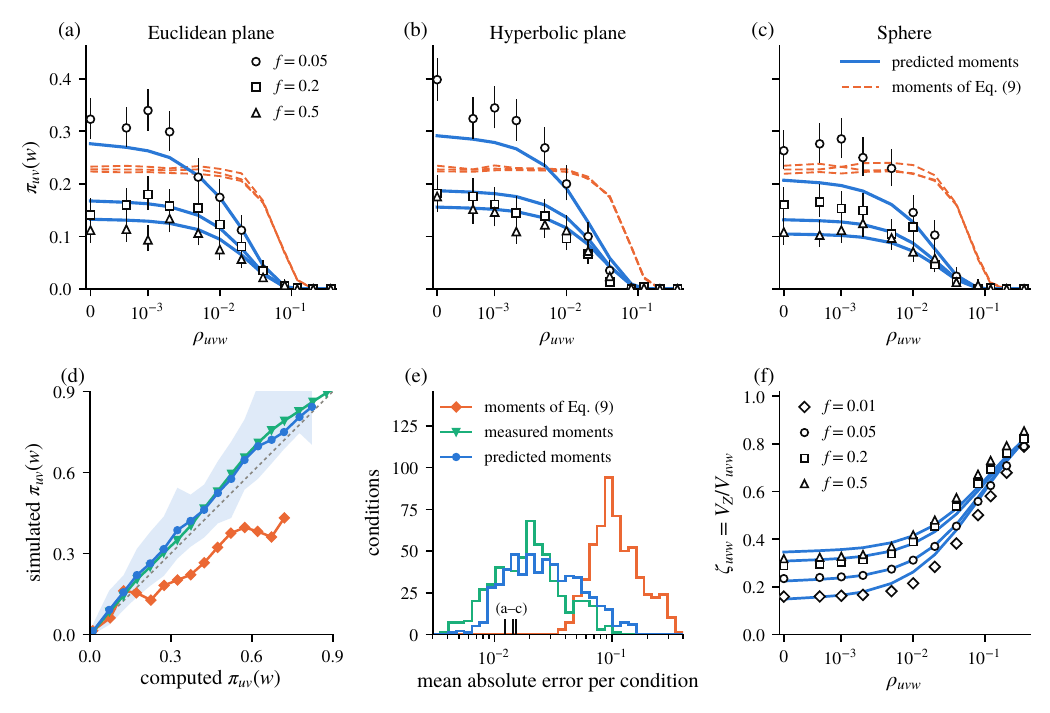}
\caption{%
\textbf{Accuracy of the Gaussian form.}
(a)--(c)~Simulated shortest-path node probability \(\pi_{uv}(w)\)
(symbols) and Eq.~\eqref{eq:interval_estimate} (lines) against the
relative geometric deviation \(\rho_{uvw}\) in soft RGGs with
\(N_b=16{,}000\) nodes, 14 expected neighbors of a node at the center of
the domain and the steepness \(\beta=8\) at the distance \(d_{uv}=1.1\).
The foot of the node \(w\) on the geodesic lies at the fraction \(f\) of
the distance \(d_{uv}\) from the nearer end node. Solid lines use the
predicted moments and dashed lines the moments of the Gaussian
estimate~\eqref{eq:gaussian_membership}. Error bars are 95\% Wilson
intervals. (d)~Mean simulated shortest-path node probability
\(\pi_{uv}(w)\) in bins of width 0.05 of the computed value for the
three choices of the moments at the 167{,}685 test positions. The shade
spans the 10th to the 90th percentile of the simulated value in each bin
of the predicted moments. The dashed line marks equality. (e)~Mean
absolute error of Eq.~\eqref{eq:interval_estimate} for the three choices
of the moments in each of the 567 conditions. Ticks mark the conditions
of (a)--(c). (f)~Correlation factor \(\zeta_{uvw}=V_Z/V_{uvw}\) against
the relative geometric deviation \(\rho_{uvw}\) at four fractions \(f\).
Symbols give the median of the measured ratio over the 189 networks and
lines the median of Eq.~\eqref{eq:correlation_factor} at the same test
positions. On the geodesic, the fraction \(f\) equals the relative
distance \(f\) of Eq.~\eqref{eq:correlation_factor}. The horizontal axes
of (a)--(c) and (f) are linear below \(\rho_{uvw}=5\times10^{-4}\) and
logarithmic above.}
\label{fig:computation}
\end{figure*}

In our simulations of the cross-space experiment,
Eq.~\eqref{eq:interval_estimate} with the moments of the Gaussian
estimate~\eqref{eq:gaussian_membership} has the mean absolute error
0.122 at the 167{,}685 test positions [Fig.~\ref{fig:computation}(d)].
In the same simulations, the measured moments give the mean absolute
error 0.024. In the same simulations, the predicted moments give the
mean absolute error 0.034. In the same simulations, the predicted
moments give a smaller mean absolute error than the moments of the
Gaussian estimate~\eqref{eq:gaussian_membership} in all 567 conditions
[Fig.~\ref{fig:computation}(e)]. Figure~\ref{fig:computation}(a)--(c)
compares the simulated and the computed profiles in one network of each
space. The predicted moments need only the positions of the three nodes,
the mean hopcount and the hopcount variance of node pairs and
Eq.~\eqref{eq:correlation_factor}, so that the computation needs no
shortest path through the node \(w\).

Table~\ref{tab:computation} gives the mean absolute error of
Eq.~\eqref{eq:interval_estimate} by space and by position. In our
simulations, the error is larger near the end nodes than near the
midpoint. In the same simulations, the predicted moments without the
correlation factor, with \(V_Z=V_{uvw}\), give the mean absolute error
0.061. In the same simulations, the mean \(m_Z\) from the linear law
with a constant intercept \(c\) and the predicted variance \(V_Z\) give
the mean absolute error 0.082. In the same simulations, where the
simulated shortest-path node probability \(\pi_{uv}(w)\) is at least
0.05, the median relative error is 0.53 with the moments of the Gaussian
estimate~\eqref{eq:gaussian_membership}, 0.19 with the predicted moments
and 0.14 with the measured moments. In the same simulations, the
frequency of \(Z_{uvw}=0\) in the training realizations predicts the
frequency in the test realizations with the mean absolute error 0.013,
which reflects the sampling noise of the frequencies.

\begin{table}[tb]
\caption{Mean absolute error of Eq.~\eqref{eq:interval_estimate} in our
simulations of the cross-space experiment, with the moments of the
Gaussian estimate~\eqref{eq:gaussian_membership}, the predicted moments
and the measured moments. Near the end nodes, the test position lies at
most 0.05 of the way from the nearer end node along the geodesic. Near
the midpoint, the test position lies between 0.35 and 0.65 of the way.}
\label{tab:computation}
\begin{ruledtabular}
\begin{tabular}{lrccc}
Test positions & Number & Eq.~\eqref{eq:gaussian_membership} & Predicted & Measured\\
\hline
All & 167{,}685 & 0.122 & 0.034 & 0.024\\
Euclidean plane & 42{,}159 & 0.120 & 0.034 & 0.024\\
Hyperbolic plane & 63{,}403 & 0.122 & 0.037 & 0.025\\
Sphere & 62{,}123 & 0.123 & 0.032 & 0.023\\
Near the end nodes & 77{,}110 & 0.133 & 0.045 & 0.029\\
Near the midpoint & 38{,}979 & 0.110 & 0.022 & 0.017\\
\end{tabular}
\end{ruledtabular}
\end{table}

\section{Quantized distribution}
\label{app:quantization}

\subsection{Selection condition}

Let \(K=k^\star(d_{uv})\) be the predicted hopcount of the pair
\((u,v)\) and \(j=k^\star(d_{uw})\) that of the pair \((u,w)\). By the
definition~\eqref{eq:predicted_hopcount},
\begin{equation}
(K-1)\nu<d_{uv}\le K\nu,\qquad (j-1)\nu<d_{uw}\le j\nu.
\label{eq:ceiling_bounds}
\end{equation}
The quantization rule~\eqref{eq:quantization_rule} requires
\(k^\star(d_{wv})=K-j\), that is, \((K-j-1)\nu<d_{wv}\le(K-j)\nu\). The
lower bound always holds, because the triangle inequality and the
bounds~\eqref{eq:ceiling_bounds} give
\(d_{wv}\ge d_{uv}-d_{uw}>(K-j-1)\nu\). The quantization
rule~\eqref{eq:quantization_rule} thus reduces to the upper bound
\(d_{wv}\le(K-j)\nu\). Conversely, the upper bound
\(d_{wv}\le(K-j)\nu\) gives \(d_{uw}\ge d_{uv}-d_{wv}>(j-1)\nu\). The
selected positions with \(k^\star(d_{uw})=j\) are therefore the positions
with \(d_{uw}\le j\nu\) and \(d_{wv}\le(K-j)\nu\), for
\(j=1,\dots,K-1\) (the values \(j=0\) and \(j=K\) give the end nodes).
These positions form the intersection of the disk of radius \(j\nu\)
around \(x_u\) with the disk of radius \((K-j)\nu\) around \(x_v\). In the
limit of Proposition~\ref{prop:quantization}, the node that a shortest
path from the source \(u\) reaches after \(j\) hops lies in the \(j\)th
intersection.

With \(d_{uv}=K\nu-W(d_{uv})\) and \(d_{uw}=j\nu-W(d_{uw})\) from the
definition~\eqref{eq:slack} of the slack and
\(d_{wv}=d_{uv}-d_{uw}+\rho_{uvw}d_{uv}\) from
Eq.~\eqref{eq:relative_deviation}, the upper bound
\(d_{wv}\le(K-j)\nu\) becomes
\begin{equation}
\rho_{uvw}\,d_{uv}+W(d_{uw})\le W(d_{uv}).
\label{eq:selection}
\end{equation}
A node \(w\) is thus selected exactly when the extra distance
\(\rho_{uvw}d_{uv}\) of the detour and the slack \(W(d_{uw})\) together do
not exceed the slack \(W(d_{uv})\).

\subsection{Separate regions}

Because \(W(d_{uw})\ge0\), every selected position has the relative
geometric deviation \(\rho_{uvw}\le W(d_{uv})/d_{uv}<\nu/d_{uv}\). In
Euclidean space, the selected positions thus lie inside an ellipse with
foci \(x_u\) and \(x_v\) (Fig.~\ref{fig:detour}). Because
\(\rho_{uvw}\ge0\), every selected position has
\(W(d_{uw})\le W(d_{uv})\), which with \(d_{uw}=j\nu-W(d_{uw})\) reads
\(j\nu-W(d_{uv})\le d_{uw}\le j\nu\). The \(j\)th intersection thus lies
in a ring around \(x_u\) of width \(W(d_{uv})\). Two neighboring rings are
separated by a ring of width \(\nu-W(d_{uv})>0\) that contains no
selected position. The selected positions therefore form \(K-1\)
separate regions, in which the shortest-path node probability
\(\pi_{uv}(w)\) tends to one, while the shortest-path node probability
\(\pi_{uv}(w)\) tends to zero elsewhere. On the geodesic, where
\(\rho_{uvw}=0\), the selection condition~\eqref{eq:selection} holds at
every position of the rings, so that each region has the width
\(W(d_{uv})\) along the geodesic. In the Euclidean plane, the
intersections of the two disks are lens-shaped petals along the geodesic
[Fig.~\ref{fig:quant_euclid}(a)]. For a small but nonzero variance
\(s_{uv}^2\), the probability \(\Pr[H_{uv}=k]\) in
Eq.~\eqref{eq:unit_bin} passes from one to zero within a window of order
\(s_{uv}\) around each integer value of \(d_{uv}/\nu\). The boundaries of
the regions are then blurred over a distance of order \(\nu s_{uv}\), so
that the transition between the regions and the rest of the space is
narrow.

\subsection{Curved space}

The derivation of the selection condition~\eqref{eq:selection} uses only
the definition~\eqref{eq:predicted_hopcount} and the triangle
inequality. The selection condition~\eqref{eq:selection} and the
separate regions therefore hold in every space with a distance.

On a sphere of radius \(R\), let \(x_{\bar u}\) be the antipode of
\(x_u\). Every position \(x_w\) lies on a great circle through \(x_u\)
and \(x_{\bar u}\), so that \(d_{uw}+d_{w\bar u}=\pi R\). The antipode
\(x_{\bar u}\) lies at the distance \(d_{\bar uv}=\pi R-d_{uv}\) from
\(x_v\). The triangle inequality \(d_{wv}\le d_{w\bar u}+d_{\bar uv}\)
then gives \(d_{uw}+d_{wv}\le2\pi R-d_{uv}\), that is,
\begin{equation}
\rho_{uvw}\,d_{uv}\le2(\pi R-d_{uv}).
\label{eq:sphere_bound}
\end{equation}
By the selection condition~\eqref{eq:selection}, every position with
\(W(d_{uw})\le W(d_{uv})-2(\pi R-d_{uv})\) is selected in every
direction from \(x_u\). When the bound~\eqref{eq:sphere_bound} is
smaller than the slack \(W(d_{uv})\), the selected regions thus
contain full circles around \(x_u\) and close into bands.

\section{Inhomogeneous model network}
\label{app:hetero}

\subsection{Node density and link rule}

The domain is the unit square with open boundaries. The nodes form a
Poisson point process with the node density
\begin{equation}
\lambda(x)=\lambda_0\left[1+\sum_{m=1}^{14}b_m
\exp\!\left(-\frac{\|x-x_m\|^2}{2\sigma_m^2}\right)\right]
\label{eq:app_hetero_density}
\end{equation}
with the background node density \(\lambda_0=100\) and with the centers
\(x_m\), the strengths \(b_m\) and the widths \(\sigma_m\) of
Table~\ref{tab:hubs}. The number of nodes is random. The
node density \(\lambda(x)\) ranges from 379 to 28{,}314 nodes per unit
area. The connection radius is
\begin{equation}
r(x)=\min\left\{0.03,\ \max\left[0.02,\
0.025\left(\frac{\lambda(x)}{\bar\lambda}\right)^{0.2}\right]\right\},
\label{eq:app_hetero_radius}
\end{equation}
where \(\bar\lambda=8{,}777\) is the mean of the node density
\(\lambda(x)\) over the square. Two nodes \(u\) and \(v\) are linked when
\(d_{uv}\le[r(x_u)+r(x_v)]/2\).

\begin{table}[tb]
\caption{Centers \(x_m=(x_m^{(1)},x_m^{(2)})\), strengths \(b_m\) and
widths \(\sigma_m\) of the 14 terms of the node density
\(\lambda(x)\) in Eq.~\eqref{eq:app_hetero_density}.}
\label{tab:hubs}
\begin{ruledtabular}
\begin{tabular}{rrrrr}
\(m\) & \(x_m^{(1)}\) & \(x_m^{(2)}\) & \(b_m\) & \(\sigma_m\)\\
\hline
1&0.05&0.08&75&0.120\\
2&0.10&0.22&80&0.110\\
3&0.15&0.38&85&0.105\\
4&0.20&0.58&105&0.100\\
5&0.30&0.78&160&0.095\\
6&0.50&0.60&110&0.095\\
7&0.65&0.47&100&0.100\\
8&0.75&0.47&150&0.105\\
9&0.86&0.45&105&0.100\\
10&0.92&0.72&80&0.110\\
11&0.96&0.92&60&0.120\\
12&0.70&0.05&60&0.200\\
13&0.90&0.05&60&0.200\\
14&0.10&0.90&60&0.200\\
\end{tabular}
\end{ruledtabular}
\end{table}

\subsection{Velocity field and travel time}

At each position \(x\), the velocity field \(\widehat\nu(x)\) of
Eq.~\eqref{eq:velocity_field} follows from
Eqs.~\eqref{eq:app_FL}--\eqref{eq:app_means} for a hard rule with the
node density \(\lambda(x)\) and the connection radius \(r(x)\). The
velocity field \(\widehat\nu(x)\) ranges from 0.0036 to 0.029. No
simulated shortest path enters the velocity field \(\widehat\nu(x)\).
The travel time is computed on a grid of \(450\times450\) points that
covers the square. Each grid point \(y\) is linked to the grid points
at the offsets \((\pm1,0)\), \((0,\pm1)\), \((\pm1,\pm1)\),
\((\pm2,\pm1)\) and \((\pm1,\pm2)\) with the travel time
\begin{equation}
\frac{2\,\|y'-y\|}{\widehat\nu(y)+\widehat\nu(y')}
\label{eq:app_grid_weight}
\end{equation}
of the link from \(y\) to \(y'\). Dijkstra's algorithm on the grid gives
the travel times from the source \(u\) and from the destination \(v\).
The grid computation divides the velocity field \(\widehat\nu(x)\) by
the maximum of the velocity field, which multiplies all travel times by
one constant and leaves the regions unchanged. Each node takes the
travel times of the nearest grid point.

\subsection{Regions, coverage and pairs}

For each pair \((u,v)\), the threshold \(t_0\) of the travel-time region
\(C^\tau\) is the quantile of the excess travel time \(t_{uvw}\) over
the grid points at the area budget \(A\). The threshold \(\rho_0\) of
the near-geodesic region \(C^E\) is the same quantile of the relative
geometric deviation \(\rho_{uvw}\). The area budgets are
\(A=0.010\), 0.015, \ldots, 0.500. The area \(A_{90}\) of a region is
the fraction of grid points whose excess is at most the
0.9-quantile of the excess at the shortest-path nodes of the pair. The
set \(\mathcal S_{uv}(G)\) of shortest-path nodes contains the end
nodes. Each node counts once.

The simulations use 20 realizations of the network. In each
realization, we take up to 36 anchor nodes in the largest component,
the nodes nearest to the centers of a \(6\times6\) grid of squares. We
use the pairs of anchor nodes with the distance \(d_{uv}\ge0.25\), the
hopcount \(h_{uv}\ge15\) and at least 30 shortest-path nodes. In our
simulations, 10{,}176 pairs meet these conditions. A median over pairs is the median over the pairs
of each realization, followed by the median over the realizations. The
95\% intervals come from 1{,}500 bootstrap samples that resample the
realizations and, within each realization, the anchor nodes.
In our simulations at the area budget \(A=0.05\), the 95\% interval of
the median paired gain \(\Delta\kappa\) is 0.15 to 0.21. In our
simulations, the median ratio \(A_{90}(C^\tau)/A_{90}(C^E)\) over all
pairs is 0.63, with the 95\% interval from 0.61 to 0.65. In our simulations over
the area budgets \(A\) from 0.01 to 0.50, at most 14\% of the pairs have
fewer shortest-path nodes in the travel-time region \(C^\tau\). From
\(A=0.125\) on, the median paired gain \(\Delta\kappa\) in our
simulations is zero, because both regions contain nearly all
shortest-path nodes.

\section{Empirical networks}
\label{app:empirical}

\subsection{Data}

Table~\ref{tab:empirical_size} lists the 22 networks. The 15 urban and
the three rural street networks are drivable streets from
OpenStreetMap, downloaded with OSMnx~\cite{boeing2017osmnx} within a
given distance of a center, except Cairo, which we extracted from the
Geofabrik file of Egypt. The Zuid-Holland network is the drivable
network of the Geofabrik file of the province. The California--Nevada
network comes from the Ninth DIMACS Implementation Challenge, the
power grid from GridKit for Europe and the airline network from
OpenFlights, with one link between two airports that are connected by
at least one flight. Each network is made undirected and without loops or
multiple links. Only the largest connected component is kept. Every
link has weight one. The 18 urban and rural networks use projected
coordinates in meters. The other four networks use great-circle
distances, with local planar coordinates for the directions and the
areas.

\begin{table}[tb]
\caption{Empirical networks. \(\widehat D\) is the largest hopcount found
by four breadth-first searches, \(n_{\rm pair}\) is the number of
sampled pairs \((u,v)\) and \(k\) is the number of nearest nodes over
which the velocity is averaged.}
\label{tab:empirical_size}
\begin{ruledtabular}
\footnotesize
\begin{tabular}{lrrrrr}
Network & Nodes & Links & \(\widehat D\) & \(n_{\rm pair}\) & \(k\)\\
\hline
Amsterdam&18{,}692&26{,}842&79&1{,}110{,}001&29\\
Barcelona&11{,}282&17{,}655&92&1{,}938{,}942&39\\
Bras\'{i}lia&12{,}891&17{,}259&160&3{,}094{,}783&20\\
Buenos Aires&15{,}390&26{,}557&120&1{,}357{,}308&52\\
Cairo&100{,}405&138{,}088&172&285{,}497&36\\
Chicago&11{,}430&18{,}789&99&1{,}850{,}554&35\\
California--Nevada&1{,}890{,}815&2{,}315{,}222&2{,}315&6{,}053&29\\
GridKit Europe&8{,}422&11{,}428&165&2{,}929{,}311&29\\
La Plata&14{,}769&24{,}697&184&2{,}070{,}557&55\\
London&40{,}625&52{,}876&267&460{,}080&31\\
Manhattan&12{,}230&20{,}513&122&2{,}711{,}115&32\\
OpenFlights&3{,}188&18{,}833&12&2{,}228{,}875&128\\
Zuid-Holland&662{,}836&705{,}286&2{,}352&24{,}390&10\\
Paris&24{,}949&37{,}610&143&1{,}117{,}769&33\\
Phoenix&11{,}820&17{,}464&119&3{,}168{,}911&31\\
Rome&23{,}464&33{,}897&161&876{,}805&29\\
Rural Iowa&10{,}606&15{,}342&104&3{,}084{,}175&27\\
Rural Kansas&4{,}031&6{,}133&82&7{,}174{,}141&23\\
Rural Nebraska&2{,}239&3{,}253&76&11{,}250{,}462&8\\
Salt Lake City&9{,}799&13{,}983&97&2{,}257{,}428&36\\
Tokyo&52{,}673&79{,}244&211&540{,}004&39\\
Venice&4{,}118&5{,}483&107&7{,}915{,}828&21\\
\end{tabular}
\end{ruledtabular}
\end{table}

\subsection{Velocity at the nodes}

At a node \(i\), let \(L_i\) be the length of the longest link of the
node \(i\) and let \(P_{i,q}\) be the largest projection
\eqref{eq:app_Pmax} of the links of the node \(i\) on the direction
\(e_q=(\cos q\pi/4,\sin q\pi/4)\), for \(q=0,\ldots,7\). The raw
estimate at the node \(i\) is
\begin{equation}
v_i=\frac12\left(L_i+\frac18\sum_{q=0}^{7}P_{i,q}\right).
\label{eq:app_node_velocity}
\end{equation}
The estimated velocity \(\widehat\nu_i\) is the mean of the raw estimate
\(v_j\) over the \(k\) nodes \(j\) nearest to the node \(i\), including
the node \(i\). The number \(k\) is the median number of nodes within
four times the median link length of a node, rounded and limited to the
range from 8 to 128. The number \(k\) is fixed before any hopcount is
compared.

\subsection{Predictions and errors}

\begin{figure*}[tb]
\centering
\subfloat[Hopcount error.\label{fig:app_empirical_rmse}]{%
\includegraphics[width=0.32\linewidth]{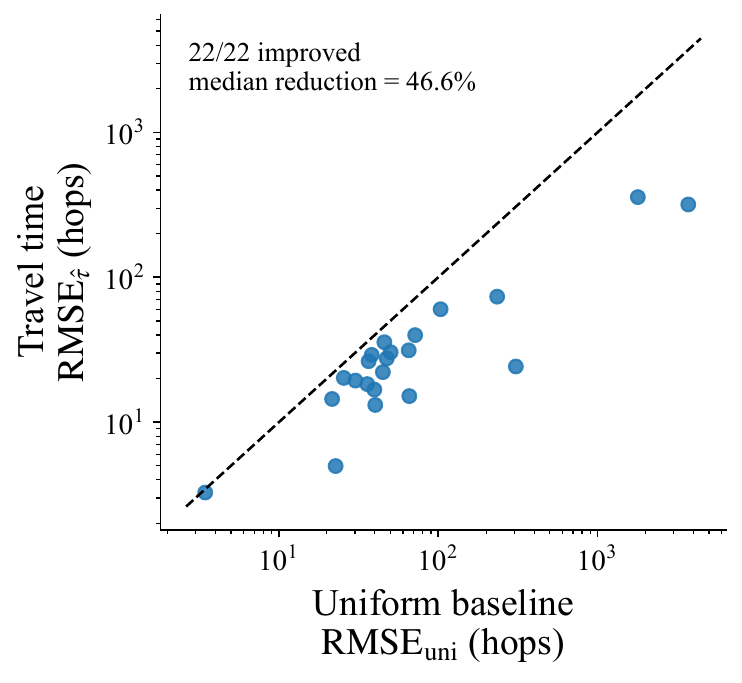}}
\hspace{0.08\linewidth}
\subfloat[Profile deviation.\label{fig:app_empirical_profiles}]{%
\includegraphics[width=0.32\linewidth]{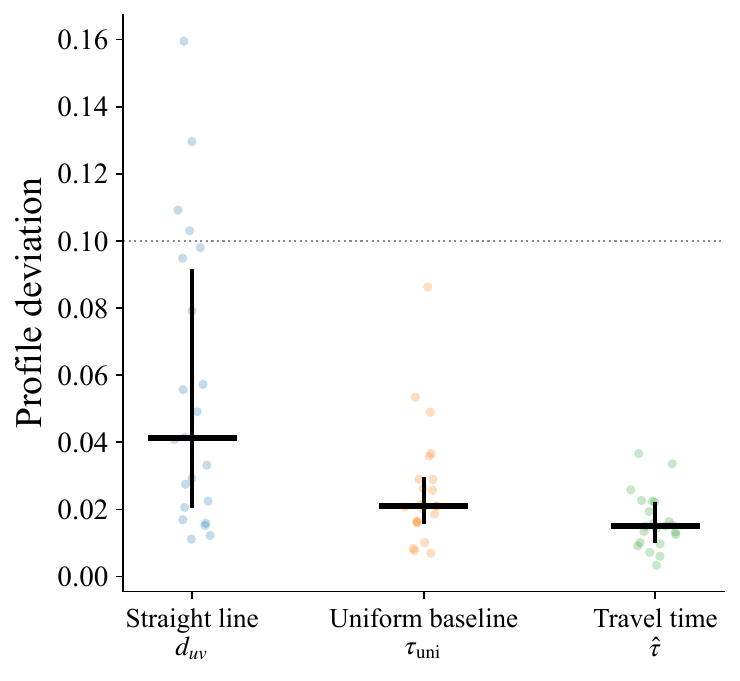}}
\caption{%
\textbf{In our tests on 22 empirical networks, the travel time
\(\widehat\tau\) improves the predicted hopcount and the agreement of
the profiles.} (a) Hopcount error of the travel time \(\widehat\tau\)
against the hopcount error of the uniform baseline \(\tau_{\rm uni}\).
Each point is one network. All points lie below the dashed line of
equal error. (b) Profile deviation in the straight-line distance
\(d_{uv}\), in the uniform baseline \(\tau_{\rm uni}\) and in the travel
time \(\widehat\tau\). Each point is one network. Horizontal bars give
the median and vertical bars the interquartile range over the 22
networks. The dotted line marks the profile deviation 0.10.}
\label{fig:app_empirical}
\end{figure*}

The travel time \(\widehat\tau(u,v)\) is the smallest sum of the link
times \(2d_{ij}/(\widehat\nu_i+\widehat\nu_j)\) along a path from
\(u\) to \(v\). The uniform baseline \(\tau_{\rm uni}(u,v)\) is the
smallest sum of the link lengths \(d_{ij}\) along a path, divided by the
velocity \(\nu_0\), the median of the estimated velocities
\(\widehat\nu_i\). The straight-line prediction is \(d_{uv}/\nu_0\). No
prediction has a constant fitted to hopcounts. For a prediction \(Q\),
the error over the sampled pairs is
\begin{equation}
\mathrm{RMSE}_Q=\left[\frac{1}{n_{\rm pair}}\sum_{(u,v)}
\bigl(Q(u,v)-h_{uv}\bigr)^2\right]^{1/2}.
\label{eq:app_rmse}
\end{equation}
The pairs are sampled with replacement. For each network except
OpenFlights, the target hopcounts are the rounded values of
\(f\widehat D\) for \(f=0.10\), 0.15, \ldots, 0.85, with at least two
hops. For OpenFlights, the target hopcounts are 2 to 11. For a target
hopcount, the source \(u\) is a uniformly random node and the
destination \(v\) is a uniformly random node within one hop of the
target hopcount from the source \(u\). Table~\ref{tab:empirical_performance}
gives the errors of the three predictions. The reduction
\begin{equation}
I_{\mathrm{RMSE}}=1-\frac{\mathrm{RMSE}_{\widehat\tau}}{\mathrm{RMSE}_{\mathrm{uni}}}
\label{eq:rmse_reduction}
\end{equation}
compares the travel time \(\widehat\tau\) with the uniform baseline
\(\tau_{\rm uni}\). In our tests, the travel time \(\widehat\tau\)
predicts the hopcount better than the uniform baseline \(\tau_{\rm uni}\)
in all 22 networks, with the median reduction
\(I_{\mathrm{RMSE}}=46.6\%\) [Fig.~\ref{fig:app_empirical}(a)]. The
uniform baseline \(\tau_{\rm uni}\) and the travel time \(\widehat\tau\)
differ only in the velocity, so that the reduction \(I_{\mathrm{RMSE}}\)
measures the effect of the velocity field \(\widehat\nu_i\). Against the
straight-line prediction \(d_{uv}/\nu_0\), the error in our tests falls
in 16 of the 22 networks, with a median reduction of 10.5\%.

\subsection{Area and coverage}

The bounding rectangle of the network is divided into \(192\times192\)
cells. Each cell that contains nodes has the same weight, shared by at
most four nodes of the cell drawn at random. The area budget \(A\) is a
fraction of the total weight of these cells. For each pair \((u,v)\)
and each of the three distances, the region contains the cells whose
excess \(Q(u,w)+Q(w,v)-Q(u,v)\) at the drawn nodes \(w\) is smallest,
up to the area budget \(A\). The coverage \(\kappa\) counts the end
nodes. For the pairs of Table~\ref{tab:empirical_size}, the coverage
\(\kappa\) of a region in a network is the mean over the pairs. The
medians of these means are taken over the 22 networks. Part of the
paired gain \(\Delta\kappa\) comes from measuring along the links. In
our tests at \(A=0.05\), regions built from the uniform baseline
\(\tau_{\rm uni}\) reach a median coverage \(\kappa\) of 0.60. The estimated velocity
\(\widehat\nu_i\) raises the coverage \(\kappa\) of these regions by a
median of 0.14, again in all 22 tested networks. Equal area does not
mean an equal number of nodes, because the node density of the empirical
networks varies strongly.

\begin{table*}[tb]
\caption{Hopcount errors and coverages \(\kappa\) at the area budget
\(A=0.05\) in our tests. The reductions of the error by the travel time
\(\widehat\tau\) are relative to the uniform baseline and to the
straight-line prediction. The coverages are means over the pairs of
Table~\ref{tab:empirical_size} and the paired gain is
\(\Delta\kappa=\kappa(C^\tau)-\kappa(C^E)\).}
\label{tab:empirical_performance}
\begin{ruledtabular}
\begin{tabular}{lrrrrrrrrr}
 & \multicolumn{3}{c}{RMSE (hops)} & \multicolumn{2}{c}{Reduction (\%)} & \multicolumn{3}{c}{Coverage \(\kappa\)} & \\
Network & Uniform & Straight line & Travel time & Uniform & Straight line & \(C^E\) & Uniform & \(C^\tau\) & \(\Delta\kappa\)\\
\hline
Amsterdam & 65.3 & 43.9 & 31.3 & 52.0 & 28.6 & 0.58 & 0.71 & 0.85 & 0.28\\
Barcelona & 30.2 & 22.3 & 19.4 & 36.0 & 13.1 & 0.46 & 0.56 & 0.71 & 0.25\\
Bras\'{i}lia & 65.8 & 23.1 & 15.1 & 77.0 & 34.6 & 0.41 & 0.77 & 0.90 & 0.48\\
Buenos Aires & 38.3 & 28.5 & 29.1 & 23.9 & $-2.3$ & 0.41 & 0.47 & 0.59 & 0.18\\
Cairo & 234 & 191 & 73.5 & 68.6 & 61.6 & 0.48 & 0.61 & 0.78 & 0.30\\
Chicago & 25.5 & 18.8 & 20.2 & 20.9 & $-7.6$ & 0.49 & 0.56 & 0.65 & 0.16\\
California--Nevada & 3{,}709 & 3{,}097 & 318 & 91.4 & 89.7 & 0.51 & 0.68 & 0.83 & 0.32\\
GridKit Europe & 103 & 45.9 & 60.2 & 41.8 & $-31.1$ & 0.32 & 0.69 & 0.78 & 0.46\\
La Plata & 21.6 & 13.4 & 14.5 & 33.0 & $-8.0$ & 0.53 & 0.60 & 0.74 & 0.21\\
London & 36.6 & 19.3 & 26.4 & 27.9 & $-36.3$ & 0.56 & 0.67 & 0.77 & 0.20\\
Manhattan & 35.9 & 19.5 & 18.3 & 48.9 & 6.0 & 0.41 & 0.59 & 0.70 & 0.29\\
OpenFlights & 3.45 & 3.40 & 3.27 & 5.2 & 3.9 & 0.39 & 0.69 & 0.74 & 0.35\\
Zuid-Holland & 1{,}789 & 1{,}033 & 357 & 80.0 & 65.4 & 0.43 & 0.72 & 1.00 & 0.56\\
Paris & 45.9 & 35.9 & 35.6 & 22.4 & 0.6 & 0.47 & 0.56 & 0.70 & 0.23\\
Phoenix & 47.6 & 25.0 & 27.6 & 42.0 & $-10.4$ & 0.44 & 0.53 & 0.70 & 0.26\\
Rome & 50.3 & 31.0 & 30.4 & 39.5 & 2.0 & 0.50 & 0.63 & 0.80 & 0.30\\
Rural Iowa & 307 & 226 & 24.2 & 92.1 & 89.3 & 0.53 & 0.57 & 0.68 & 0.15\\
Rural Kansas & 39.7 & 21.7 & 16.8 & 57.8 & 22.5 & 0.53 & 0.58 & 0.67 & 0.14\\
Rural Nebraska & 22.7 & 11.1 & 4.98 & 78.1 & 55.1 & 0.50 & 0.61 & 0.78 & 0.27\\
Salt Lake City & 44.9 & 24.1 & 22.2 & 50.7 & 8.0 & 0.47 & 0.61 & 0.78 & 0.31\\
Tokyo & 71.6 & 54.8 & 39.9 & 44.3 & 27.2 & 0.45 & 0.53 & 0.70 & 0.25\\
Venice & 40.2 & 25.9 & 13.1 & 67.3 & 49.3 & 0.38 & 0.54 & 0.80 & 0.42\\
\hline
Median & & & & 46.6 & 10.5 & 0.47 & 0.60 & 0.75 & 0.27\\
\end{tabular}
\end{ruledtabular}
\end{table*}

\subsection{Second sample of pairs}

The pairs of Table~\ref{tab:empirical_size} give the hopcount errors,
the profiles and the coverages of Table~\ref{tab:empirical_performance}
as sums over the pairs. The statistics of single pairs in
Sec.~\ref{sec:empirical} and Fig.~\ref{fig:empirical} come from a
second sample of 1{,}000 distinct pairs per network, drawn with the same
rule and the same target hopcounts. For each pair of the second sample,
we store the coverages \(\kappa\) of the travel-time region \(C^\tau\)
and the near-geodesic region \(C^E\) at every area budget \(A\) and the
areas \(A_{90}\) of both regions. Within each network, the pairs are
divided into thirds of the distance \(d_{uv}\). In our tests, the median
over the networks of the mean coverage \(\kappa\) at \(A=0.05\) is 0.47
for the near-geodesic region \(C^E\) and 0.75 for the travel-time region
\(C^\tau\) in both samples.

\subsection{Profiles}

A uniform sample of at most 40{,}000 nodes of each network gives the
test nodes \(w\). For each of the three coordinates, the relative excess
of the test node \(w\) is the excess
\(Q(u,w)+Q(w,v)-Q(u,v)\) divided by \(Q(u,v)\). Within a window of the
target hopcount and a bin \(b\) of the relative excess, the frequency
\(p_b\) is the fraction of triples \((u,v,w)\) with
\(h_{uw}+h_{wv}=h_{uv}\). To compare the windows, we use the pairs whose
source \(u\) lies in 26 of 64 equal tiles of the bounding rectangle to
measure the mean \(m_b\) and the variance \(V_b\) of the hopcount excess
\(Z_{uvw}\) in each bin \(b\). The frequency \(p_b\) is measured on the
pairs whose source \(u\) lies in the other 38 tiles. With
\(z_b=m_b/\sqrt{V_b}\) and the normal probability
\(g_b=\Phi[(1/2-m_b)/\sqrt{V_b}]-\Phi[(-1/2-m_b)/\sqrt{V_b}]\) of the
value zero, the standardized profile is
\begin{equation}
Y_b=\frac{(p_b/g_b)\,e^{-z_b^2/2}}{(p_{b_0}/g_{b_0})\,e^{-z_{b_0}^2/2}}
\quad\text{against}\quad
\xi_b=\sqrt{z_b^2-z_{b_0}^2},
\label{eq:app_standardized_profile}
\end{equation}
where \(b_0\) is the first bin of the window and \(\Phi\) is the
standard normal distribution function. The profile deviation of a
window is the root-mean-square difference between the standardized
profile \(Y_b\) and the median profile of the network. The windows are
the target hopcounts \(f\widehat D\) with \(f=0.15\), 0.20, \ldots, 0.45
and the target hopcounts 2 to 6 for OpenFlights. The value of a network
in Fig.~\ref{fig:app_empirical}(b) is the median profile deviation over
the windows. In our tests, the shortest-path frequencies of different
pairs in one network agree better in travel-time coordinates. In our
tests, the median profile deviation over the 22 networks is 0.041 for
the straight-line distance \(d_{uv}\), 0.021 for the uniform baseline
\(\tau_{\rm uni}\) and 0.015 for the travel time \(\widehat\tau\)
[Fig.~\ref{fig:app_empirical}(b)].

\footnotesize
\bibliography{apssamp}

\begin{thebibliography}{50}%
\makeatletter
\providecommand \@ifxundefined [1]{%
 \@ifx{#1\undefined}
}%
\providecommand \@ifnum [1]{%
 \ifnum #1\expandafter \@firstoftwo
 \else \expandafter \@secondoftwo
 \fi
}%
\providecommand \@ifx [1]{%
 \ifx #1\expandafter \@firstoftwo
 \else \expandafter \@secondoftwo
 \fi
}%
\providecommand \natexlab [1]{#1}%
\providecommand \enquote  [1]{``#1''}%
\providecommand \bibnamefont  [1]{#1}%
\providecommand \bibfnamefont [1]{#1}%
\providecommand \citenamefont [1]{#1}%
\providecommand \href@noop [0]{\@secondoftwo}%
\providecommand \href [0]{\begingroup \@sanitize@url \@href}%
\providecommand \@href[1]{\@@startlink{#1}\@@href}%
\providecommand \@@href[1]{\endgroup#1\@@endlink}%
\providecommand \@sanitize@url [0]{\catcode `\\12\catcode `\$12\catcode
  `\&12\catcode `\#12\catcode `\^12\catcode `\_12\catcode `\%12\relax}%
\providecommand \@@startlink[1]{}%
\providecommand \@@endlink[0]{}%
\providecommand \url  [0]{\begingroup\@sanitize@url \@url }%
\providecommand \@url [1]{\endgroup\@href {#1}{\urlprefix }}%
\providecommand \urlprefix  [0]{URL }%
\providecommand \Eprint [0]{\href }%
\providecommand \doibase [0]{https://doi.org/}%
\providecommand \selectlanguage [0]{\@gobble}%
\providecommand \bibinfo  [0]{\@secondoftwo}%
\providecommand \bibfield  [0]{\@secondoftwo}%
\providecommand \translation [1]{[#1]}%
\providecommand \BibitemOpen [0]{}%
\providecommand \bibitemStop [0]{}%
\providecommand \bibitemNoStop [0]{.\EOS\space}%
\providecommand \EOS [0]{\spacefactor3000\relax}%
\providecommand \BibitemShut  [1]{\csname bibitem#1\endcsname}%
\let\auto@bib@innerbib\@empty
%</preamble>
\bibitem [{\citenamefont {Homer}(1900)}]{homer1900odyssey}%
  \BibitemOpen
  \bibfield  {author} {\bibinfo {author} {\bibnamefont {Homer}},\ }\href@noop
  {} {\emph {\bibinfo {title} {The {O}dyssey}}}\ (\bibinfo  {publisher}
  {Longmans, Green, and Co.},\ \bibinfo {address} {London},\ \bibinfo {year}
  {1900})\ \bibinfo {note} {translated by Samuel Butler}\BibitemShut {NoStop}%
\bibitem [{\citenamefont {Kurose}\ and\ \citenamefont
  {Ross}(2021)}]{kurose2021computer}%
  \BibitemOpen
  \bibfield  {author} {\bibinfo {author} {\bibfnamefont {J.~F.}\ \bibnamefont
  {Kurose}}\ and\ \bibinfo {author} {\bibfnamefont {K.~W.}\ \bibnamefont
  {Ross}},\ }\href@noop {} {\emph {\bibinfo {title} {Computer Networking: A
  Top-Down Approach}}},\ \bibinfo {edition} {8th}\ ed.\ (\bibinfo  {publisher}
  {Pearson},\ \bibinfo {address} {Hoboken, NJ},\ \bibinfo {year}
  {2021})\BibitemShut {NoStop}%
\bibitem [{\citenamefont {Stamatiou}\ and\ \citenamefont
  {Haenggi}(2014)}]{stamatiou2014delay}%
  \BibitemOpen
  \bibfield  {author} {\bibinfo {author} {\bibfnamefont {K.}~\bibnamefont
  {Stamatiou}}\ and\ \bibinfo {author} {\bibfnamefont {M.}~\bibnamefont
  {Haenggi}},\ }\bibfield  {title} {\bibinfo {title} {Delay characterization of
  multihop transmission in a {Poisson} field of interference},\ }\href@noop {}
  {\bibfield  {journal} {\bibinfo  {journal} {IEEE/ACM Transactions on
  Networking}\ }\textbf {\bibinfo {volume} {22}},\ \bibinfo {pages} {1794}
  (\bibinfo {year} {2014})}\BibitemShut {NoStop}%
\bibitem [{\citenamefont {Malkin}(1998)}]{rfc2453}%
  \BibitemOpen
  \bibfield  {author} {\bibinfo {author} {\bibfnamefont {G.~S.}\ \bibnamefont
  {Malkin}},\ }\href@noop {} {\bibinfo {title} {{RIP} version 2}},\ \bibinfo
  {howpublished} {Request for Comments 2453, Internet Engineering Task Force}
  (\bibinfo {year} {1998})\BibitemShut {NoStop}%
\bibitem [{\citenamefont {Clausen}\ and\ \citenamefont
  {Jacquet}(2003)}]{rfc3626}%
  \BibitemOpen
  \bibfield  {author} {\bibinfo {author} {\bibfnamefont {T.}~\bibnamefont
  {Clausen}}\ and\ \bibinfo {author} {\bibfnamefont {P.}~\bibnamefont
  {Jacquet}},\ }\href@noop {} {\bibinfo {title} {Optimized link state routing
  protocol ({OLSR})}},\ \bibinfo {howpublished} {Request for Comments 3626,
  Internet Engineering Task Force} (\bibinfo {year} {2003})\BibitemShut
  {NoStop}%
\bibitem [{\citenamefont {Dijkstra}(1959)}]{dijkstra1959note}%
  \BibitemOpen
  \bibfield  {author} {\bibinfo {author} {\bibfnamefont {E.~W.}\ \bibnamefont
  {Dijkstra}},\ }\bibfield  {title} {\bibinfo {title} {A note on two problems
  in connexion with graphs},\ }\href {https://doi.org/10.1007/BF01386390}
  {\bibfield  {journal} {\bibinfo  {journal} {Numerische Mathematik}\ }\textbf
  {\bibinfo {volume} {1}},\ \bibinfo {pages} {269} (\bibinfo {year}
  {1959})}\BibitemShut {NoStop}%
\bibitem [{\citenamefont {Ko}\ and\ \citenamefont
  {Vaidya}(2000)}]{ko2000location}%
  \BibitemOpen
  \bibfield  {author} {\bibinfo {author} {\bibfnamefont {Y.-B.}\ \bibnamefont
  {Ko}}\ and\ \bibinfo {author} {\bibfnamefont {N.~H.}\ \bibnamefont
  {Vaidya}},\ }\bibfield  {title} {\bibinfo {title} {Location-aided routing
  ({LAR}) in mobile ad hoc networks},\ }\href@noop {} {\bibfield  {journal}
  {\bibinfo  {journal} {Wireless Networks}\ }\textbf {\bibinfo {volume} {6}},\
  \bibinfo {pages} {307} (\bibinfo {year} {2000})}\BibitemShut {NoStop}%
\bibitem [{\citenamefont {Papadimitriou}\ and\ \citenamefont
  {Yannakakis}(1991)}]{papadimitriou1991shortest}%
  \BibitemOpen
  \bibfield  {author} {\bibinfo {author} {\bibfnamefont {C.~H.}\ \bibnamefont
  {Papadimitriou}}\ and\ \bibinfo {author} {\bibfnamefont {M.}~\bibnamefont
  {Yannakakis}},\ }\bibfield  {title} {\bibinfo {title} {Shortest paths without
  a map},\ }\href@noop {} {\bibfield  {journal} {\bibinfo  {journal}
  {Theoretical Computer Science}\ }\textbf {\bibinfo {volume} {84}},\ \bibinfo
  {pages} {127} (\bibinfo {year} {1991})}\BibitemShut {NoStop}%
\bibitem [{\citenamefont {Gilbert}(1961)}]{gilbert1961random}%
  \BibitemOpen
  \bibfield  {author} {\bibinfo {author} {\bibfnamefont {E.~N.}\ \bibnamefont
  {Gilbert}},\ }\bibfield  {title} {\bibinfo {title} {Random plane networks},\
  }\href@noop {} {\bibfield  {journal} {\bibinfo  {journal} {Journal of the
  Society for Industrial and Applied Mathematics}\ }\textbf {\bibinfo {volume}
  {9}},\ \bibinfo {pages} {533} (\bibinfo {year} {1961})}\BibitemShut {NoStop}%
\bibitem [{\citenamefont {Penrose}(2003)}]{penrose2003random}%
  \BibitemOpen
  \bibfield  {author} {\bibinfo {author} {\bibfnamefont {M.}~\bibnamefont
  {Penrose}},\ }\href@noop {} {\emph {\bibinfo {title} {Random Geometric
  Graphs}}},\ \bibinfo {series} {Oxford Studies in Probability}, Vol.~\bibinfo
  {volume} {5}\ (\bibinfo  {publisher} {Oxford University Press},\ \bibinfo
  {address} {Oxford},\ \bibinfo {year} {2003})\BibitemShut {NoStop}%
\bibitem [{\citenamefont {Dettmann}\ and\ \citenamefont
  {Georgiou}(2016)}]{dettmann2016random}%
  \BibitemOpen
  \bibfield  {author} {\bibinfo {author} {\bibfnamefont {C.~P.}\ \bibnamefont
  {Dettmann}}\ and\ \bibinfo {author} {\bibfnamefont {O.}~\bibnamefont
  {Georgiou}},\ }\bibfield  {title} {\bibinfo {title} {Random geometric graphs
  with general connection functions},\ }\href@noop {} {\bibfield  {journal}
  {\bibinfo  {journal} {Physical Review E}\ }\textbf {\bibinfo {volume} {93}},\
  \bibinfo {pages} {032313} (\bibinfo {year} {2016})}\BibitemShut {NoStop}%
\bibitem [{\citenamefont {Penrose}(2016)}]{penrose2016connectivity}%
  \BibitemOpen
  \bibfield  {author} {\bibinfo {author} {\bibfnamefont {M.~D.}\ \bibnamefont
  {Penrose}},\ }\bibfield  {title} {\bibinfo {title} {Connectivity of soft
  random geometric graphs},\ }\href {https://doi.org/10.1214/15-AAP1110}
  {\bibfield  {journal} {\bibinfo  {journal} {The Annals of Applied
  Probability}\ }\textbf {\bibinfo {volume} {26}},\ \bibinfo {pages} {986}
  (\bibinfo {year} {2016})}\BibitemShut {NoStop}%
\bibitem [{\citenamefont {Yao}\ \emph {et~al.}(2011)\citenamefont {Yao},
  \citenamefont {Chen},\ and\ \citenamefont {Guo}}]{yao2011large}%
  \BibitemOpen
  \bibfield  {author} {\bibinfo {author} {\bibfnamefont {C.-L.}\ \bibnamefont
  {Yao}}, \bibinfo {author} {\bibfnamefont {G.}~\bibnamefont {Chen}},\ and\
  \bibinfo {author} {\bibfnamefont {T.-D.}\ \bibnamefont {Guo}},\ }\bibfield
  {title} {\bibinfo {title} {Large deviations for the graph distance in
  supercritical continuum percolation},\ }\href
  {https://doi.org/10.1239/jap/1300198142} {\bibfield  {journal} {\bibinfo
  {journal} {Journal of Applied Probability}\ }\textbf {\bibinfo {volume}
  {48}},\ \bibinfo {pages} {154} (\bibinfo {year} {2011})}\BibitemShut
  {NoStop}%
\bibitem [{\citenamefont {D{\'\i}az}\ \emph {et~al.}(2016)\citenamefont
  {D{\'\i}az}, \citenamefont {Mitsche}, \citenamefont {Perarnau},\ and\
  \citenamefont {P{\'e}rez-Gim{\'e}nez}}]{diaz2016relation}%
  \BibitemOpen
  \bibfield  {author} {\bibinfo {author} {\bibfnamefont {J.}~\bibnamefont
  {D{\'\i}az}}, \bibinfo {author} {\bibfnamefont {D.}~\bibnamefont {Mitsche}},
  \bibinfo {author} {\bibfnamefont {G.}~\bibnamefont {Perarnau}},\ and\
  \bibinfo {author} {\bibfnamefont {X.}~\bibnamefont {P{\'e}rez-Gim{\'e}nez}},\
  }\bibfield  {title} {\bibinfo {title} {On the relation between graph distance
  and {Euclidean} distance in random geometric graphs},\ }\href
  {https://doi.org/10.1017/apr.2016.31} {\bibfield  {journal} {\bibinfo
  {journal} {Advances in Applied Probability}\ }\textbf {\bibinfo {volume}
  {48}},\ \bibinfo {pages} {848} (\bibinfo {year} {2016})}\BibitemShut
  {NoStop}%
\bibitem [{\citenamefont {Hirsch}\ \emph {et~al.}(2015)\citenamefont {Hirsch},
  \citenamefont {Neuh{\"a}user}, \citenamefont {Gloaguen},\ and\ \citenamefont
  {Schmidt}}]{hirsch2015first}%
  \BibitemOpen
  \bibfield  {author} {\bibinfo {author} {\bibfnamefont {C.}~\bibnamefont
  {Hirsch}}, \bibinfo {author} {\bibfnamefont {D.}~\bibnamefont
  {Neuh{\"a}user}}, \bibinfo {author} {\bibfnamefont {C.}~\bibnamefont
  {Gloaguen}},\ and\ \bibinfo {author} {\bibfnamefont {V.}~\bibnamefont
  {Schmidt}},\ }\bibfield  {title} {\bibinfo {title} {First passage percolation
  on random geometric graphs and an application to shortest-path trees},\
  }\href {https://doi.org/10.1239/aap/1435236978} {\bibfield  {journal}
  {\bibinfo  {journal} {Advances in Applied Probability}\ }\textbf {\bibinfo
  {volume} {47}},\ \bibinfo {pages} {328} (\bibinfo {year} {2015})}\BibitemShut
  {NoStop}%
\bibitem [{\citenamefont {L{\"u}chtrath}(2024)}]{luchtrath2024all}%
  \BibitemOpen
  \bibfield  {author} {\bibinfo {author} {\bibfnamefont {L.}~\bibnamefont
  {L{\"u}chtrath}},\ }\href@noop {} {\bibinfo {title} {All spatial random
  graphs with weak long-range effects have chemical distance comparable to
  {Euclidean} distance}} (\bibinfo {year} {2024}),\ \Eprint
  {https://arxiv.org/abs/2412.12796} {arXiv:2412.12796 [math.PR]} \BibitemShut
  {NoStop}%
\bibitem [{\citenamefont {Biskup}(2004)}]{biskup2004scaling}%
  \BibitemOpen
  \bibfield  {author} {\bibinfo {author} {\bibfnamefont {M.}~\bibnamefont
  {Biskup}},\ }\bibfield  {title} {\bibinfo {title} {On the scaling of the
  chemical distance in long-range percolation models},\ }\href
  {https://doi.org/10.1214/009117904000000577} {\bibfield  {journal} {\bibinfo
  {journal} {The Annals of Probability}\ }\textbf {\bibinfo {volume} {32}},\
  \bibinfo {pages} {2938} (\bibinfo {year} {2004})}\BibitemShut {NoStop}%
\bibitem [{\citenamefont {Freeman}(1977)}]{freeman1977betweenness}%
  \BibitemOpen
  \bibfield  {author} {\bibinfo {author} {\bibfnamefont {L.~C.}\ \bibnamefont
  {Freeman}},\ }\bibfield  {title} {\bibinfo {title} {A set of measures of
  centrality based on betweenness},\ }\href@noop {} {\bibfield  {journal}
  {\bibinfo  {journal} {Sociometry}\ }\textbf {\bibinfo {volume} {40}},\
  \bibinfo {pages} {35} (\bibinfo {year} {1977})}\BibitemShut {NoStop}%
\bibitem [{\citenamefont {Brandes}(2001)}]{brandes2001faster}%
  \BibitemOpen
  \bibfield  {author} {\bibinfo {author} {\bibfnamefont {U.}~\bibnamefont
  {Brandes}},\ }\bibfield  {title} {\bibinfo {title} {A faster algorithm for
  betweenness centrality},\ }\href@noop {} {\bibfield  {journal} {\bibinfo
  {journal} {The Journal of Mathematical Sociology}\ }\textbf {\bibinfo
  {volume} {25}},\ \bibinfo {pages} {163} (\bibinfo {year} {2001})}\BibitemShut
  {NoStop}%
\bibitem [{\citenamefont {Barth{\'e}lemy}(2004)}]{barthelemy2004betweenness}%
  \BibitemOpen
  \bibfield  {author} {\bibinfo {author} {\bibfnamefont {M.}~\bibnamefont
  {Barth{\'e}lemy}},\ }\bibfield  {title} {\bibinfo {title} {Betweenness
  centrality in large complex networks},\ }\href@noop {} {\bibfield  {journal}
  {\bibinfo  {journal} {The European Physical Journal B}\ }\textbf {\bibinfo
  {volume} {38}},\ \bibinfo {pages} {163} (\bibinfo {year} {2004})}\BibitemShut
  {NoStop}%
\bibitem [{\citenamefont {Kirkley}\ \emph {et~al.}(2018)\citenamefont
  {Kirkley}, \citenamefont {Barbosa}, \citenamefont {Barthelemy},\ and\
  \citenamefont {Ghoshal}}]{kirkley2018betweenness}%
  \BibitemOpen
  \bibfield  {author} {\bibinfo {author} {\bibfnamefont {A.}~\bibnamefont
  {Kirkley}}, \bibinfo {author} {\bibfnamefont {H.}~\bibnamefont {Barbosa}},
  \bibinfo {author} {\bibfnamefont {M.}~\bibnamefont {Barthelemy}},\ and\
  \bibinfo {author} {\bibfnamefont {G.}~\bibnamefont {Ghoshal}},\ }\bibfield
  {title} {\bibinfo {title} {From the betweenness centrality in street networks
  to structural invariants in random planar graphs},\ }\href@noop {} {\bibfield
   {journal} {\bibinfo  {journal} {Nature Communications}\ }\textbf {\bibinfo
  {volume} {9}},\ \bibinfo {pages} {2501} (\bibinfo {year} {2018})}\BibitemShut
  {NoStop}%
\bibitem [{\citenamefont {Kartun-Giles}\ \emph {et~al.}(2019)\citenamefont
  {Kartun-Giles}, \citenamefont {Barthelemy},\ and\ \citenamefont
  {Dettmann}}]{kartun2019shape}%
  \BibitemOpen
  \bibfield  {author} {\bibinfo {author} {\bibfnamefont {A.~P.}\ \bibnamefont
  {Kartun-Giles}}, \bibinfo {author} {\bibfnamefont {M.}~\bibnamefont
  {Barthelemy}},\ and\ \bibinfo {author} {\bibfnamefont {C.~P.}\ \bibnamefont
  {Dettmann}},\ }\bibfield  {title} {\bibinfo {title} {Shape of shortest paths
  in random spatial networks},\ }\href@noop {} {\bibfield  {journal} {\bibinfo
  {journal} {Physical Review E}\ }\textbf {\bibinfo {volume} {100}},\ \bibinfo
  {pages} {032315} (\bibinfo {year} {2019})}\BibitemShut {NoStop}%
\bibitem [{\citenamefont {Van~Mieghem}(2014)}]{vanmieghem2014performance}%
  \BibitemOpen
  \bibfield  {author} {\bibinfo {author} {\bibfnamefont {P.}~\bibnamefont
  {Van~Mieghem}},\ }\href@noop {} {\emph {\bibinfo {title} {Performance
  Analysis of Complex Networks and Systems}}}\ (\bibinfo  {publisher}
  {Cambridge University Press},\ \bibinfo {address} {Cambridge},\ \bibinfo
  {year} {2014})\BibitemShut {NoStop}%
\bibitem [{\citenamefont {Barth{\'e}lemy}(2011)}]{barthelemy2011spatial}%
  \BibitemOpen
  \bibfield  {author} {\bibinfo {author} {\bibfnamefont {M.}~\bibnamefont
  {Barth{\'e}lemy}},\ }\bibfield  {title} {\bibinfo {title} {Spatial
  networks},\ }\href@noop {} {\bibfield  {journal} {\bibinfo  {journal}
  {Physics Reports}\ }\textbf {\bibinfo {volume} {499}},\ \bibinfo {pages} {1}
  (\bibinfo {year} {2011})}\BibitemShut {NoStop}%
\bibitem [{\citenamefont {Barth{\'e}lemy}(2018)}]{barthelemy2018morphogenesis}%
  \BibitemOpen
  \bibfield  {author} {\bibinfo {author} {\bibfnamefont {M.}~\bibnamefont
  {Barth{\'e}lemy}},\ }\href@noop {} {\emph {\bibinfo {title} {Morphogenesis of
  Spatial Networks}}},\ Lecture Notes in Morphogenesis\ (\bibinfo  {publisher}
  {Springer},\ \bibinfo {address} {Cham},\ \bibinfo {year} {2018})\BibitemShut
  {NoStop}%
\bibitem [{\citenamefont {Kleinberg}(2000)}]{kleinberg2000navigation}%
  \BibitemOpen
  \bibfield  {author} {\bibinfo {author} {\bibfnamefont {J.~M.}\ \bibnamefont
  {Kleinberg}},\ }\bibfield  {title} {\bibinfo {title} {Navigation in a small
  world},\ }\href@noop {} {\bibfield  {journal} {\bibinfo  {journal} {Nature}\
  }\textbf {\bibinfo {volume} {406}},\ \bibinfo {pages} {845} (\bibinfo {year}
  {2000})}\BibitemShut {NoStop}%
\bibitem [{\citenamefont {Bogu{\~n}{\'a}}\ \emph {et~al.}(2009)\citenamefont
  {Bogu{\~n}{\'a}}, \citenamefont {Krioukov},\ and\ \citenamefont
  {Claffy}}]{boguna2009navigability}%
  \BibitemOpen
  \bibfield  {author} {\bibinfo {author} {\bibfnamefont {M.}~\bibnamefont
  {Bogu{\~n}{\'a}}}, \bibinfo {author} {\bibfnamefont {D.}~\bibnamefont
  {Krioukov}},\ and\ \bibinfo {author} {\bibfnamefont {K.~C.}\ \bibnamefont
  {Claffy}},\ }\bibfield  {title} {\bibinfo {title} {Navigability of complex
  networks},\ }\href@noop {} {\bibfield  {journal} {\bibinfo  {journal} {Nature
  Physics}\ }\textbf {\bibinfo {volume} {5}},\ \bibinfo {pages} {74} (\bibinfo
  {year} {2009})}\BibitemShut {NoStop}%
\bibitem [{\citenamefont {Bogu{\~n}{\'a}}\ \emph {et~al.}(2010)\citenamefont
  {Bogu{\~n}{\'a}}, \citenamefont {Papadopoulos},\ and\ \citenamefont
  {Krioukov}}]{boguna2010sustaining}%
  \BibitemOpen
  \bibfield  {author} {\bibinfo {author} {\bibfnamefont {M.}~\bibnamefont
  {Bogu{\~n}{\'a}}}, \bibinfo {author} {\bibfnamefont {F.}~\bibnamefont
  {Papadopoulos}},\ and\ \bibinfo {author} {\bibfnamefont {D.}~\bibnamefont
  {Krioukov}},\ }\bibfield  {title} {\bibinfo {title} {Sustaining the
  {I}nternet with hyperbolic mapping},\ }\href@noop {} {\bibfield  {journal}
  {\bibinfo  {journal} {Nature Communications}\ }\textbf {\bibinfo {volume}
  {1}},\ \bibinfo {pages} {62} (\bibinfo {year} {2010})}\BibitemShut {NoStop}%
\bibitem [{\citenamefont {Krioukov}\ \emph {et~al.}(2010)\citenamefont
  {Krioukov}, \citenamefont {Papadopoulos}, \citenamefont {Kitsak},
  \citenamefont {Vahdat},\ and\ \citenamefont
  {Bogu{\~n}{\'a}}}]{krioukov2010hyperbolic}%
  \BibitemOpen
  \bibfield  {author} {\bibinfo {author} {\bibfnamefont {D.}~\bibnamefont
  {Krioukov}}, \bibinfo {author} {\bibfnamefont {F.}~\bibnamefont
  {Papadopoulos}}, \bibinfo {author} {\bibfnamefont {M.}~\bibnamefont
  {Kitsak}}, \bibinfo {author} {\bibfnamefont {A.}~\bibnamefont {Vahdat}},\
  and\ \bibinfo {author} {\bibfnamefont {M.}~\bibnamefont {Bogu{\~n}{\'a}}},\
  }\bibfield  {title} {\bibinfo {title} {Hyperbolic geometry of complex
  networks},\ }\href@noop {} {\bibfield  {journal} {\bibinfo  {journal}
  {Physical Review E}\ }\textbf {\bibinfo {volume} {82}},\ \bibinfo {pages}
  {036106} (\bibinfo {year} {2010})}\BibitemShut {NoStop}%
\bibitem [{\citenamefont {Bogu{\~n}{\'a}}\ \emph {et~al.}(2021)\citenamefont
  {Bogu{\~n}{\'a}}, \citenamefont {Bonamassa}, \citenamefont {De~Domenico},
  \citenamefont {Havlin}, \citenamefont {Krioukov},\ and\ \citenamefont
  {Serrano}}]{boguna2021network}%
  \BibitemOpen
  \bibfield  {author} {\bibinfo {author} {\bibfnamefont {M.}~\bibnamefont
  {Bogu{\~n}{\'a}}}, \bibinfo {author} {\bibfnamefont {I.}~\bibnamefont
  {Bonamassa}}, \bibinfo {author} {\bibfnamefont {M.}~\bibnamefont
  {De~Domenico}}, \bibinfo {author} {\bibfnamefont {S.}~\bibnamefont {Havlin}},
  \bibinfo {author} {\bibfnamefont {D.}~\bibnamefont {Krioukov}},\ and\
  \bibinfo {author} {\bibfnamefont {M.~{\'A}.}\ \bibnamefont {Serrano}},\
  }\bibfield  {title} {\bibinfo {title} {Network geometry},\ }\href@noop {}
  {\bibfield  {journal} {\bibinfo  {journal} {Nature Reviews Physics}\ }\textbf
  {\bibinfo {volume} {3}},\ \bibinfo {pages} {114} (\bibinfo {year}
  {2021})}\BibitemShut {NoStop}%
\bibitem [{\citenamefont {Serrano}\ and\ \citenamefont
  {Bogu{\~n}{\'a}}(2022)}]{serrano2022shortest}%
  \BibitemOpen
  \bibfield  {author} {\bibinfo {author} {\bibfnamefont {M.~{\'A}.}\
  \bibnamefont {Serrano}}\ and\ \bibinfo {author} {\bibfnamefont
  {M.}~\bibnamefont {Bogu{\~n}{\'a}}},\ }\href@noop {} {\emph {\bibinfo {title}
  {The Shortest Path to Network Geometry: A Practical Guide to Basic Models and
  Applications}}},\ Elements in the Structure and Dynamics of Complex Networks\
  (\bibinfo  {publisher} {Cambridge University Press},\ \bibinfo {address}
  {Cambridge},\ \bibinfo {year} {2022})\BibitemShut {NoStop}%
\bibitem [{\citenamefont {Kitsak}\ \emph {et~al.}(2023)\citenamefont {Kitsak},
  \citenamefont {Ganin}, \citenamefont {Elmokashfi}, \citenamefont {Cui},
  \citenamefont {Eisenberg}, \citenamefont {Alderson}, \citenamefont {Korkin},\
  and\ \citenamefont {Linkov}}]{kitsak2023finding}%
  \BibitemOpen
  \bibfield  {author} {\bibinfo {author} {\bibfnamefont {M.}~\bibnamefont
  {Kitsak}}, \bibinfo {author} {\bibfnamefont {A.}~\bibnamefont {Ganin}},
  \bibinfo {author} {\bibfnamefont {A.}~\bibnamefont {Elmokashfi}}, \bibinfo
  {author} {\bibfnamefont {H.}~\bibnamefont {Cui}}, \bibinfo {author}
  {\bibfnamefont {D.~A.}\ \bibnamefont {Eisenberg}}, \bibinfo {author}
  {\bibfnamefont {D.~L.}\ \bibnamefont {Alderson}}, \bibinfo {author}
  {\bibfnamefont {D.}~\bibnamefont {Korkin}},\ and\ \bibinfo {author}
  {\bibfnamefont {I.}~\bibnamefont {Linkov}},\ }\bibfield  {title} {\bibinfo
  {title} {Finding shortest and nearly shortest path nodes in large
  substantially incomplete networks by hyperbolic mapping},\ }\href@noop {}
  {\bibfield  {journal} {\bibinfo  {journal} {Nature Communications}\ }\textbf
  {\bibinfo {volume} {14}},\ \bibinfo {pages} {186} (\bibinfo {year}
  {2023})}\BibitemShut {NoStop}%
\bibitem [{\citenamefont {Qiu}\ \emph {et~al.}(2026)\citenamefont {Qiu},
  \citenamefont {Balogh}, \citenamefont {Liu}, \citenamefont {Van~Mieghem},\
  and\ \citenamefont {Kitsak}}]{qiu2026geometric}%
  \BibitemOpen
  \bibfield  {author} {\bibinfo {author} {\bibfnamefont {Z.}~\bibnamefont
  {Qiu}}, \bibinfo {author} {\bibfnamefont {S.~G.}\ \bibnamefont {Balogh}},
  \bibinfo {author} {\bibfnamefont {X.}~\bibnamefont {Liu}}, \bibinfo {author}
  {\bibfnamefont {P.}~\bibnamefont {Van~Mieghem}},\ and\ \bibinfo {author}
  {\bibfnamefont {M.}~\bibnamefont {Kitsak}},\ }\bibfield  {title} {\bibinfo
  {title} {Geometric organization and inference of shortest path nodes in soft
  random geometric graphs},\ }\href {https://doi.org/10.1103/dq2j-z8k6}
  {\bibfield  {journal} {\bibinfo  {journal} {Physical Review Research}\
  }\textbf {\bibinfo {volume} {8}},\ \bibinfo {pages} {033149} (\bibinfo {year}
  {2026})}\BibitemShut {NoStop}%
\bibitem [{\citenamefont {Cannistraci}\ and\ \citenamefont
  {Muscoloni}(2022)}]{cannistraci2022geometrical}%
  \BibitemOpen
  \bibfield  {author} {\bibinfo {author} {\bibfnamefont {C.~V.}\ \bibnamefont
  {Cannistraci}}\ and\ \bibinfo {author} {\bibfnamefont {A.}~\bibnamefont
  {Muscoloni}},\ }\bibfield  {title} {\bibinfo {title} {Geometrical congruence,
  greedy navigability and myopic transfer in complex networks and brain
  connectomes},\ }\href {https://doi.org/10.1038/s41467-022-34634-6} {\bibfield
   {journal} {\bibinfo  {journal} {Nature Communications}\ }\textbf {\bibinfo
  {volume} {13}},\ \bibinfo {pages} {7308} (\bibinfo {year}
  {2022})}\BibitemShut {NoStop}%
\bibitem [{\citenamefont {Verbavatz}\ and\ \citenamefont
  {Barthelemy}(2022)}]{verbavatz2022betweenness}%
  \BibitemOpen
  \bibfield  {author} {\bibinfo {author} {\bibfnamefont {V.}~\bibnamefont
  {Verbavatz}}\ and\ \bibinfo {author} {\bibfnamefont {M.}~\bibnamefont
  {Barthelemy}},\ }\bibfield  {title} {\bibinfo {title} {Betweenness centrality
  in dense spatial networks},\ }\href
  {https://doi.org/10.1103/PhysRevE.105.054303} {\bibfield  {journal} {\bibinfo
   {journal} {Physical Review E}\ }\textbf {\bibinfo {volume} {105}},\ \bibinfo
  {pages} {054303} (\bibinfo {year} {2022})}\BibitemShut {NoStop}%
\bibitem [{\citenamefont {Born}\ and\ \citenamefont
  {Wolf}(1999)}]{born1999principles}%
  \BibitemOpen
  \bibfield  {author} {\bibinfo {author} {\bibfnamefont {M.}~\bibnamefont
  {Born}}\ and\ \bibinfo {author} {\bibfnamefont {E.}~\bibnamefont {Wolf}},\
  }\href@noop {} {\emph {\bibinfo {title} {Principles of Optics}}},\ \bibinfo
  {edition} {7th}\ ed.\ (\bibinfo  {publisher} {Cambridge University Press},\
  \bibinfo {address} {Cambridge},\ \bibinfo {year} {1999})\BibitemShut
  {NoStop}%
\bibitem [{\citenamefont {Sethian}(1996)}]{sethian1996fast}%
  \BibitemOpen
  \bibfield  {author} {\bibinfo {author} {\bibfnamefont {J.~A.}\ \bibnamefont
  {Sethian}},\ }\bibfield  {title} {\bibinfo {title} {A fast marching level set
  method for monotonically advancing fronts},\ }\href@noop {} {\bibfield
  {journal} {\bibinfo  {journal} {Proceedings of the National Academy of
  Sciences}\ }\textbf {\bibinfo {volume} {93}},\ \bibinfo {pages} {1591}
  (\bibinfo {year} {1996})}\BibitemShut {NoStop}%
\bibitem [{\citenamefont {Sethian}(1999)}]{sethian1999level}%
  \BibitemOpen
  \bibfield  {author} {\bibinfo {author} {\bibfnamefont {J.~A.}\ \bibnamefont
  {Sethian}},\ }\href@noop {} {\emph {\bibinfo {title} {Level Set Methods and
  Fast Marching Methods: Evolving Interfaces in Computational Geometry, Fluid
  Mechanics, Computer Vision, and Materials Science}}},\ \bibinfo {edition}
  {2nd}\ ed.,\ \bibinfo {series} {Cambridge Monographs on Applied and
  Computational Mathematics}, Vol.~\bibinfo {volume} {3}\ (\bibinfo
  {publisher} {Cambridge University Press},\ \bibinfo {address} {Cambridge},\
  \bibinfo {year} {1999})\BibitemShut {NoStop}%
\bibitem [{\citenamefont {Van~Mieghem}(2023)}]{vanmieghem2023graph}%
  \BibitemOpen
  \bibfield  {author} {\bibinfo {author} {\bibfnamefont {P.}~\bibnamefont
  {Van~Mieghem}},\ }\href@noop {} {\emph {\bibinfo {title} {Graph Spectra for
  Complex Networks}}},\ \bibinfo {edition} {2nd}\ ed.\ (\bibinfo  {publisher}
  {Cambridge University Press},\ \bibinfo {address} {Cambridge},\ \bibinfo
  {year} {2023})\BibitemShut {NoStop}%
\bibitem [{\citenamefont {Haenggi}\ \emph {et~al.}(2009)\citenamefont
  {Haenggi}, \citenamefont {Andrews}, \citenamefont {Baccelli}, \citenamefont
  {Dousse},\ and\ \citenamefont {Franceschetti}}]{haenggi2009stochastic}%
  \BibitemOpen
  \bibfield  {author} {\bibinfo {author} {\bibfnamefont {M.}~\bibnamefont
  {Haenggi}}, \bibinfo {author} {\bibfnamefont {J.~G.}\ \bibnamefont
  {Andrews}}, \bibinfo {author} {\bibfnamefont {F.}~\bibnamefont {Baccelli}},
  \bibinfo {author} {\bibfnamefont {O.}~\bibnamefont {Dousse}},\ and\ \bibinfo
  {author} {\bibfnamefont {M.}~\bibnamefont {Franceschetti}},\ }\bibfield
  {title} {\bibinfo {title} {Stochastic geometry and random graphs for the
  analysis and design of wireless networks},\ }\href@noop {} {\bibfield
  {journal} {\bibinfo  {journal} {IEEE Journal on Selected Areas in
  Communications}\ }\textbf {\bibinfo {volume} {27}},\ \bibinfo {pages} {1029}
  (\bibinfo {year} {2009})}\BibitemShut {NoStop}%
\bibitem [{\citenamefont {Benjamini}\ \emph {et~al.}(2003)\citenamefont
  {Benjamini}, \citenamefont {Kalai},\ and\ \citenamefont
  {Schramm}}]{benjamini2003first}%
  \BibitemOpen
  \bibfield  {author} {\bibinfo {author} {\bibfnamefont {I.}~\bibnamefont
  {Benjamini}}, \bibinfo {author} {\bibfnamefont {G.}~\bibnamefont {Kalai}},\
  and\ \bibinfo {author} {\bibfnamefont {O.}~\bibnamefont {Schramm}},\
  }\bibfield  {title} {\bibinfo {title} {First passage percolation has
  sublinear distance variance},\ }\href@noop {} {\bibfield  {journal} {\bibinfo
   {journal} {The Annals of Probability}\ }\textbf {\bibinfo {volume} {31}},\
  \bibinfo {pages} {1970} (\bibinfo {year} {2003})}\BibitemShut {NoStop}%
\bibitem [{\citenamefont {Damron}\ \emph {et~al.}(2015)\citenamefont {Damron},
  \citenamefont {Hanson},\ and\ \citenamefont {Sosoe}}]{damron2015sublinear}%
  \BibitemOpen
  \bibfield  {author} {\bibinfo {author} {\bibfnamefont {M.}~\bibnamefont
  {Damron}}, \bibinfo {author} {\bibfnamefont {J.}~\bibnamefont {Hanson}},\
  and\ \bibinfo {author} {\bibfnamefont {P.}~\bibnamefont {Sosoe}},\ }\bibfield
   {title} {\bibinfo {title} {Sublinear variance in first-passage percolation
  for general distributions},\ }\href@noop {} {\bibfield  {journal} {\bibinfo
  {journal} {Probability Theory and Related Fields}\ }\textbf {\bibinfo
  {volume} {163}},\ \bibinfo {pages} {223} (\bibinfo {year}
  {2015})}\BibitemShut {NoStop}%
\bibitem [{\citenamefont {Dembin}(2024)}]{dembin2024variance}%
  \BibitemOpen
  \bibfield  {author} {\bibinfo {author} {\bibfnamefont {B.}~\bibnamefont
  {Dembin}},\ }\bibfield  {title} {\bibinfo {title} {The variance of the graph
  distance in the infinite cluster of percolation is sublinear},\ }\href
  {https://doi.org/10.30757/ALEA.v21-13} {\bibfield  {journal} {\bibinfo
  {journal} {ALEA Latin American Journal of Probability and Mathematical
  Statistics}\ }\textbf {\bibinfo {volume} {21}},\ \bibinfo {pages} {307}
  (\bibinfo {year} {2024})}\BibitemShut {NoStop}%
\bibitem [{\citenamefont {Auffinger}\ \emph {et~al.}(2017)\citenamefont
  {Auffinger}, \citenamefont {Damron},\ and\ \citenamefont
  {Hanson}}]{auffinger2017fifty}%
  \BibitemOpen
  \bibfield  {author} {\bibinfo {author} {\bibfnamefont {A.}~\bibnamefont
  {Auffinger}}, \bibinfo {author} {\bibfnamefont {M.}~\bibnamefont {Damron}},\
  and\ \bibinfo {author} {\bibfnamefont {J.}~\bibnamefont {Hanson}},\
  }\href@noop {} {\emph {\bibinfo {title} {50 Years of First-Passage
  Percolation}}},\ \bibinfo {series} {University Lecture Series}, Vol.~\bibinfo
  {volume} {68}\ (\bibinfo  {publisher} {American Mathematical Society},\
  \bibinfo {address} {Providence, RI},\ \bibinfo {year} {2017})\BibitemShut
  {NoStop}%
\bibitem [{\citenamefont {Newton}(1726)}]{newton1726principia}%
  \BibitemOpen
  \bibfield  {author} {\bibinfo {author} {\bibfnamefont {I.}~\bibnamefont
  {Newton}},\ }\href@noop {} {\emph {\bibinfo {title} {Philosophi{\ae}
  Naturalis Principia Mathematica}}},\ \bibinfo {edition} {3rd}\ ed.\ (\bibinfo
   {publisher} {Guil. \& Joh. Innys},\ \bibinfo {address} {London},\ \bibinfo
  {year} {1726})\BibitemShut {NoStop}%
\bibitem [{\citenamefont {Tang}\ \emph {et~al.}(2008)\citenamefont {Tang},
  \citenamefont {Wang},\ and\ \citenamefont {Van~Mieghem}}]{tang2008effect}%
  \BibitemOpen
  \bibfield  {author} {\bibinfo {author} {\bibfnamefont {S.}~\bibnamefont
  {Tang}}, \bibinfo {author} {\bibfnamefont {H.}~\bibnamefont {Wang}},\ and\
  \bibinfo {author} {\bibfnamefont {P.}~\bibnamefont {Van~Mieghem}},\
  }\bibfield  {title} {\bibinfo {title} {The effect of peer selection with
  hopcount or delay constraint on peer-to-peer networking},\ }in\ \href
  {https://doi.org/10.1007/978-3-540-79549-0_31} {\emph {\bibinfo {booktitle}
  {NETWORKING 2008 Ad Hoc and Sensor Networks, Wireless Networks, Next
  Generation Internet}}},\ \bibinfo {series} {Lecture Notes in Computer
  Science}, Vol.\ \bibinfo {volume} {4982}\ (\bibinfo  {publisher} {Springer},\
  \bibinfo {address} {Berlin, Heidelberg},\ \bibinfo {year} {2008})\ pp.\
  \bibinfo {pages} {358--365}\BibitemShut {NoStop}%
\bibitem [{\citenamefont {Pinto}\ \emph {et~al.}(2012)\citenamefont {Pinto},
  \citenamefont {Thiran},\ and\ \citenamefont {Vetterli}}]{pinto2012locating}%
  \BibitemOpen
  \bibfield  {author} {\bibinfo {author} {\bibfnamefont {P.~C.}\ \bibnamefont
  {Pinto}}, \bibinfo {author} {\bibfnamefont {P.}~\bibnamefont {Thiran}},\ and\
  \bibinfo {author} {\bibfnamefont {M.}~\bibnamefont {Vetterli}},\ }\bibfield
  {title} {\bibinfo {title} {Locating the source of diffusion in large-scale
  networks},\ }\href@noop {} {\bibfield  {journal} {\bibinfo  {journal}
  {Physical Review Letters}\ }\textbf {\bibinfo {volume} {109}},\ \bibinfo
  {pages} {068702} (\bibinfo {year} {2012})}\BibitemShut {NoStop}%
\bibitem [{\citenamefont
  {Chatterjee}(2014)}]{chatterjee2014superconcentration}%
  \BibitemOpen
  \bibfield  {author} {\bibinfo {author} {\bibfnamefont {S.}~\bibnamefont
  {Chatterjee}},\ }\href@noop {} {\emph {\bibinfo {title} {Superconcentration
  and Related Topics}}},\ Springer Monographs in Mathematics\ (\bibinfo
  {publisher} {Springer},\ \bibinfo {address} {Cham},\ \bibinfo {year}
  {2014})\BibitemShut {NoStop}%
\bibitem [{\citenamefont {Ahlberg}\ \emph {et~al.}(2023)\citenamefont
  {Ahlberg}, \citenamefont {Deijfen},\ and\ \citenamefont
  {Sfragara}}]{ahlberg2023chaos}%
  \BibitemOpen
  \bibfield  {author} {\bibinfo {author} {\bibfnamefont {D.}~\bibnamefont
  {Ahlberg}}, \bibinfo {author} {\bibfnamefont {M.}~\bibnamefont {Deijfen}},\
  and\ \bibinfo {author} {\bibfnamefont {M.}~\bibnamefont {Sfragara}},\
  }\href@noop {} {\bibinfo {title} {Chaos, concentration and multiple valleys
  in first-passage percolation}} (\bibinfo {year} {2023}),\ \Eprint
  {https://arxiv.org/abs/2302.11367} {arXiv:2302.11367 [math.PR]} \BibitemShut
  {NoStop}%
\bibitem [{\citenamefont {Boeing}(2017)}]{boeing2017osmnx}%
  \BibitemOpen
  \bibfield  {author} {\bibinfo {author} {\bibfnamefont {G.}~\bibnamefont
  {Boeing}},\ }\bibfield  {title} {\bibinfo {title} {{OSMnx}: New methods for
  acquiring, constructing, analyzing, and visualizing complex street
  networks},\ }\href {https://doi.org/10.1016/j.compenvurbsys.2017.05.004}
  {\bibfield  {journal} {\bibinfo  {journal} {Computers, Environment and Urban
  Systems}\ }\textbf {\bibinfo {volume} {65}},\ \bibinfo {pages} {126}
  (\bibinfo {year} {2017})}\BibitemShut {NoStop}%
\end{thebibliography}%

\end{document}